\RequirePackage{fix-cm}
\documentclass[smallextended]{svjour3}       
\smartqed  
\usepackage{graphicx}
\usepackage{bm}
\usepackage{caption}
\usepackage{amsmath}
\usepackage{amssymb}  
\usepackage{mathtools}
\usepackage{mathptmx}
\usepackage{subcaption}
\usepackage[table]{xcolor} 
\usepackage{float} 

\usepackage[colorlinks=true, linkcolor=blue, citecolor=blue, urlcolor=blue]{hyperref}
\journalname{Nonlinear Dynamics}

\begin{document}

\title{Data-Driven Prediction of Chaotic Transition in Periapsis Poincaré Maps}







\author{
Shanshan Pan\textsuperscript{*\dag} 
\and Taiki Urashi\textsuperscript{*}
\and Mai Bando\textsuperscript{\dag} 
\and Yasuhiro Yoshimura
\and Hongru Chen
\and Toshiya Hanada
}

\institute{
\textsuperscript{*}These authors contributed equally.\\
\textsuperscript{\dag}Corresponding authors: \email{shanshan.pan@aero.kyushu-u.ac.jp,mbando@aero.kyushu-u.ac.jp} \\
Shanshan Pan, Taiki Urashi, Mai Bando, Yasuhiro Yoshimura, Toshiya Hanada,  \at
Department of Aeronautics and Astronautics, Kyushu University, 819-0395, Japan
\and
Hongru Chen \at
Department of Mechanical and Aerospace Engineering, University of Texas at Arlington, TX 76019, United States
}

\date{Received: date / Accepted: date}

\maketitle

\begin{abstract}

Chaotic trajectories in multi-body dynamical systems play a crucial role in designing low-energy trajectories in astrodynamics. However, predicting these trajectories is inherently difficult, as small errors in initial conditions can grow exponentially, making long-term predictions unreliable. This study introduces a novel methodology using Dynamic Mode Decomposition (DMD) to predict chaotic transitions in the periapsis Poincaré map of the circular restricted three-body problem (CRTBP). Unlike standard DMD approaches that model continuous equations of motion, the proposed method approximates deformations in a low-dimensional Poincaré map, enabling trajectory prediction and revealing transition structures. Two approaches are developed: the Local Deformation Map-based DMD (LDMD) and the Global Deformation Map-based DMD (GDMD). LDMD constructs discrete maps to track local deformations of periapsis sets, while GDMD captures global deformations using widely distributed data. A key advantage of this framework is that it approximates nonlinear chaotic transport using a linear operator, which enables fast prediction of periapsis evolution via matrix powers and direct access to geometric structures. To validate the proposed method, the deformation map is applied to design ballistic transfer trajectories to the Moon using a targeting strategy, demonstrating its practical relevance in astrodynamics. This work highlights the potential of data-driven modeling to bridge chaotic dynamics with systematic trajectory design.

\end{abstract}

\section{Introduction}\label{intro}

Traditional interplanetary trajectory design has long relied on the patched conic method, which decomposes a multi-body system into a series of two-body Keplerian arcs~\cite{Prussing2013}. While computationally efficient, this method fails to account for the complex gravitational interactions present in multi-body regimes, especially when long-duration, low-energy trajectories are desired. In recent decades, the use of low-energy, non-Keplerian trajectories, particularly those influenced by chaotic dynamics, has proved advantageous for lunar and interplanetary missions~\cite{belbruno1993sun,koon2000dynamical,koon2001low}. These trajectories leverage the intricate geometry of phase space in multi-body systems, enabling fuel-efficient transfers that exploit natural dynamical channels such as invariant manifolds and resonant gravity assists.

Poincaré maps provide a powerful framework for capturing the global structure of chaotic transport in the circular restricted three-body problem (CRTBP) and have been extensively applied in previous studies~\cite{Koon2000,Gomez2001,howell2001temporary}. To further support trajectory design, alternative representations are often desirable. One such approach defines the surface of section at periapsis, yielding a map expressed directly in position space. This construction, known as the periapsis Poincaré map (PPM), was first introduced by Villac and Scheeres~\cite{Villac2003} and later extended in subsequent studies~\cite{Paskowitz2006A,Paskowitz2006B}. By discretizing phase space at successive periapsis events, PPMs reduce the dimensionality of the system and reveal key dynamical features, including resonance structures, invariant manifolds, and lobe dynamics~\cite{ross2007multiple,koon2000dynamical,hiraiwa2024designing}. Their applications in astrodynamics are diverse: PPMs have enabled the design of trajectories exploiting resonant gravity assists and chaotic transport mechanisms~\cite{belbruno2008resonance,ross2007multiple}, supporting complex mission scenarios in the Earth–Moon system~\cite{dellnitz2005transport,topputo2005earth,bonasera2021applying,urashi2021aas}. Beyond spacecraft mission design, they have also been employed to investigate the long-term chaotic evolution of small bodies, such as comets and asteroids~\cite{liu1994chaotic,malyshkin1999keplerian,chirikov1989chaotic,boekholt2016origin}.

However, predicting the evolution of chaotic trajectories on the PPM remains a formidable challenge due to the sensitive dependence on initial conditions, where small perturbations can lead to exponentially diverging outcomes. \textcolor{black}{Crucially, while PPMs provide a powerful visualization of chaotic transport, their direct use in trajectory design is often limited by a computational bottleneck. In the absence of an analytical representation of the discrete periapsis-to-periapsis mapping, state transitions must be evaluated through repeated numerical integrations, which is computationally expensive.}

Recent advances in data-driven modeling have enabled the prediction of complex, high-dimensional dynamics without requiring explicit knowledge of the governing equations~\cite{pathak2017using,HOFMANN2025314}. 
This study focuses on Dynamic Mode Decomposition (DMD), a projection-based technique that provides fully data-driven linear approximations of nonlinear dynamics~\cite{kutz2016dynamic}. 
DMD constructs finite-dimensional approximations of the linear operator from measurement snapshots, capturing nonlinear behavior through linear matrix factorization and separating dynamic patterns according to intrinsic timescales. The method enables rapid prediction of state evolution via matrix powers and directly identifies geometric structures such as transport boundaries and manifold separatrices, offering insight into underlying mechanisms of resonance and scattering. Its versatility has been demonstrated across diverse applications~\cite{grosek2014,kutz2015,Wustner2022,Wustner2025}. 

In this study, a novel DMD-based framework is introduced to model and predict chaotic transitions on the PPM within the CRTBP. 
\textcolor{black}{In data-driven modeling of spacecraft dynamics, the primary objective is typically to approximate or predict the evolution of individual trajectories originating from given initial conditions, using models constructed from trajectory data~\cite{HOFMANN2025314,servadio2022dynamics,servadio2023koopman}.
In contrast, chaotic transport on the PPM is not governed by the evolution of isolated trajectories, but by the collective deformation of nearby states in phase space. Transport boundaries and separatrices emerge from how ensembles of periapsis points stretch, fold, and separate across distinct dynamical regions.
When standard pointwise applications of DMD are employed in the vicinity of such transport boundaries, trajectories associated with distinct transport behaviors are simultaneously approximated by a single linear operator.
As a result, the least-squares solution averages over locally incompatible dynamics, leading to an inability to accurately capture separation structures near separatrices.
To address this limitation, the present approach reformulates DMD by explicitly modeling the deformation of a set of periapsis points. In this formulation, the consistency of the mapping across a neighborhood is implicitly enforced in the least-squares identification of the operator.}

\textcolor{black}{Based on this formulation, two complementary approaches are introduced: the Local Deformation Map-based DMD (LDMD), which resolves fine-scale deformations within localized regions of phase space, and the Global Deformation Map-based DMD (GDMD), which captures large-scale transport patterns across the phase space. Both approaches employ the same DMD-based set-deformation mapping algorithm and differ only in the spatial extent and sampling density of the data used to construct the linear operator.
LDMD is constructed from densely sampled data within selected local regions, enabling high-resolution characterization of local phase-space deformation and making it particularly suited for precise trajectory targeting. In contrast, GDMD is constructed using sparsely sampled data distributed over a wide region of phase space, allowing it to efficiently capture global transport structures and making it suitable for preliminary mission design and the identification of large-scale transport patterns.} 
These discrete models not only enable finite-time prediction of chaotic transitions but also offer geometric insight into the underlying transport structures. The proposed methodology is further demonstrated through the design of a ballistic lunar transfer by targeting a specific region within the PPM, thereby underscoring its relevance for practical trajectory optimization. This work thus advances a new data-driven paradigm for probing and exploiting chaotic dynamics in space mission design.

\section{Data-Driven Modeling Procedure for Chaotic Trajectories}\label{sec_II}

This section first introduces the dynamical system, followed by its key dynamical structures, including periodic orbits and invariant manifolds. Next, the periapsis Poincaré map used in this study is presented. Finally, the data-driven modeling method is described.

\subsection{The Planar Circular Restricted Three-Body Problem}\label{sec_II_crtbp}

The planar Circular Restricted Three-Body Problem (PCRTBP) describes the motion of an infinitesimal-mass object, $P_0$, moving under the gravitational influence of two massive bodies: the primary $P_1$ with mass $m_1$, and the secondary $P_2$ with mass $m_2$ ($m_2 < m_1$). These primaries revolve in circular orbits around their common center of mass. In this study, the Earth-Moon system is considered, where $P_1$ represents the Earth and $P_2$ the Moon. The PCRTBP is formulated in a dimensionless, planar rotating coordinate system where the distance between the primaries is normalized to one, the total mass is unity, and the system rotates with unit angular velocity. The equations of motion for $P_0$ are given by~\cite{Prussing2013}:

\begin{equation}\label{eq_crtbp_eoms}
\left\{
\begin{aligned}
\ddot x - 2\dot{y} &= \frac{\partial U}{\partial x} \\
\ddot y + 2\dot{x} &= \frac{\partial U}{\partial y}
\end{aligned}
\right.
\end{equation}
where $U$ is the effective potential:
\begin{equation}
U = \frac{1}{2}(x^2 + y^2) + \frac{1 - \mu}{r_1} + \frac{\mu}{r_2} + \frac{1}{2} \mu (1 - \mu) \label{eq_crtbp_eff_potential}
\end{equation}
in which $\mu$ denotes the mass ratio, defined as $\mu=m_2/(m_1+m_2)$. In PCRTBP, the Earth and Moon are fixed at $(-\mu, 0)$ and $(1-\mu,0)$, respectively, with $r_1$ and $r_2$ denoting the distances of the spacecraft from them, given by:
\begin{equation}
r_1 = \sqrt{(x+\mu)^2 + y^2}, \quad
r_2 = \sqrt{(x-1+\mu)^2 + y^2} \label{eq_crtbp_r1_r2}
\end{equation}

The PCRTBP admits an integral of motion, the Jacobi constant, given by:
\begin{align}
	C= 2U - (\dot{x}^2+\dot{y}^2) \label{eq_crtbp_jc}
\end{align}

From the definition of the Jacobi constant, the zero velocity curves (ZVCs) are obtained by the condition $2U-C=0$.

\subsection{Periodic Orbits and Invariant Manifolds}\label{sec_II_crtbp_po}

\begin{figure}[htbp]
    \centering
    \begin{minipage}[t]{0.48\textwidth}
        \centering
        \includegraphics[width=\textwidth]{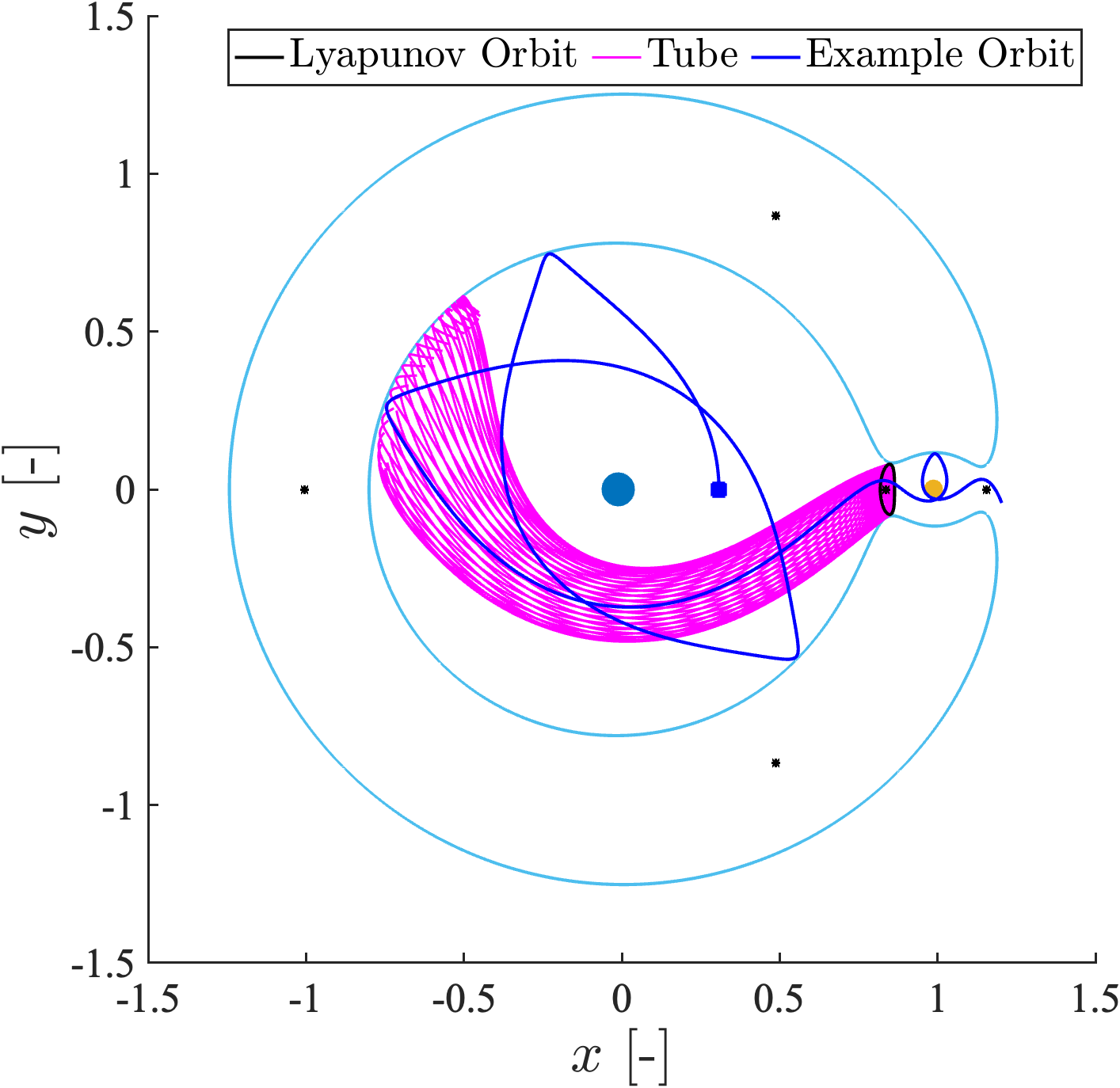}
        \subcaption{Lyapunov orbit near $L_1$, showing stable manifolds (magenta) and an example trajectory (blue) passing through the manifold.}
        \label{fig_tube}
    \end{minipage}
    \hfill
    \begin{minipage}[t]{0.48\textwidth}
        \centering
        \includegraphics[width=\textwidth]{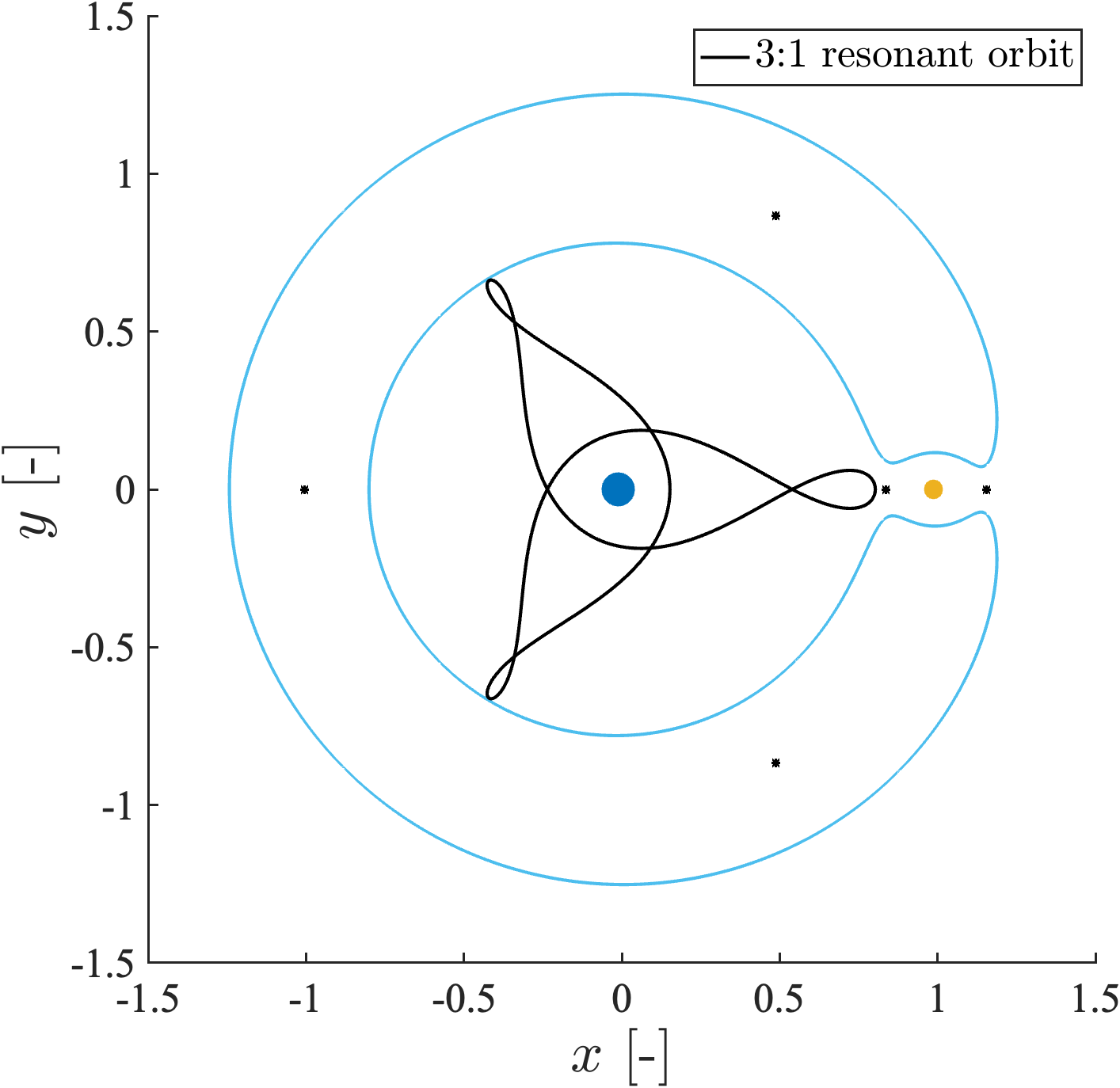}
        \subcaption{Example resonant orbit with a 3:1 resonance ratio.}
        \label{fig_reso31}
    \end{minipage}
    \caption{Dynamical structures in the Earth–Moon PCRTBP (rotating frame) for Jacobi constant $C$=3.172602661563305: Earth (blue), Moon (yellow), Lagrange points (black), and zero-velocity curves (cyan).}\label{fig_crtbp_orbits}
\end{figure}

The PCRTBP has ﬁve equilibrium points, known as Lagrange points. Three of these points---$L_1$, $L_2$ and $L_3$---are collinear. $L_1$ and $L_2$ lie near the Moon. This study introduces data-driven techniques for reconstructing transport structures associated with the \( L_1 \) region without explicitly relying on the Lyapunov orbit or its analytical manifold representation. These manifolds, or "tubes," govern the flow in phase space, separating transit orbits (inside the tube) from non-transit orbits. The stable manifolds consist of trajectories that asymptotically approach the Lyapunov orbit in forward time, while the unstable manifolds consist of trajectories that approach it in backward time. These manifolds serve as a transport pathway between the Earth and Moon regions in planar motion~\cite{koon2001low}. Figure~\ref{fig_crtbp_orbits} illustrates some example dynamical structures for a fixed Jacobi constant $C$ in the Earth-Moon PCRTBP, where the Earth is shown as a blue point, the Moon as a yellow point, and the five Lagrange points as black. The ZVCs are depicted in cyan. Specifically, Fig.~\ref{fig_tube} highlights the Lyapunov orbit near $L_1$, with its stable manifolds in magenta and an example trajectory traversing the tube in blue. The narrow passage around the $L_1$ and $L_2$, enclosed by the ZVCs, forms the "neck" region, allowing trajectories to transition between the Earth and the Moon. All states within the stable manifold pass through the neck region before reaching the Moon region, making it essential for low-energy trajectory design.

To enter the Moon region via the tube structure, a spacecraft must first reach the altitude at which the tube exists. This can be achieved using resonant gravity assists~\cite{ross2007multiple,campagnola2012three}. In a resonant orbit, the spacecraft’s orbital period maintains a simple integer ratio $p:q$ with the Moon's orbital period, known as the resonance ratio \cite{belbruno2008resonance}. Here, $p:q$ indicates that the spacecraft completes $p$ orbits around the Earth for every $q$ orbits of the Moon. This relationship enables multiple gravity assists over a short period, gradually increasing the spacecraft's semi-major axis, raising its energy to the required level for tube entry, and facilitating transfer to the Moon region. Figure~\ref{fig_reso31} shows an example resonant orbit with a 3:1 resonant orbit.

\subsection{Periapsis Poincaré Map}\label{sec_II_crtbp_ppm}

In the PCRTBP, the combination of a fixed Jacobi constant $C$ and the periapsis surface of section enables a reduction of the four-dimensional phase space to a two-dimensional representation, referred to as the Periapsis Poincaré Map (PPM)~\cite{Villac2003,Haapala2011}. The periapsis, defined as the closest point of the trajectory to the primary body (Earth in this study), occurs when the radial velocity vanishes ($\dot{r} = 0$) and the radial acceleration is positive ($\ddot{r} > 0$). In the PPM, periapsis states are parameterized by the pair $(\theta, a)$, where $a$ is the semi-major axis and $\theta \in [-\pi, \pi]$ denotes the phase angle in the rotating frame, measured from the positive $x$-line pointing from Earth to the Moon. Here, $\theta = 0$ corresponds to the positive $x$-line, $\theta > 0$ indicates angles in the direction of the Moon, and $\theta < 0$ represents angles opposite to the Moon direction. For a given $(\theta, a)$ and fixed $C$, the corresponding eccentricity $e$ is found by solving the Jacobi integral equation through numerical iteration~\cite{belbruno2008resonance}. The corresponding Cartesian state at periapsis is computed by:
\begin{align}\label{crtbp_atheta_xy_transformation}
    \left[ \begin{array}{cccc}
    x\\
    y\\
    \dot{x}\\
    \dot{y} \end{array}
  \right]=\left[
    \begin{array}{cccc}r_p\cos{\theta}-\mu\\
    r_p\sin{\theta}\\
    (r_p-v_p)\sin{\theta}\\
    (v_p-r_p)\cos{\theta} \end{array}
  \right]
\end{align}
where $r_p$ and $v_p$ are the position and velocity at periapsis, given by:
\begin{equation}
    \begin{aligned}\label{eq_ppm_rp_vp}
        r_p=a(1-e), \quad  v_p=\sqrt{\cfrac{1-\mu}{a}\left(\cfrac{1+e}{1-e}\right)} 
    \end{aligned}
\end{equation}

\begin{figure}[h!]
    \centering
    \includegraphics[width=0.85\linewidth,clip]{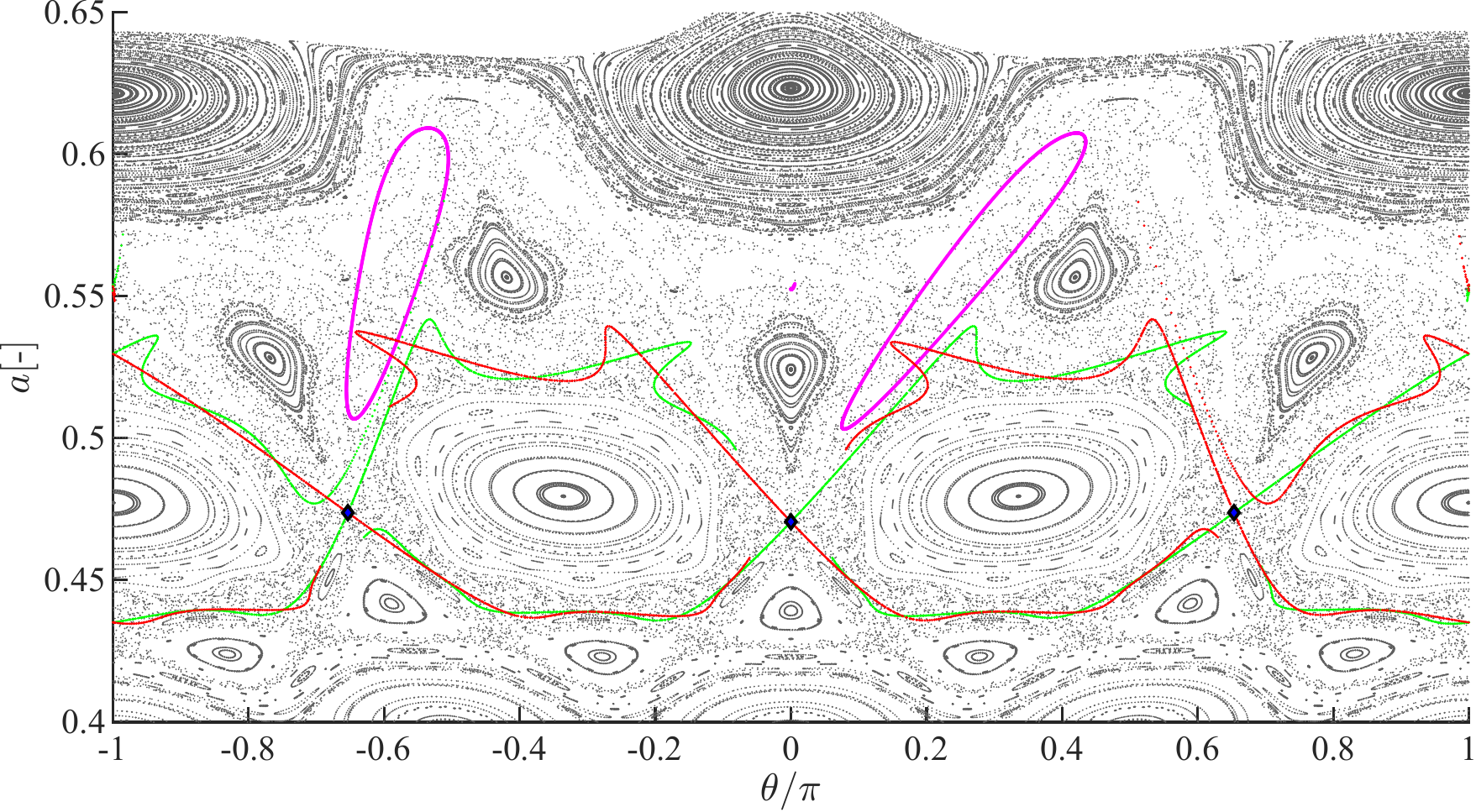}
    \caption{The PPM for Jacobi constant $C=3.172602661563305$.}\label{fig_ppm_Jacobi_candidate}
\end{figure}

Figure~\ref{fig_ppm_Jacobi_candidate} shows the PPM for $C=3.172602661563305$. Black dots represent the periapsis from 1000 randomly sampled trajectories, revealing torus (quasi-periodic motion) and chaotic regions (where periapsis parameters vary significantly). The magenta points, forming closed curves, represent the first (left) and second (right) perigees of the stable manifold emanating from the $L_1$ Lyapunov orbit in Fig.~\ref{fig_tube}, indicating the region of influence for low-energy transfers. The green and red points represent the stable and unstable manifolds of the 3:1 resonant orbit shown in Fig.~\ref{fig_reso31}, respectively. The overlapping region between these manifolds forms a "lobe" structure, which governs the transport between resonance zones~\cite{Scott2010,Oshima2015}. The lobe boundaries, associated with homoclinic and heteroclinic tangles, play a fundamental role in enabling chaotic transitions across phase-space barriers~\cite{dellnitz2005transport,wiggins2013chaotic,hiraiwa2024designing}. These dynamics are critical for adjusting a spacecraft’s semi-major axis through lunar gravity assists.

\subsection{Dynamic Mode Decomposition}\label{sec_II_dmd}

Dynamic Mode Decomposition (DMD) is a data-driven technique for extracting coherent spatiotemporal structures from nonlinear dynamical systems using time-series data~\cite{kutz2016dynamic}. By approximating the nonlinear evolution with a linear model, DMD enables analysis, prediction, and control via classical linear systems theory. Specifically, DMD assumes that the system state of a discrete dynamical system $\bm{x}_k$ evolves linearly to $\bm{x}_{k+1}$ through an operator $A$:
\begin{align}\label{eq_dmd_xk1_A_xk}
\bm{x}_{k+1}=A\bm{x}_k
\end{align}
where $\bm{x}_k \in \mathbb{R}^{n}$ is the system state at time step $k$, and $A \in \mathbb{R}^{n\times n}$ is the linear operator estimated from time-series data. In this study, DMD is applied to sequences of periapsis events. Each $\bm{x}_k$ is defined as a stacked vector of periapsis state:
\begin{align}\label{eq_dmd_xk_atheta}
        {\bm{x}_{k}} &= {\left[
    \begin{array}{c}
      \theta^1\\
      a^1 \\
      \theta^2\\
      a^2 \\
      \vdots\\
      \theta^n\\
      a^n \\
      \end{array}
  \right]_k}
\end{align}
Here, $(\theta^i, a^i)$ denotes the phase angle and semi-major axis of the $i$-th orbit at the $k$-th periapsis. The full state $\bm{x}_k$ thus represents the periapsis configuration of all $n$ orbits at step $k$. Time-series data are collected by sampling periapsis events and arranged into two data matrices:
\begin{align}\label{eq_dmd_X_Xprime}
    X = \left[
    \begin{array}{cccc}
      \mid & \mid & \quad & \mid \\
      \bm{x}_1 & \bm{x}_2 & \ldots & \bm{x}_{m-1} \\
      \mid & \mid & \quad & \mid
    \end{array}
  \right], \qquad 
    X' = \left[
    \begin{array}{cccc}
      \mid & \mid & \quad & \mid \\
      \bm{x}_2 & \bm{x}_3 & \ldots & \bm{x}_m \\
      \mid & \mid & \quad & \mid
    \end{array}
  \right]
\end{align}

\begin{figure}[htbp]
    \centering
    \includegraphics[width=0.5\linewidth,clip]{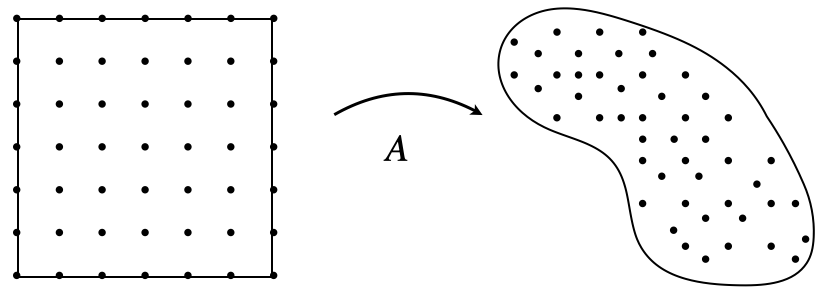}
    \caption{Schematic representation of the discrete map $A$, illustrating the deformation of the periapsis set.}\label{risanzu}
\end{figure}

Assuming the linear relation $\bm{x}_{k+1} = A\bm{x}_k$, the best-fit operator $A$ satisfies:
\begin{align}
X' = AX 
\end{align} 
and is computed via least-squares as:
\begin{align}
A = X' X^\dagger\label{eq_dmd_compute_A}
\end{align}
where ${X^\dagger}$ denotes the Moore-Penrose pseudo-inverse of $X$, typically computed using singular value decomposition (SVD). Figure~\ref{risanzu} illustrates how periapsis set deformation is represented as a discrete transformation using $A$.
Once the mapping $A$ is obtained, future periapsis states can be predicted by iteratively applying $A$:
\begin{align}
\bm{x}_{k+1} = A^k \bm{x}_1\label{eq:eq_dmd_prediction}
\end{align}

This discrete-time formulation enables the efficient reconstruction of periapsis evolution without the need for numerical integration of the full equations of motion. 

\section{Numerical Simulations}\label{sec_III}

\subsection{Local Deformation Map}\label{sec_III_lDMD}

This section presents the construction of a Local Deformation Map-based DMD (LDMD) to model chaotic periapsis evolution using a data-driven discrete mapping framework. The LDMD approach builds a linear operator $A$ that characterizes the evolution of periapsis states in a localized region of the PPM.

\begin{figure}[htbp]
    \centering
    \begin{subfigure}[b]{0.24\textwidth}
        \includegraphics[width=\textwidth]{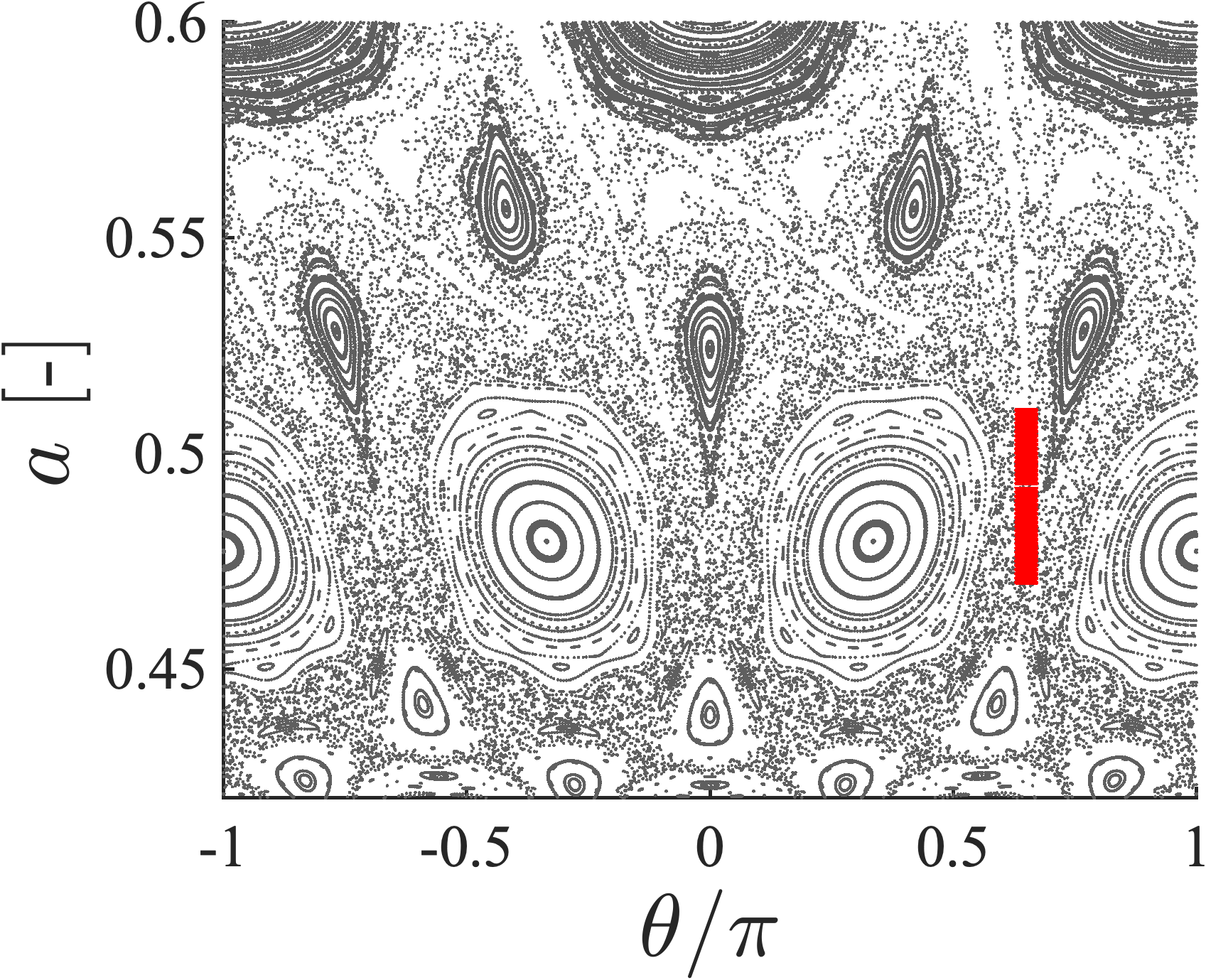}
        \caption{$k=1$}
    \end{subfigure}
    \begin{subfigure}[b]{0.24\textwidth}
        \includegraphics[width=\textwidth]{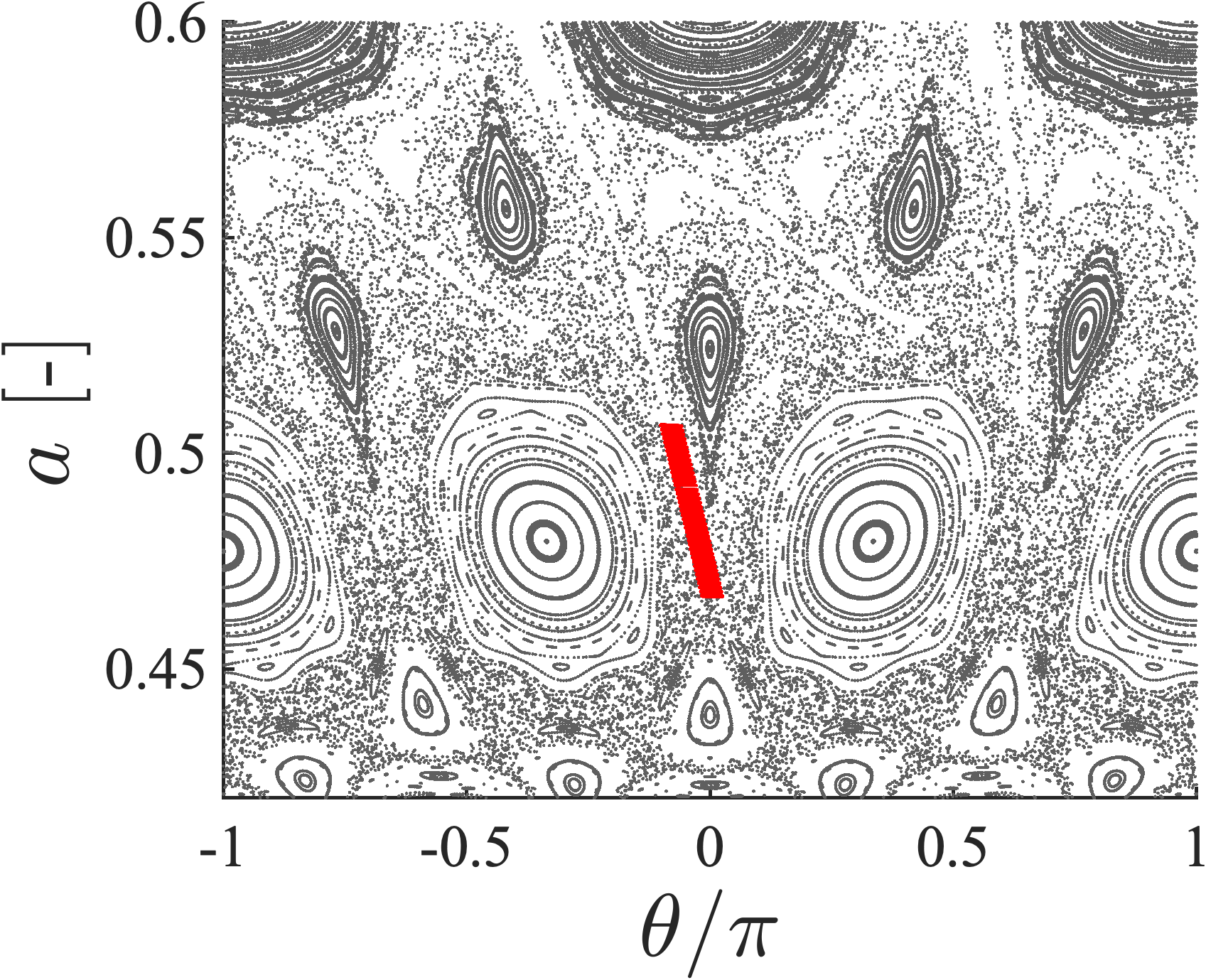}
        \caption{$k=2$}
    \end{subfigure}
    \begin{subfigure}[b]{0.24\textwidth}
        \includegraphics[width=\textwidth]{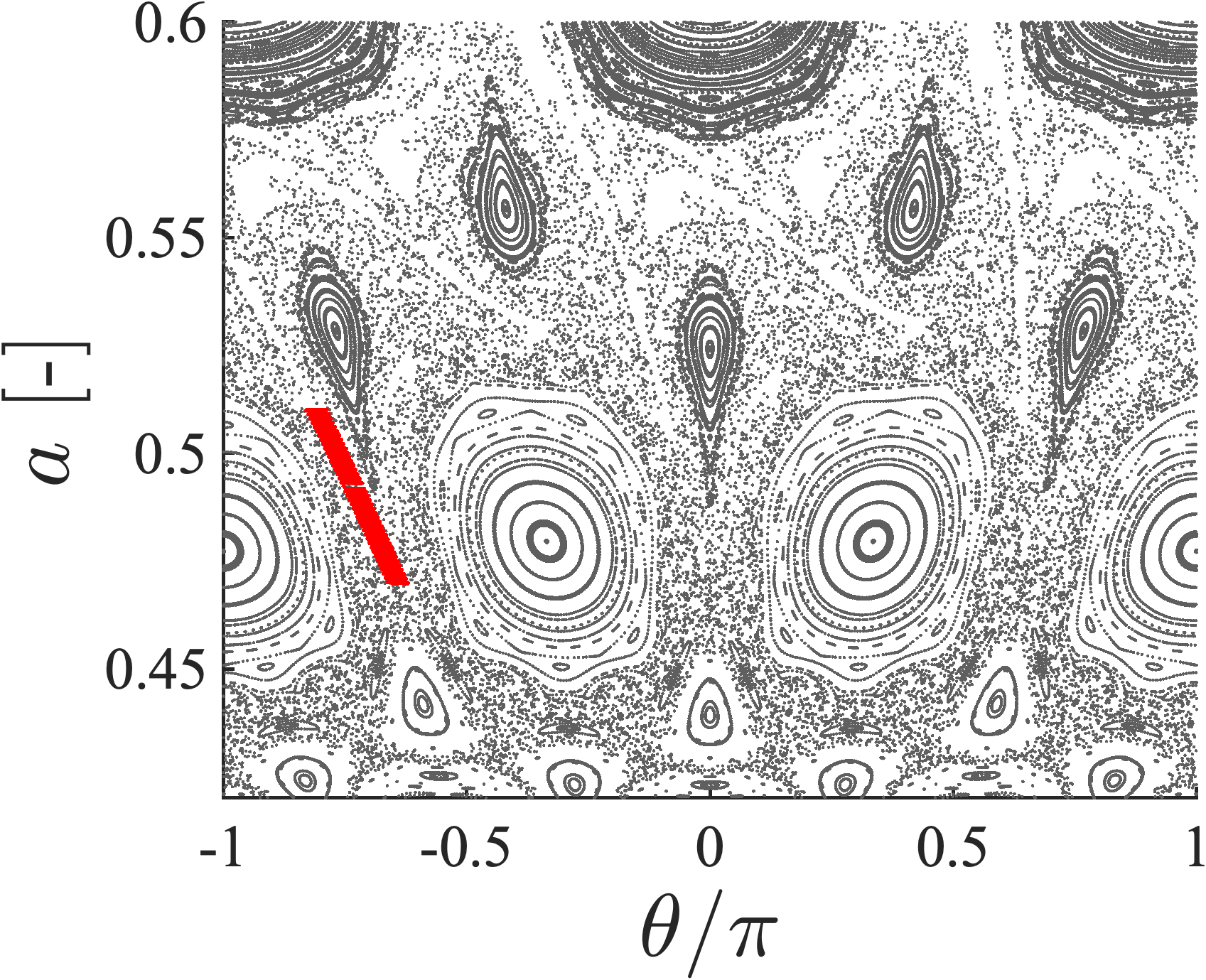}
        \caption{$k=3$}
    \end{subfigure}
    \begin{subfigure}[b]{0.24\textwidth}
        \includegraphics[width=\textwidth]{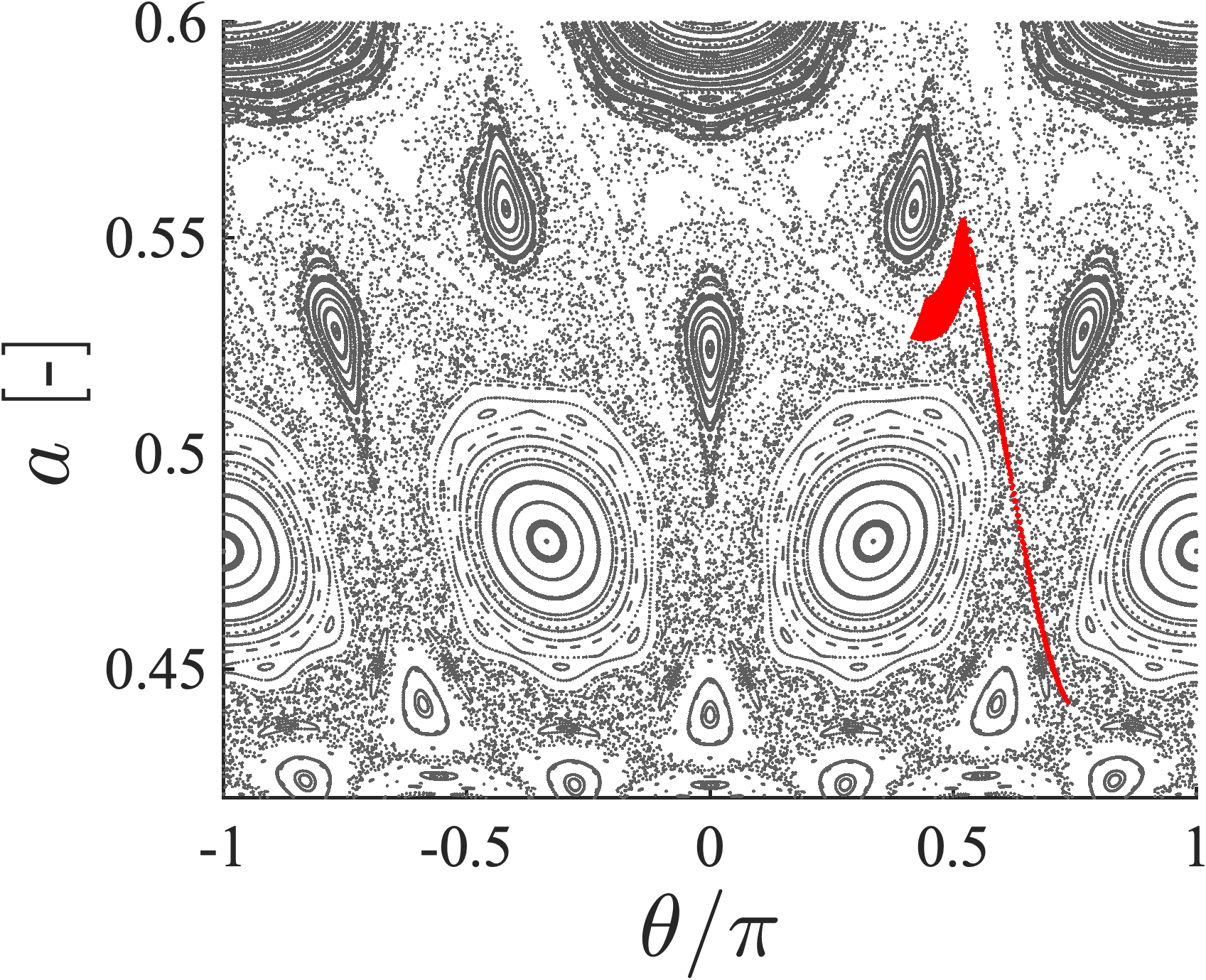}
        \caption{$k=4$}
    \end{subfigure}

    \vspace{1em} 

    \begin{subfigure}[b]{0.24\textwidth}
        \includegraphics[width=\textwidth]{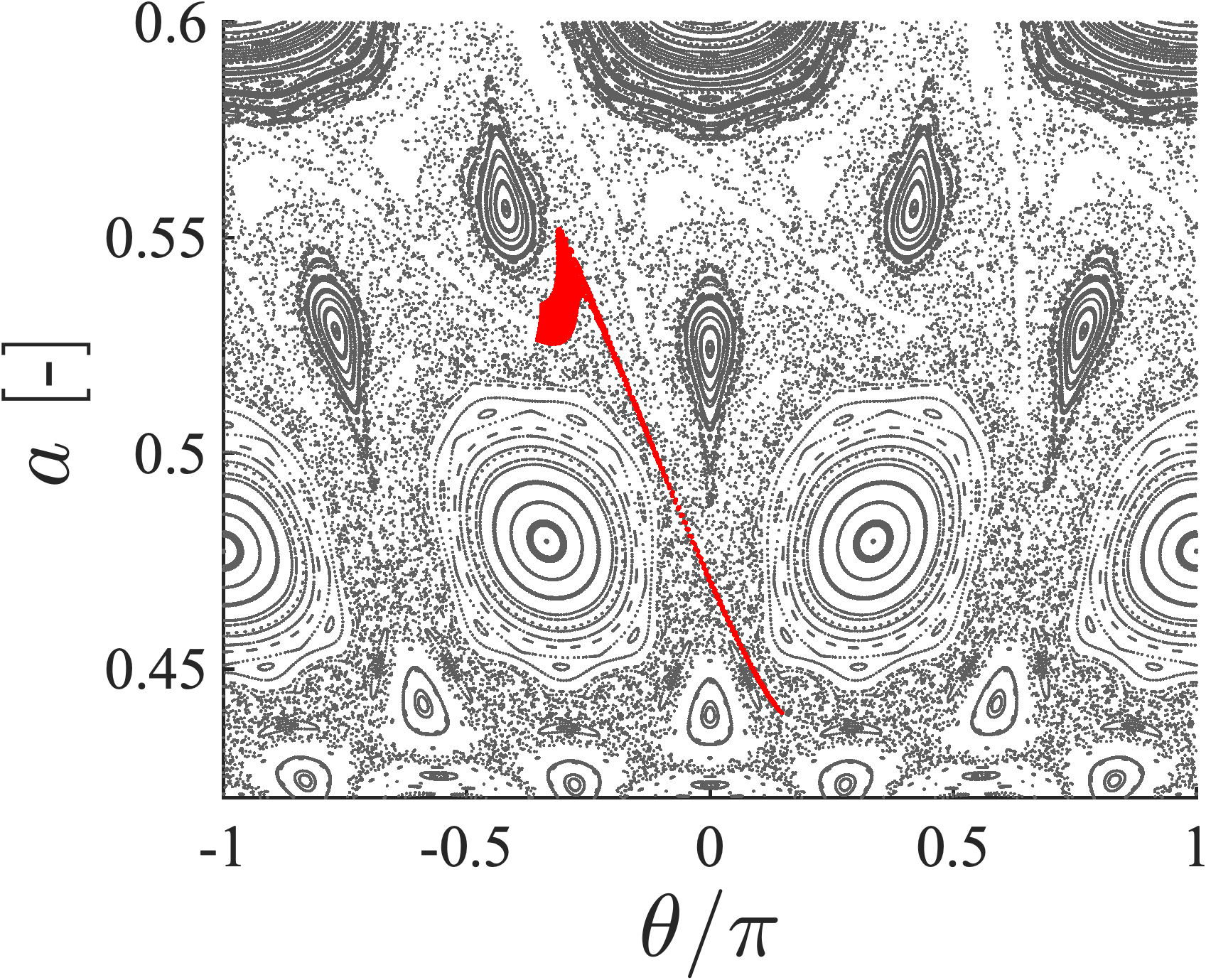}
        \caption{$k=5$}
    \end{subfigure}
    \begin{subfigure}[b]{0.24\textwidth}
        \includegraphics[width=\textwidth]{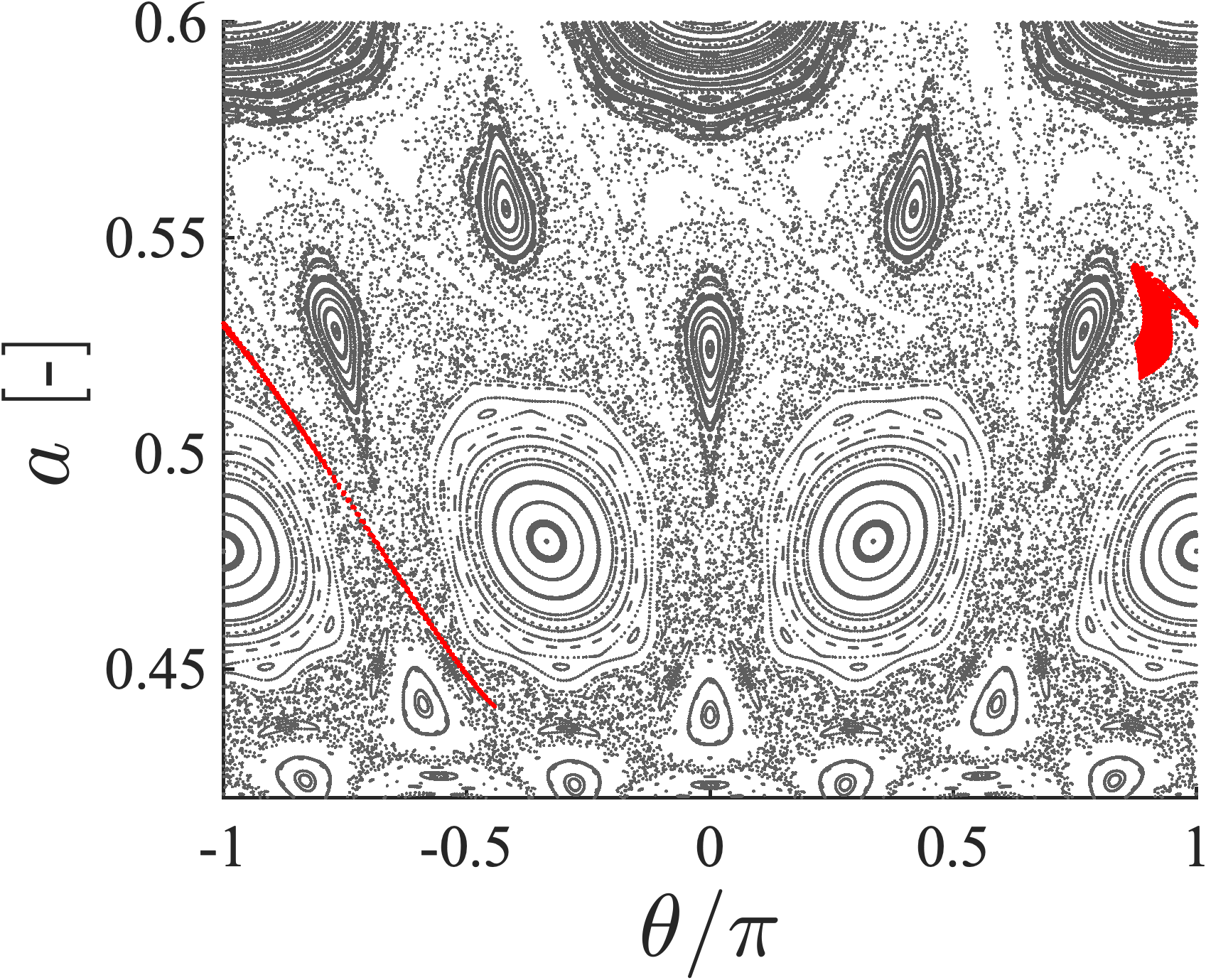}
        \caption{$k=6$}
    \end{subfigure}
    \begin{subfigure}[b]{0.24\textwidth}
        \includegraphics[width=\textwidth]{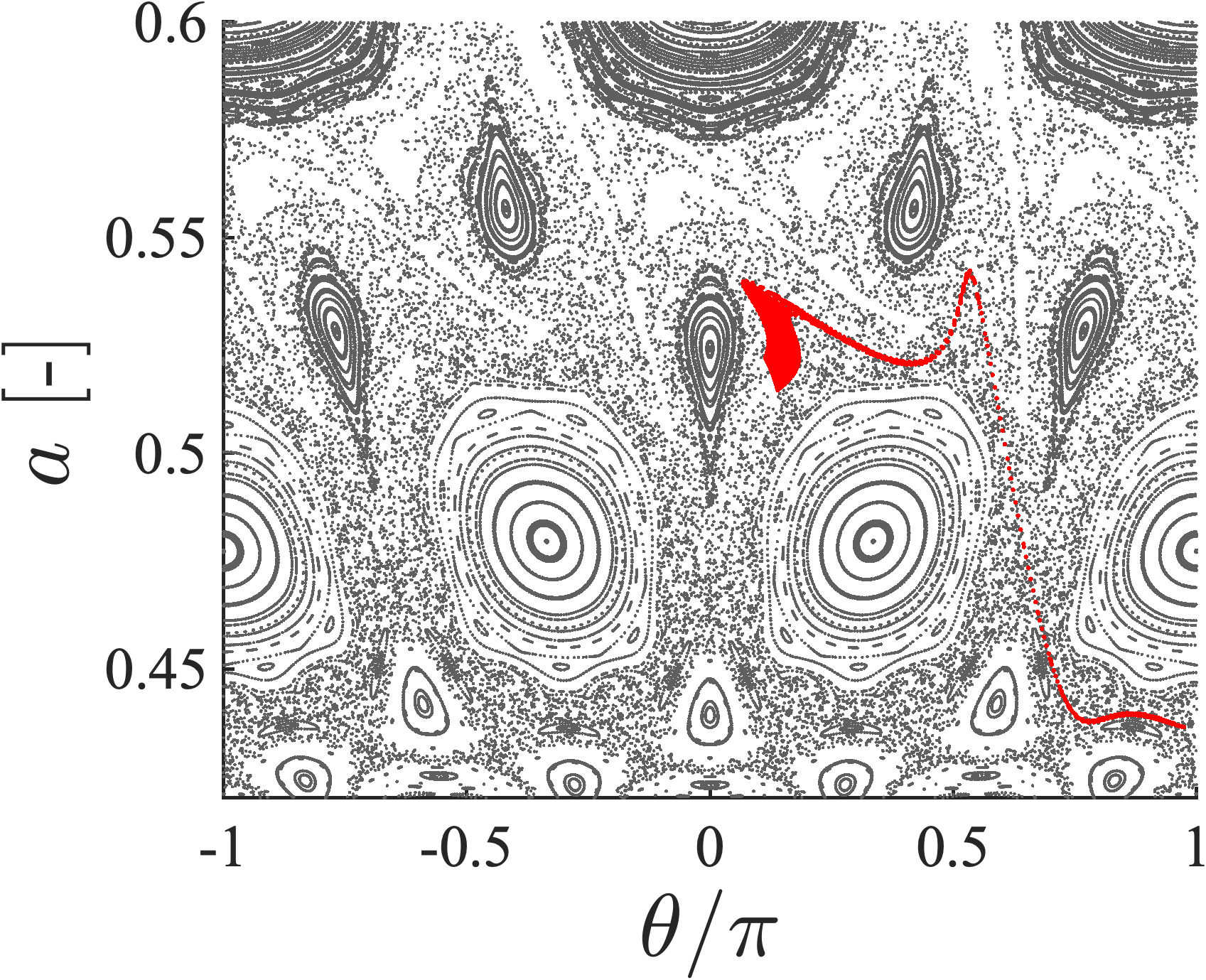}
        \caption{$k=7$}
    \end{subfigure}
    \begin{subfigure}[b]{0.24\textwidth}
        \includegraphics[width=\textwidth]{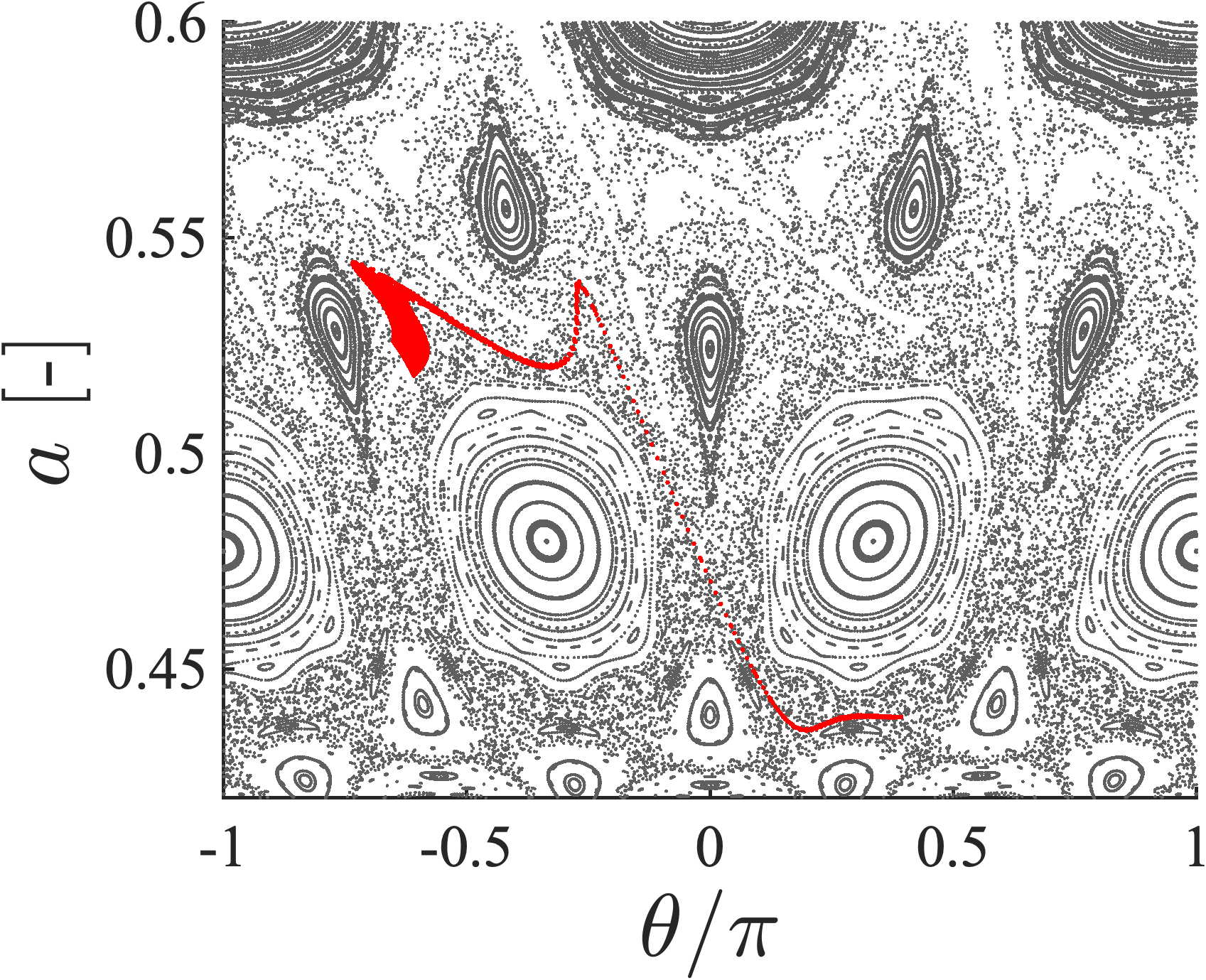}
        \caption{$k=8$}
    \end{subfigure}

    \caption{Time evolution of periapsis sets obtained from numerical integration.}
    \label{numerical_time_evolution}
\end{figure}

\textcolor{black}{The procedure for constructing the model is as follows: First, an initial set of periapsis points is selected, with the corresponding conditions summarized in Table~\ref{table:refdata}. These initial periapsis states define the dataset $\bm{x}_1$. From each initial condition, the orbit is numerically propagated forward, generating periapsis points for time steps $k = 2, 3, \dots, 8$. The resulting periapsis evolution is shown in Fig.~\ref{numerical_time_evolution}, where red dots represent the ($\theta, a$) values across all steps. At each step $k$, all $(\theta, a)$ pairs are concatenated to form the system state vector $\bm{x}_k$, as defined in Eq.~\eqref{eq_dmd_xk_atheta}. Using the sequence $\bm{x}_1, \bm{x}_2, \dots, \bm{x}_8$, the data matrices $X$ and $X'$ are constructed following Eq.~\eqref{eq_dmd_X_Xprime}. Finally, the discrete mapping $A_{LDMD}$ is computed using the least-squares formulation in Eq.~\eqref{eq:eq_dmd_prediction}, enabling the prediction of future periapsis states.}


\begin{table}[htbp]
  \centering
  \caption{\textcolor{black}{Training data conditions in the LDMD method.}}
  \label{table:refdata}
  \begin{tabular}{ll}
    \hline\noalign{\smallskip}
    Range of training data $\bm{x}_1$ & $0.63 \pi \leq\theta\leq 0.67\pi$, $0.47\leq{a}\leq0.51$ \\
    \noalign{\smallskip}\hline\noalign{\smallskip}
    Periapsis sampling interval  & $1.0\times10^{-3}$ \\
    Number of periapsis $m$ & 8 \\
    Number of orbits $n$  &  1681 \\
    \noalign{\smallskip}\hline
  \end{tabular}
\end{table}

\subsubsection{Data Recovery by LDMD}
\label{sec_III_lDMD_data_recovery}

\begin{figure}[htbp]
    \centering
    \begin{subfigure}[b]{0.24\textwidth}
        \includegraphics[width=\textwidth]{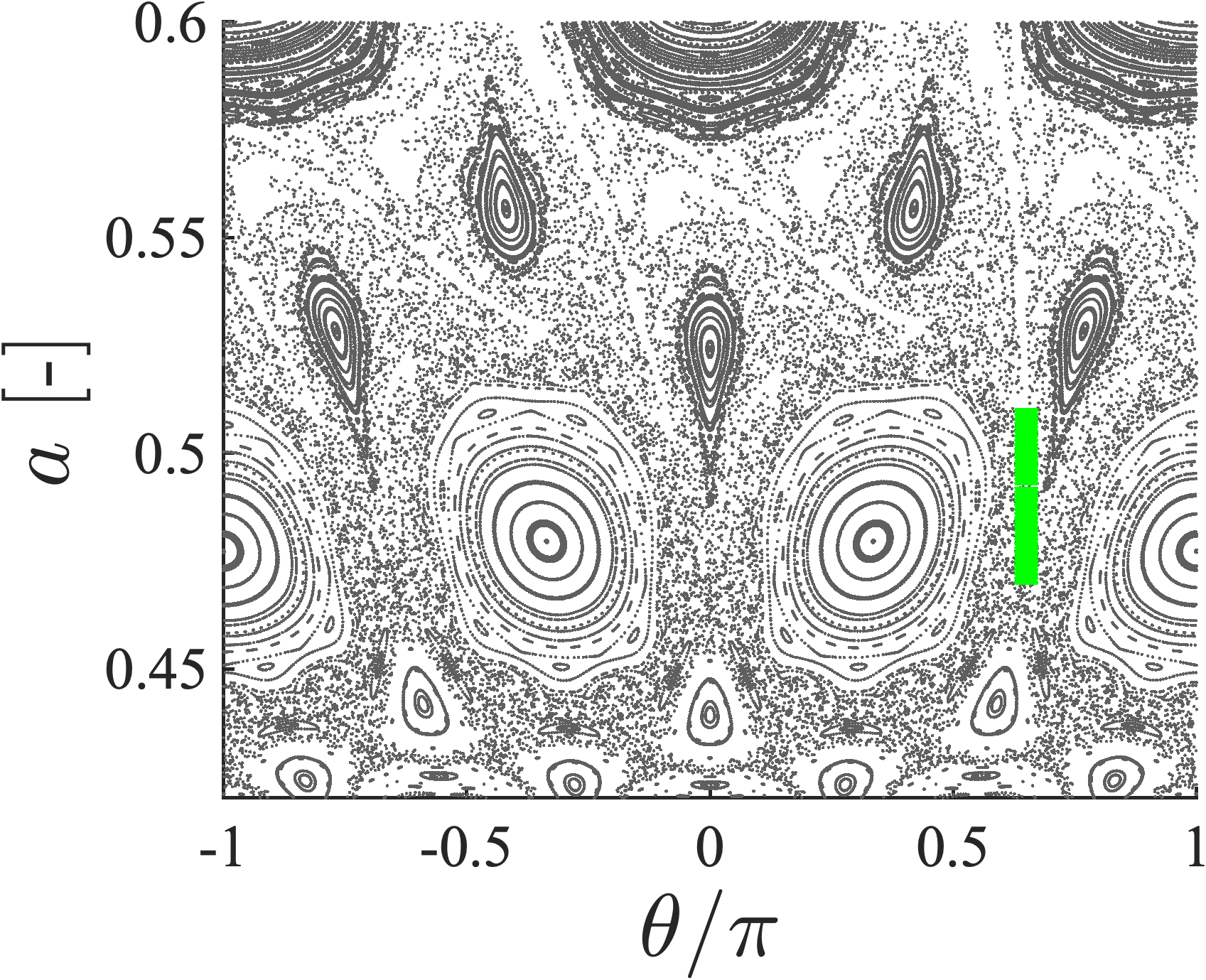}
        \caption{$k=1$}
    \end{subfigure}
    \begin{subfigure}[b]{0.24\textwidth}
        \includegraphics[width=\textwidth]{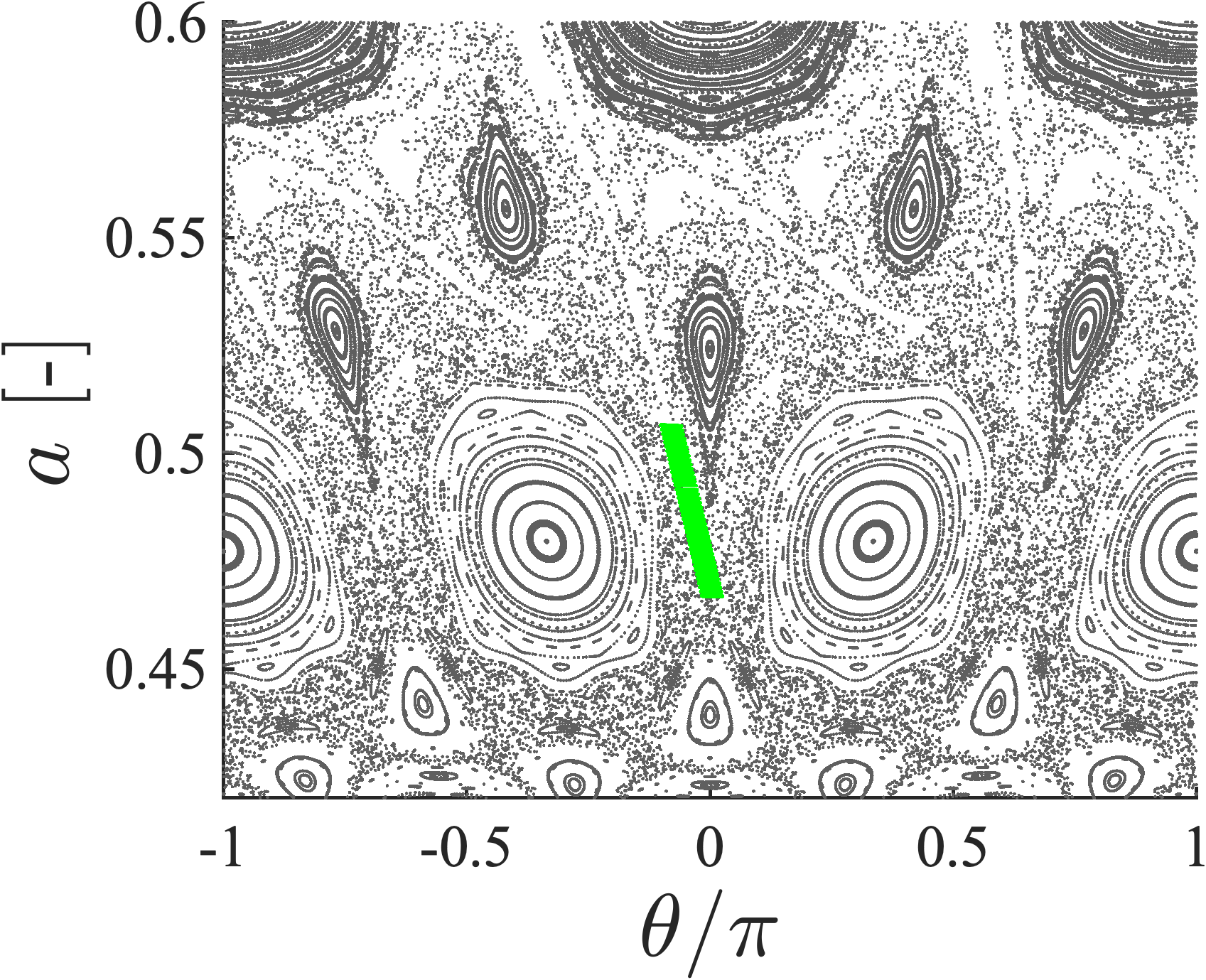}
        \caption{$k=2$}
    \end{subfigure}
    \begin{subfigure}[b]{0.24\textwidth}
        \includegraphics[width=\textwidth]{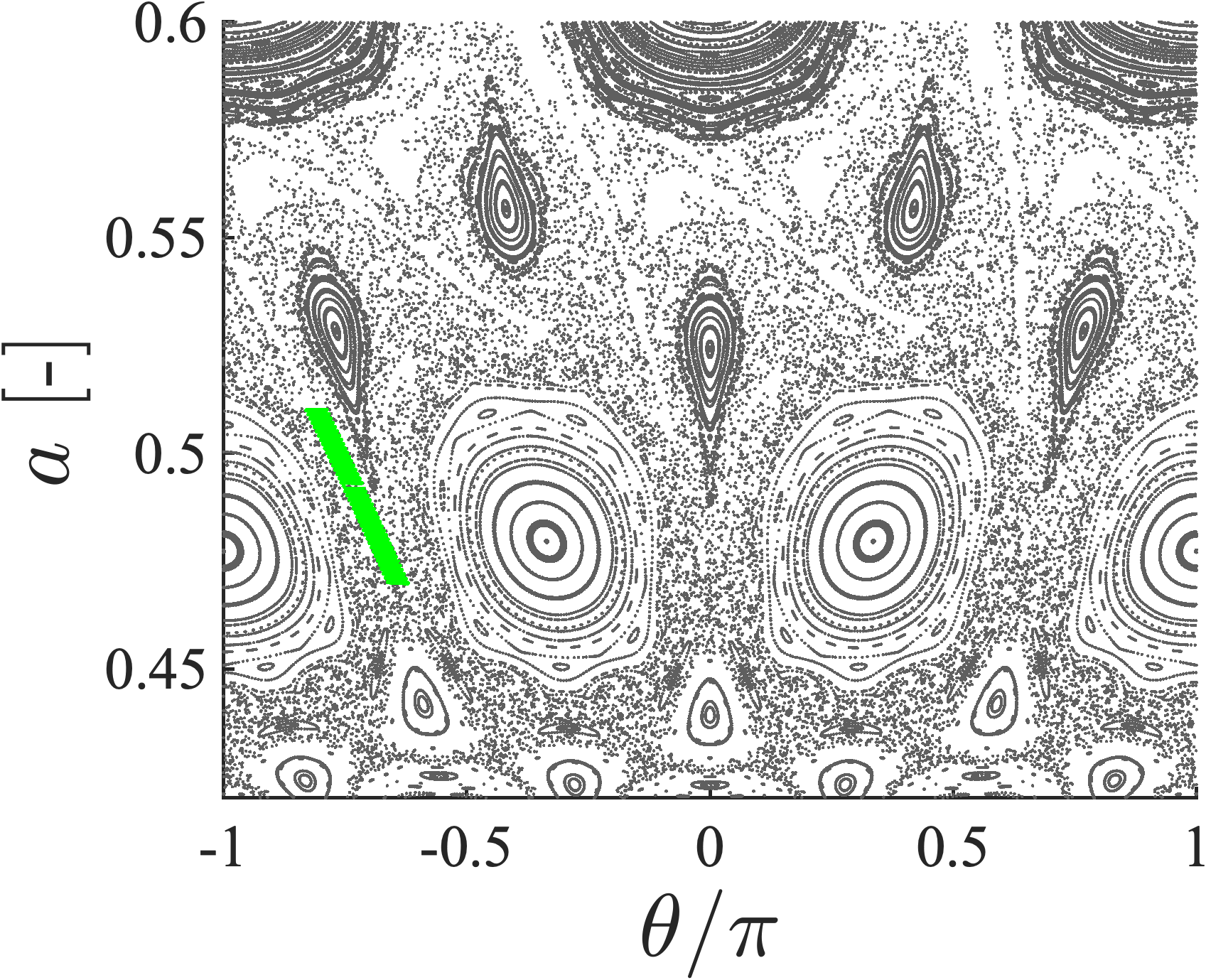}
        \caption{$k=3$}
    \end{subfigure}
    \begin{subfigure}[b]{0.24\textwidth}
        \includegraphics[width=\textwidth]{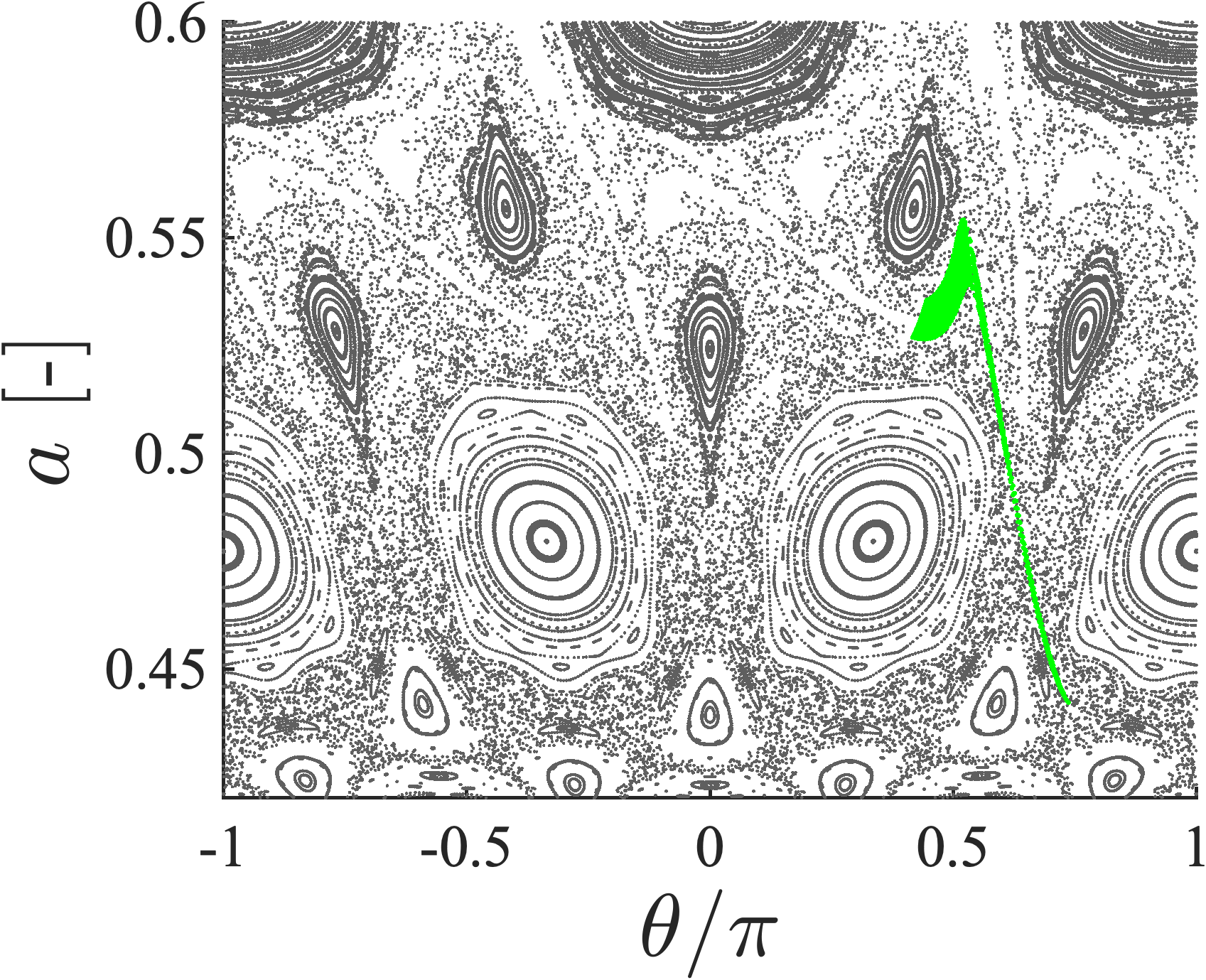}
        \caption{$k=4$}
    \end{subfigure}

    \vspace{1em} 

    \begin{subfigure}[b]{0.24\textwidth}
        \includegraphics[width=\textwidth]{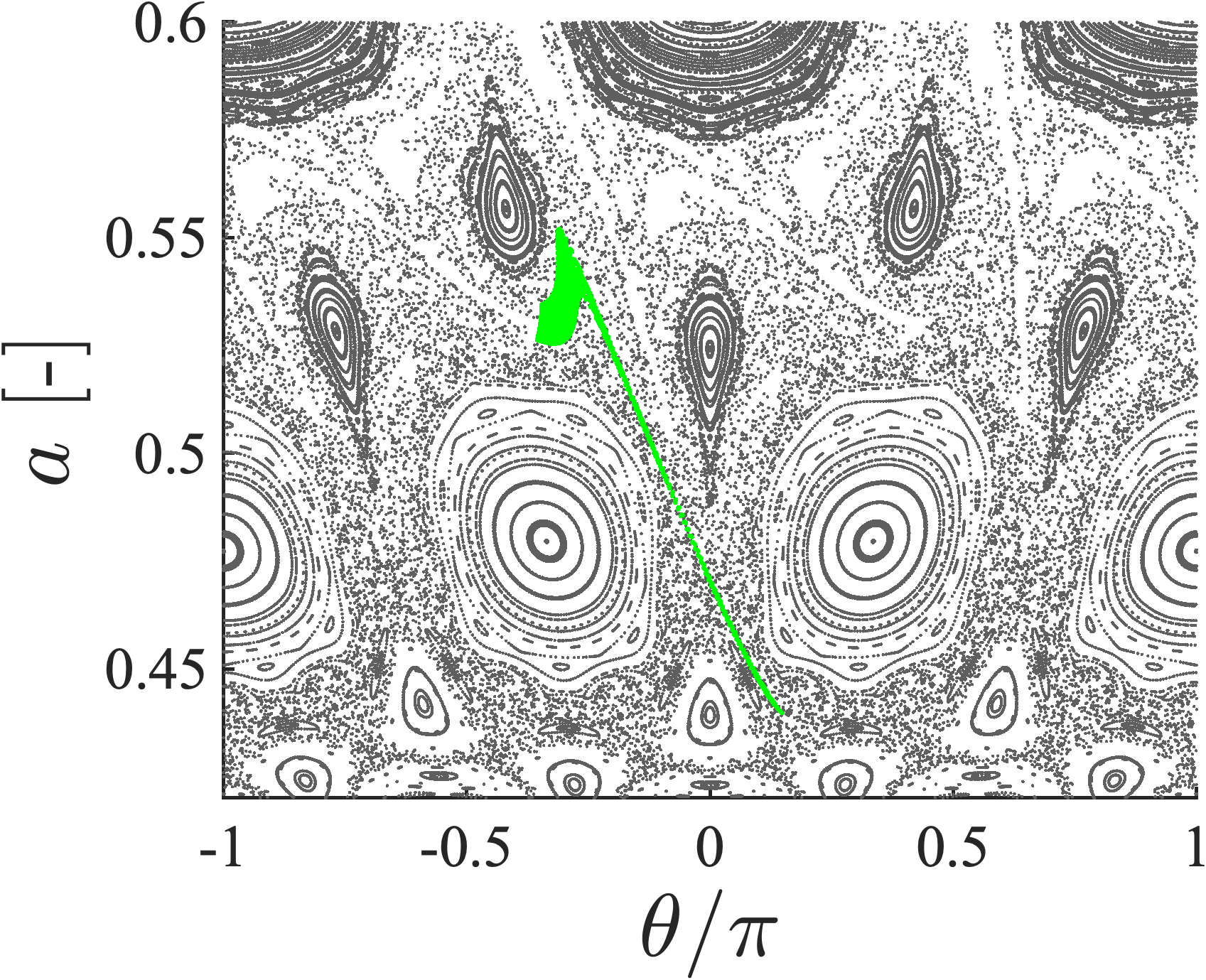}
        \caption{$k=5$}
    \end{subfigure}
    \begin{subfigure}[b]{0.24\textwidth}
        \includegraphics[width=\textwidth]{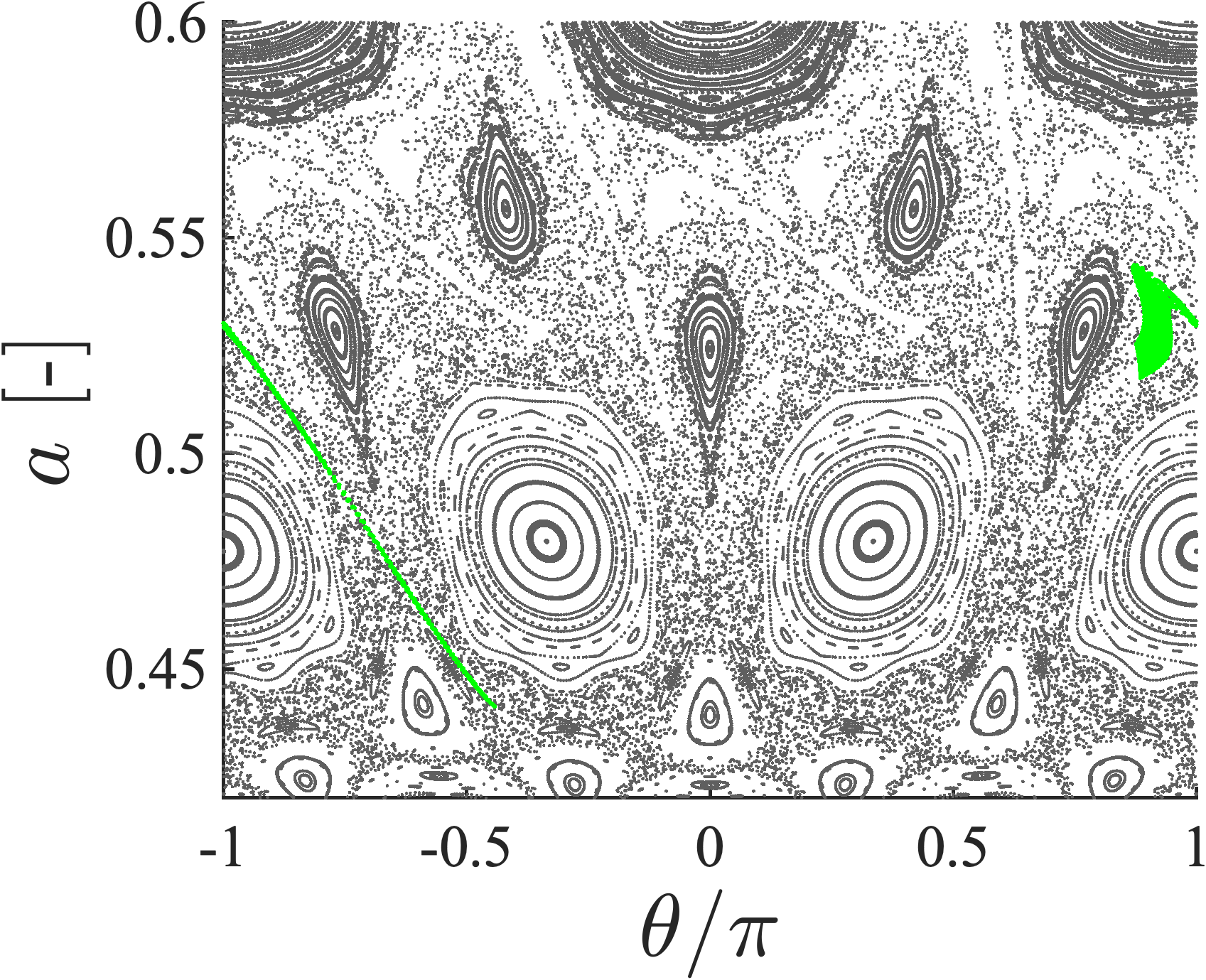}
        \caption{$k=6$}
    \end{subfigure}
    \begin{subfigure}[b]{0.24\textwidth}
        \includegraphics[width=\textwidth]{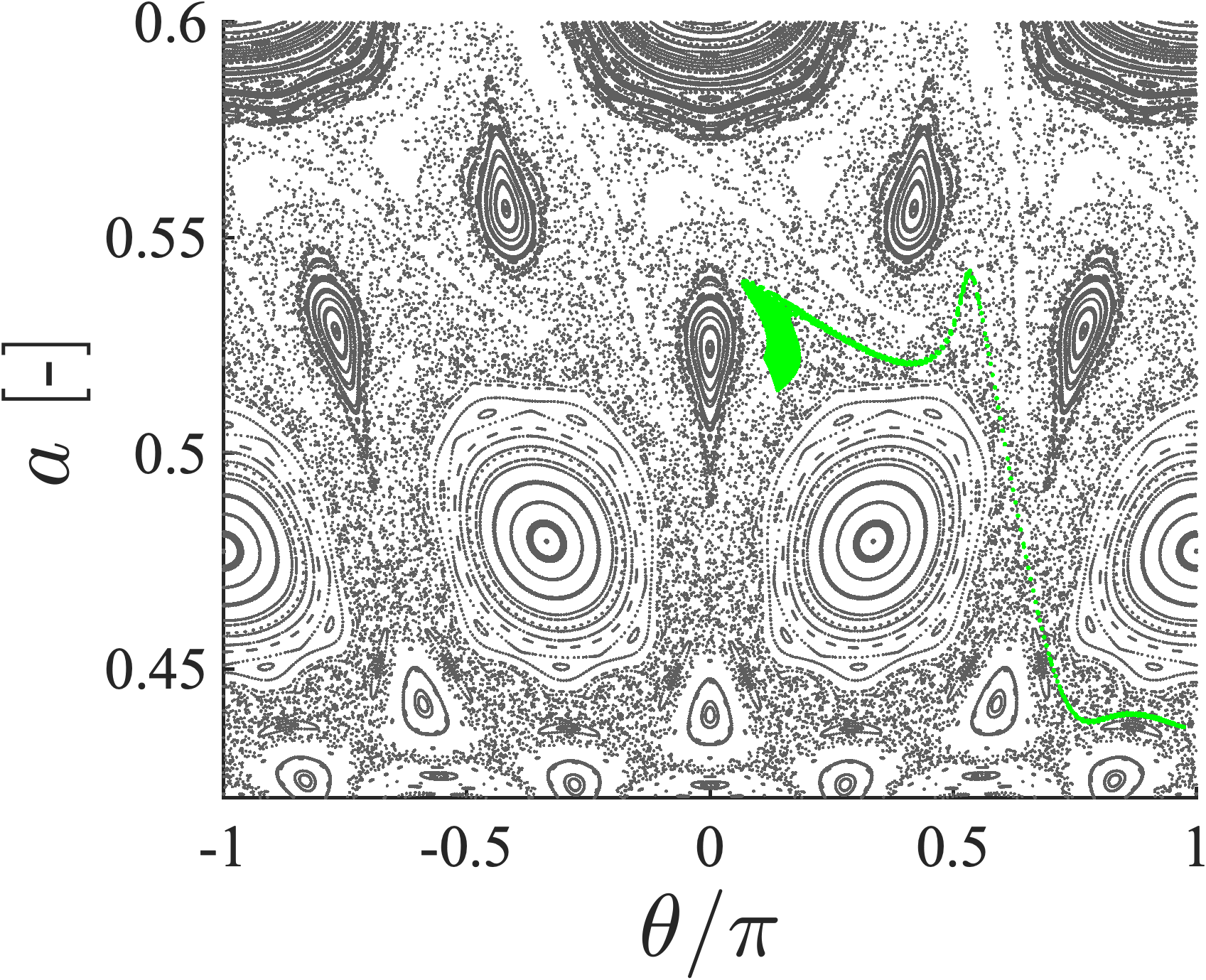}
        \caption{$k=7$}
    \end{subfigure}
    \begin{subfigure}[b]{0.24\textwidth}
        \includegraphics[width=\textwidth]{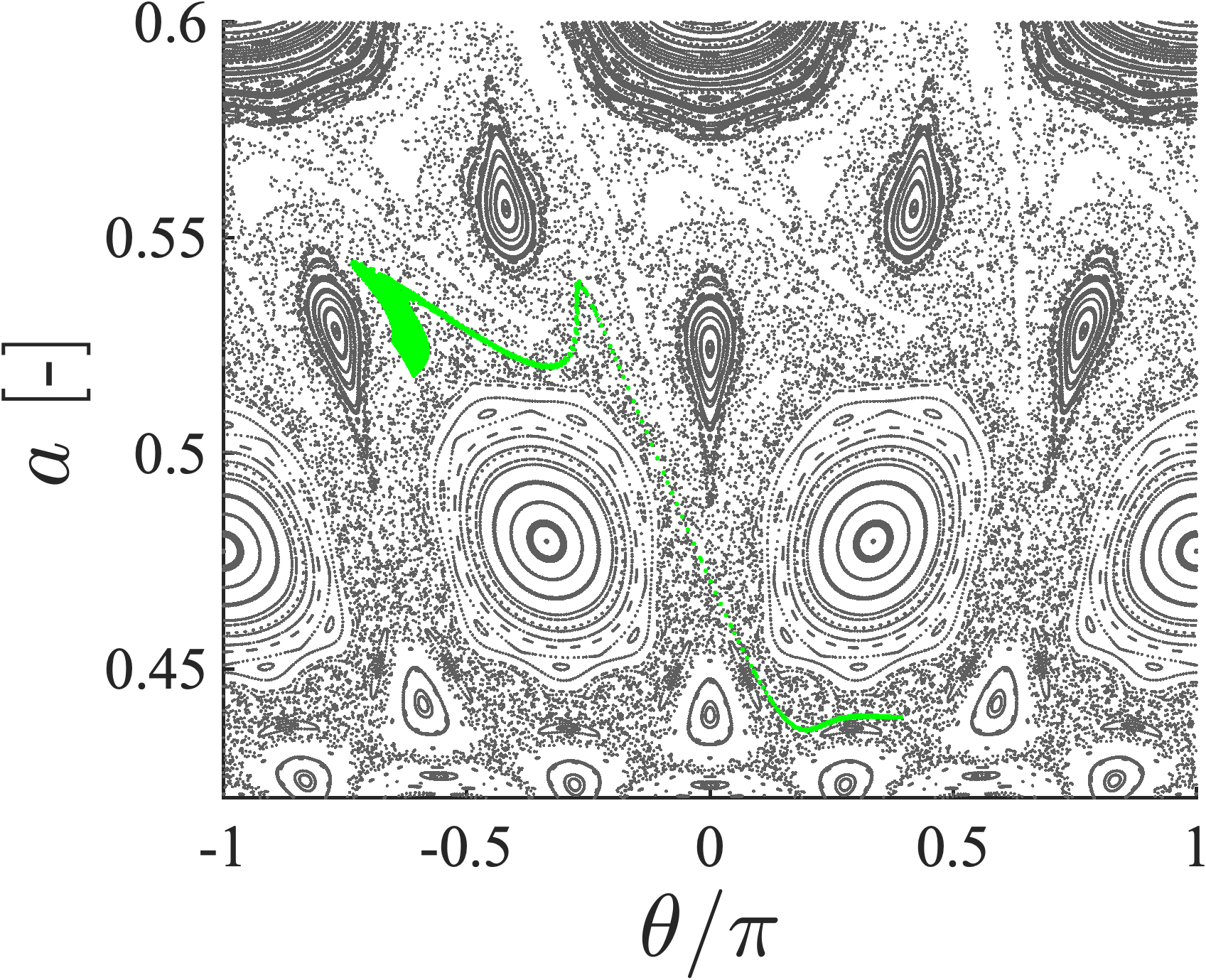}
        \caption{$k=8$}
    \end{subfigure}

    \caption{Time evolution of periapsis sets obtained from LDMD.}
    \label{LDMD_time_evolution}
\end{figure}

\begin{figure}[h!]
  \begin{minipage}[b]{0.48\hsize}
    \centering
    \includegraphics[width=\textwidth]{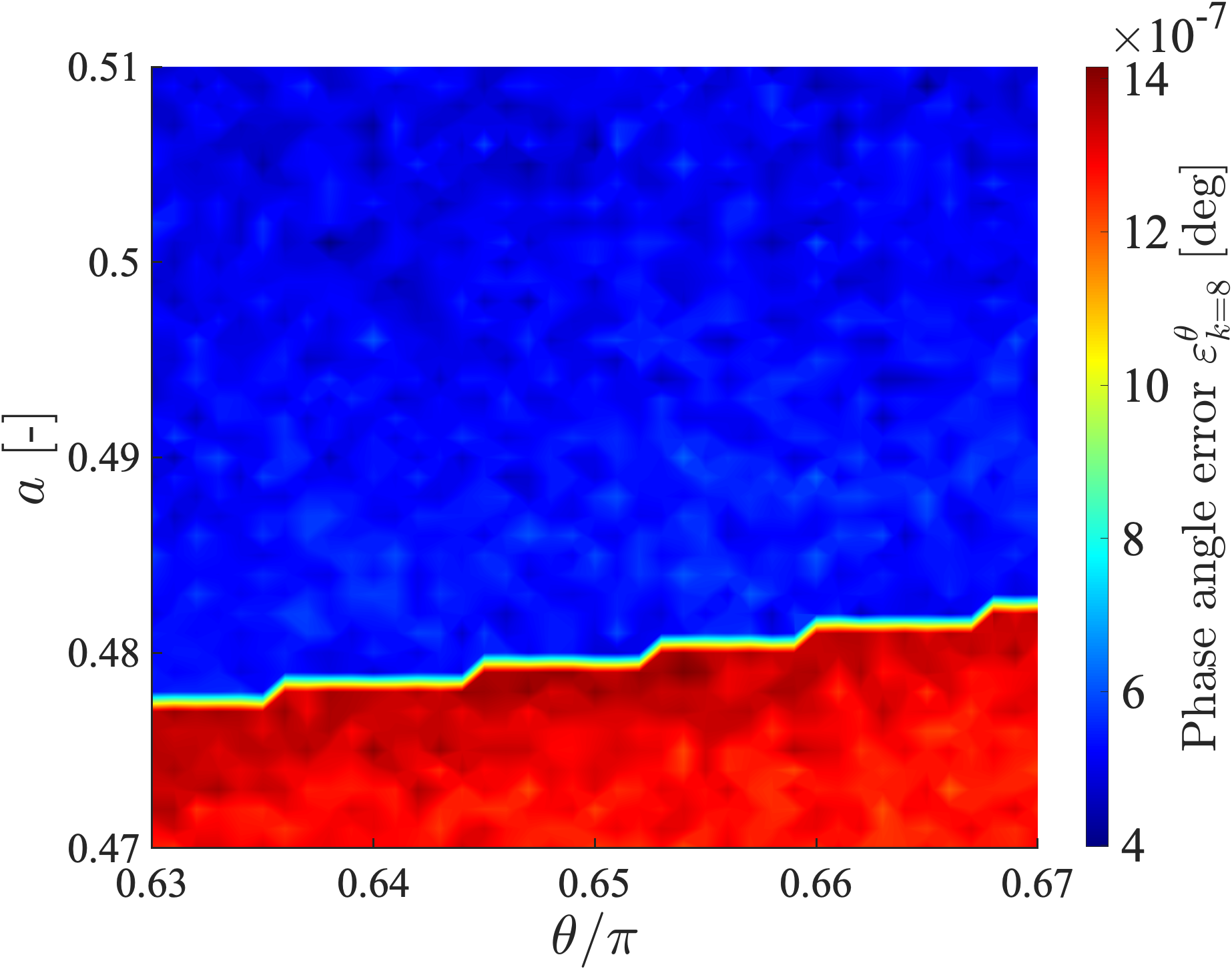}
    \subcaption{Approximation error $\epsilon^{\theta}_{k=8}$.}
    \end{minipage}
    \begin{minipage}[b]{0.48\hsize}
    \centering
    \includegraphics[width=\textwidth]{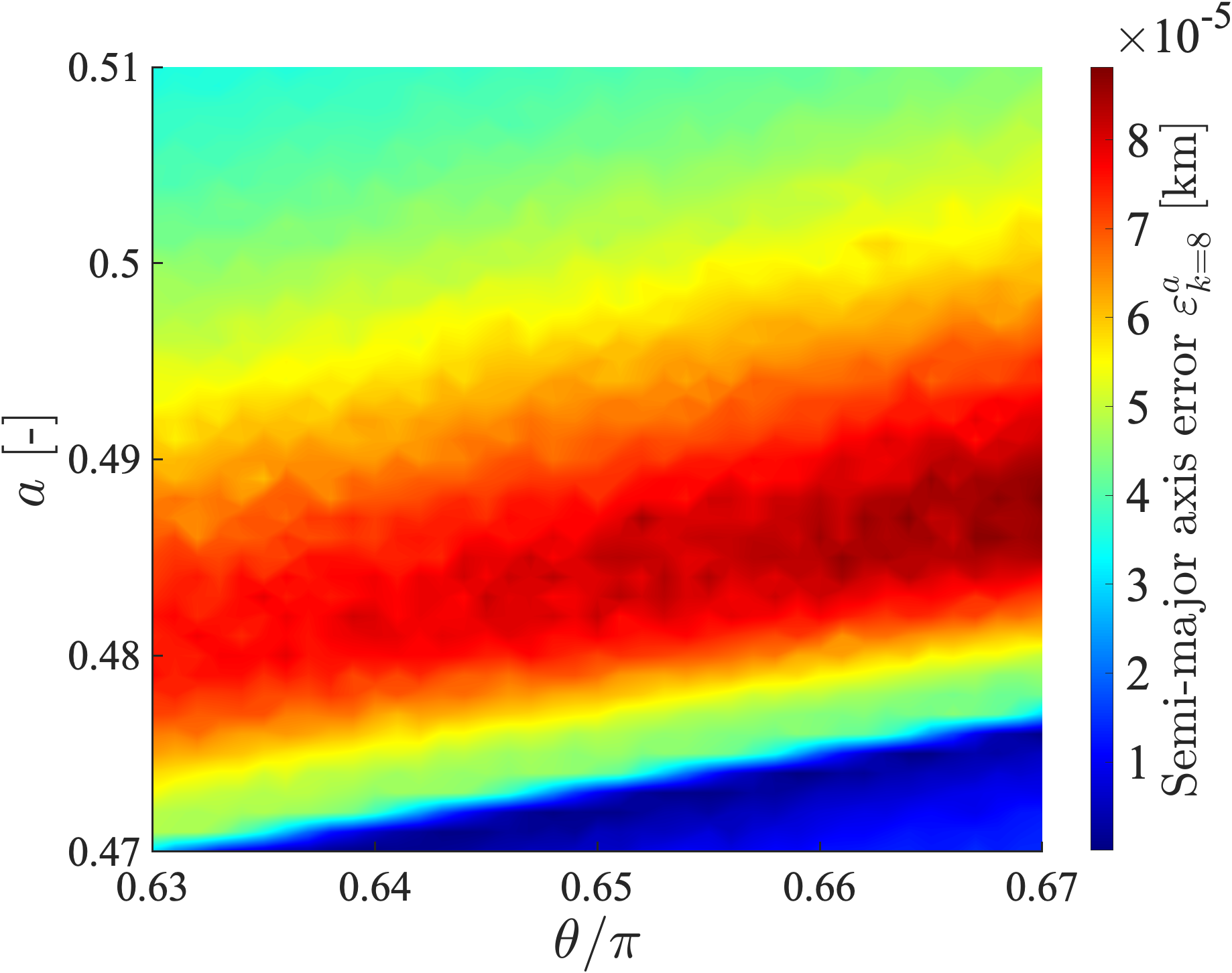}
    \subcaption{Approximation error $\epsilon^{a}_{k=8}$.}
    \end{minipage}
    \caption{Approximation error between numerical data and LDMD at prediction step $k=8$.}
    \label{fig_error_step8_1681}
\end{figure}

Figure~\ref{LDMD_time_evolution} presents the LDMD prediction (green) in the PPM using $A_{LDMD}$. To evaluate the accuracy of LDMD, the approximation errors between the LDMD predicted and numerically integrated periapsis states are defined as:
\begin{equation}
  \begin{aligned}\label{eq_ldmd_approximation_error}
    {\epsilon^{\theta}}_k=|{\theta_k}^\prime-{\theta_k}|, \quad 
    {\epsilon^{a}}_k=|{a_k}^\prime-{a_k}|
  \end{aligned}
\end{equation}
where ${\theta}^\prime_k$ and ${a}^\prime_k$ denote the predicted values by LDMD at the $k$-th step using Eq.~\eqref{eq:eq_dmd_prediction}, while $\theta_k$ and $a_k$ correspond to the reference values obtained from numerical integration. The approximation errors at step $k=8$ are visualized in Fig.~\ref{fig_error_step8_1681}. The horizontal and vertical axes correspond to the initial periapsis parameters, namely the phase angle $\theta_{k=1}$ and the semi-major axis $a_{k=1}$, expressed in non-dimensional units. The left panel displays the distribution of phase angle error $\epsilon^{\theta}_{k=8}$ in degrees, while the right panel shows the semi-major axis error $\epsilon^{a}_{k=8}$ in kilometers.

\begin{figure}[h!]
    \begin{minipage}[b]{0.48\hsize}
    \centering
    \includegraphics[width=\textwidth]{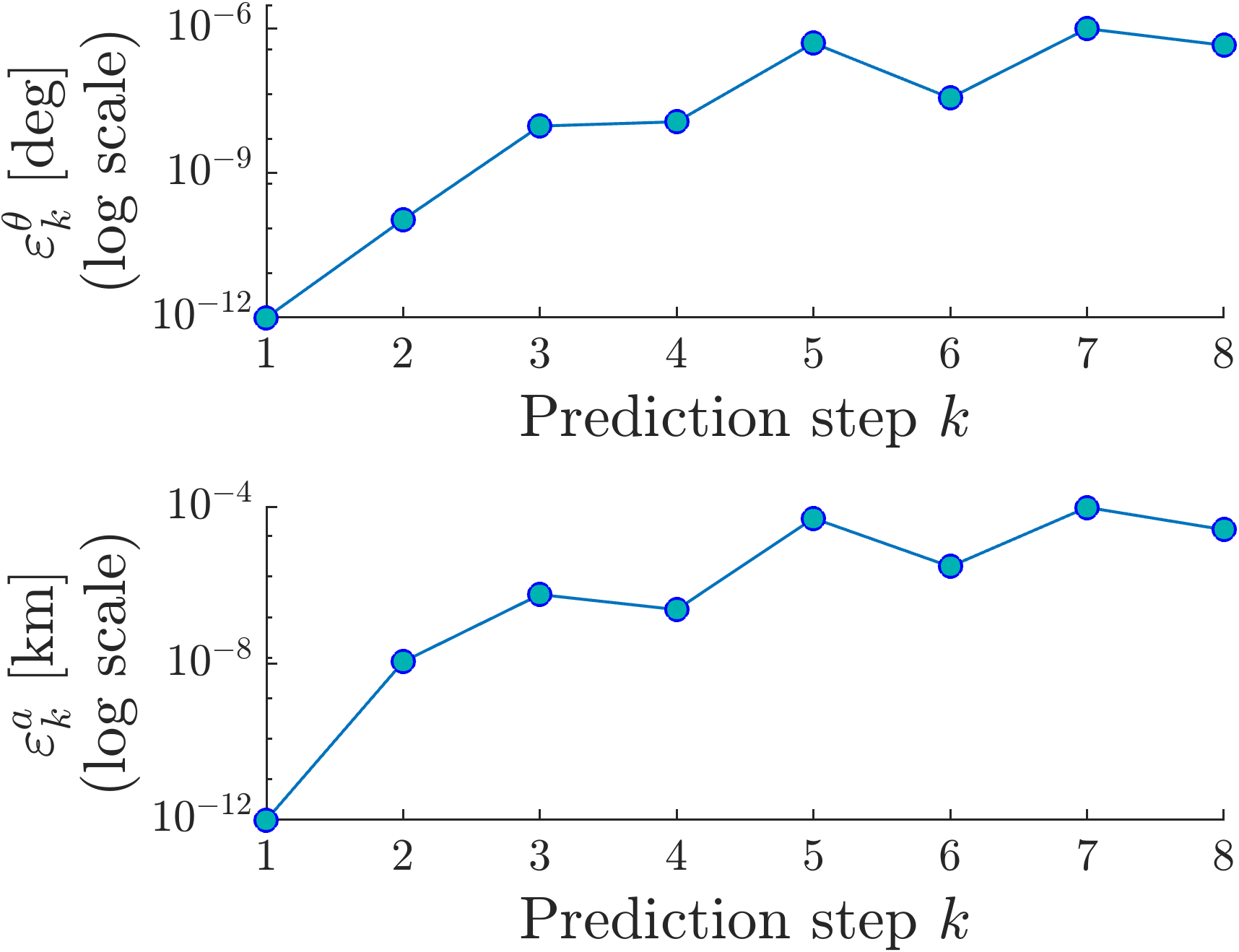}
    \subcaption{Error between numerical and LDMD over 8 steps.}
    \end{minipage}
    \begin{minipage}[b]{0.48\hsize}
    \centering
    \includegraphics[width=\textwidth]{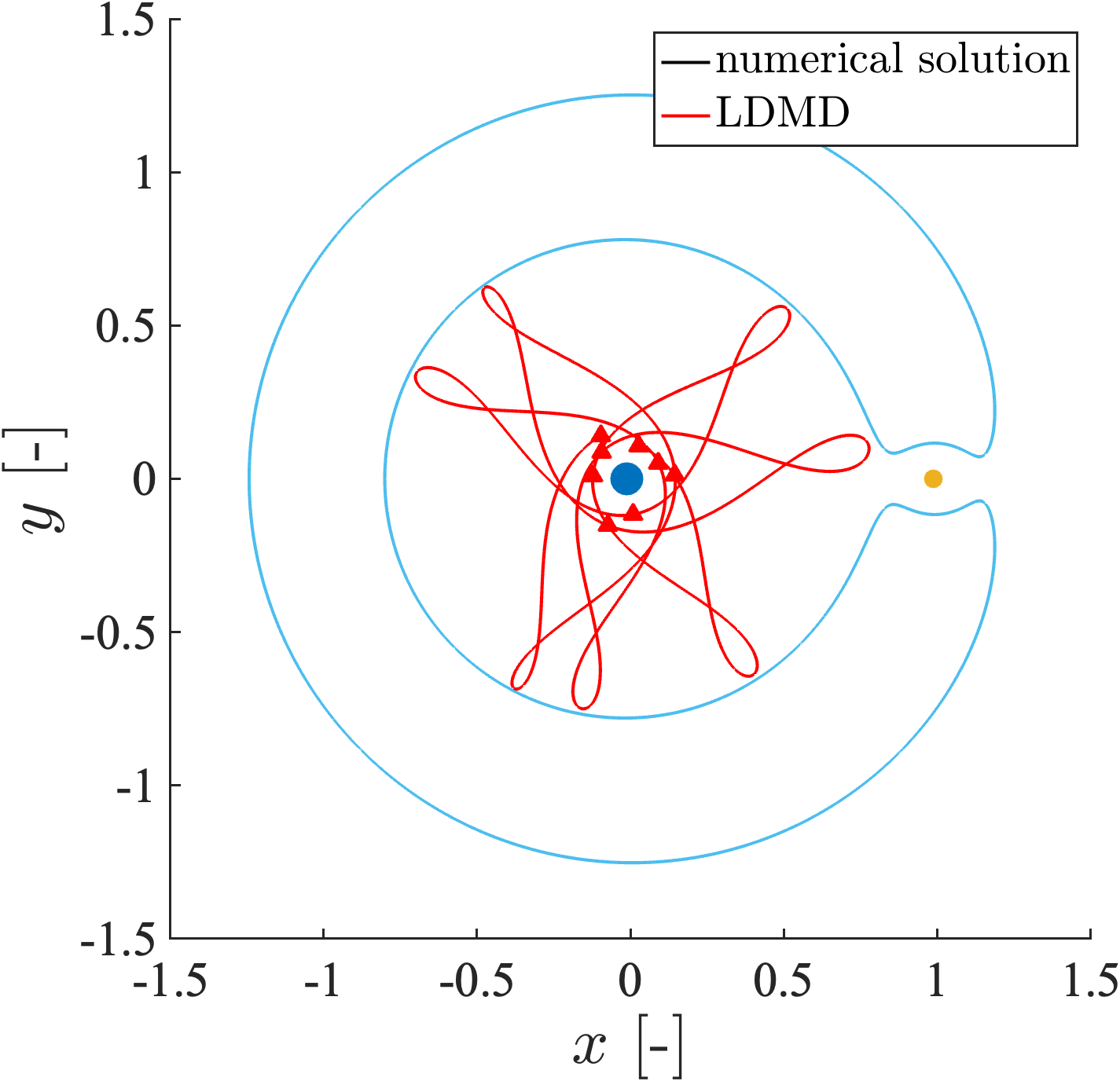}
    \subcaption{Trajectory in the rotating frame.}
    \end{minipage}
    \caption{LDMD-recovered trajectory compared with numerical integration. (a): approximation error at each step; (b): corresponding trajectories in the Earth-Moon rotating frame.}
    \label{fig_Data_Recovery_trajectory_one_example}
\end{figure}

Notably, the error distributions in Fig.~\ref{fig_error_step8_1681} exhibit sharp boundaries that resemble separatrix-like structures in phase space. These boundaries may reflect the presence of underlying invariant manifolds, which are known to govern transport in chaotic regions. The increased prediction errors near these regions suggest sensitivity to initial conditions, and imply that LDMD captures not only local deformation but also aspects of the global geometric features that influence trajectory evolution. The results indicate that LDMD achieves accurate recovery of periapsis states, with maximum approximation errors at \(k = 8\) of \(\epsilon^{\theta}_{k=8} = 1.4 \times 10^{-6}\) degrees and \(\epsilon^{a}_{k=8} = 8.8 \times 10^{-5}\) km. These predicted periapsis parameters can be transformed into full state vectors (position and velocity), allowing for the reconstruction of entire trajectories in the rotating frame.
Figure~\ref{fig_Data_Recovery_trajectory_one_example} illustrates an example trajectory recovered using LDMD, comparing the predicted periapsis states with numerically integrated data. The left panel depicts the approximation error between numerical data and LDMD predictions at each step, while the right panel shows the corresponding trajectories in the rotating frame. This comparison confirms that LDMD effectively reconstructs the trajectory, demonstrating its capability for high-accuracy predictions of chaotic orbits.

\subsubsection{\textcolor{black}{Prediction of Periapsis Not in the Data}}
\label{sec_III_lDMD_data_predictions}

This section examines the ability of LDMD to generalize beyond its training dataset and accurately predict the evolution of periapsis states not included in the original data. The discrete map $A_{LDMD}$ is applied to forecast the time evolution of these out-of-sample periapsis states. Here, the original dataset corresponds to $\bm{x}_1$ in Table~\ref{table:refdata}, while the test conditions for the additional periapsis states are provided in Table~\ref{table:siken}.


\begin{table}[htbp]
  \centering
  \caption{Initial conditions for test data.}
  \label{table:siken}
  \begin{tabular}{ll}
    \hline\noalign{\smallskip}
    Region of data & $0.63 \pi \leq\theta\leq 0.67 \pi $, $0.47\leq{a}\leq0.51$\\
    \noalign{\smallskip}\hline\noalign{\smallskip}
    Periapsis sampling interval   & $2.0\times10^{-4}$ \\
    Number of orbits   &  $40401$ \\
    \noalign{\smallskip}\hline
  \end{tabular}
\end{table}

For periapsis states outside the original dataset $\bm{x}_1$, it is assumed that states located close to each other in the $(\theta, a)$ space will exhibit similar dynamical behavior. Under this assumption, a new initial state \((\theta^*, a^*)\) is approximated by its nearest neighbor \((\theta_i, a_i)\) $\in$ \(\bm{x}_1\), identified by computing the Euclidean distance between \((\theta^*, a^*)\) and each \((\theta, a)\) pair in \(\bm{x}_1\) as follows:
\begin{align}
    d=\sqrt{(\theta^*-\theta)^2+(a^*-a)^2}\label{eq_distance_replacement}
\end{align}

\begin{figure}[h!]
    \setlength{\abovecaptionskip}{-20pt}   
    \setlength{\belowcaptionskip}{-10pt}   
    \centering
    \includegraphics[width=0.8\linewidth,clip]{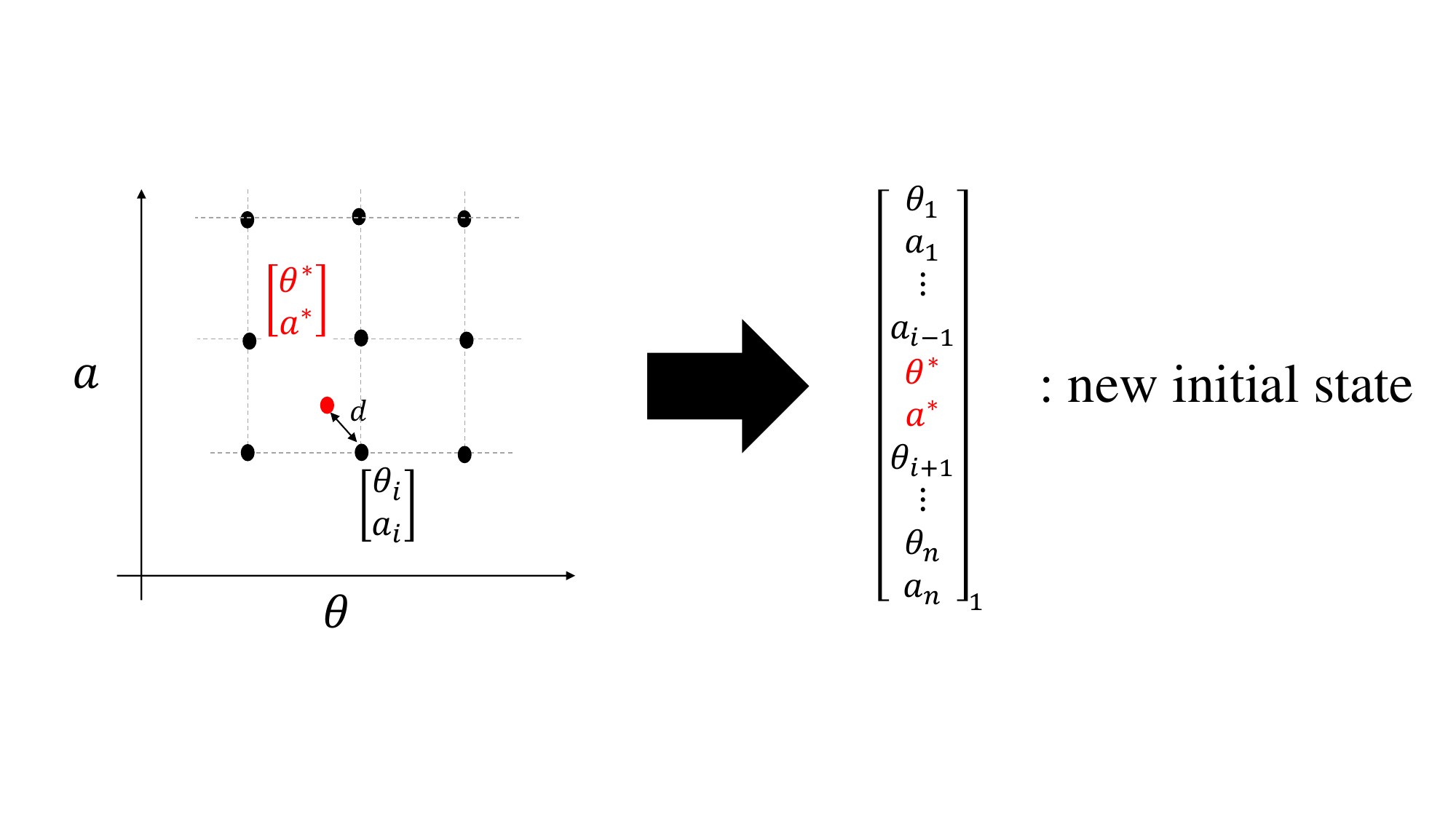}
    \caption{Process of initial-value replacement.}
    \label{irekae}
\end{figure}

The selected \((\theta_i, a_i)\) $\in$ \(\bm{x}_1\) is replaced by the target state \((\theta^*, a^*)\), as illustrated in Fig.~\ref{irekae}. This update yields a modified dataset \( \bm{x}_1' \), which is then used with \( A_{LDMD} \) via Eq.~\eqref{eq:eq_dmd_prediction} to predict the \( k \)-th step evolution of $(\theta^*, a^*)$.

\begin{figure}[h!]
  \begin{minipage}[b]{0.48\hsize}
    \centering
    \includegraphics[width=\textwidth]{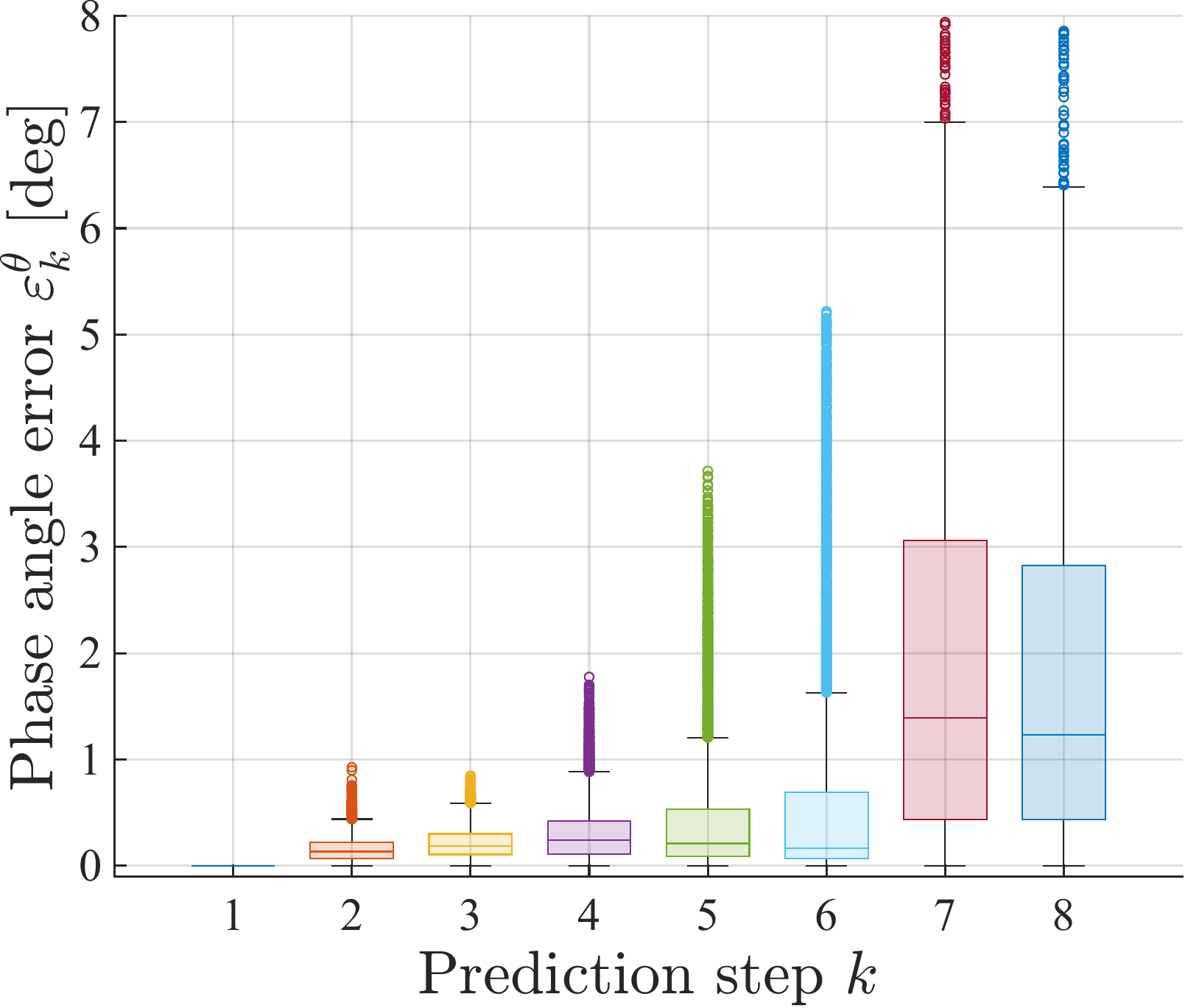}
    \subcaption{Error $\epsilon^{\theta}_{k=8}$ distribution.}
    \label{fig_ldmd_error_distribution_a}
    \end{minipage}
    \begin{minipage}[b]{0.48\hsize}
    \centering
    \includegraphics[width=\textwidth]{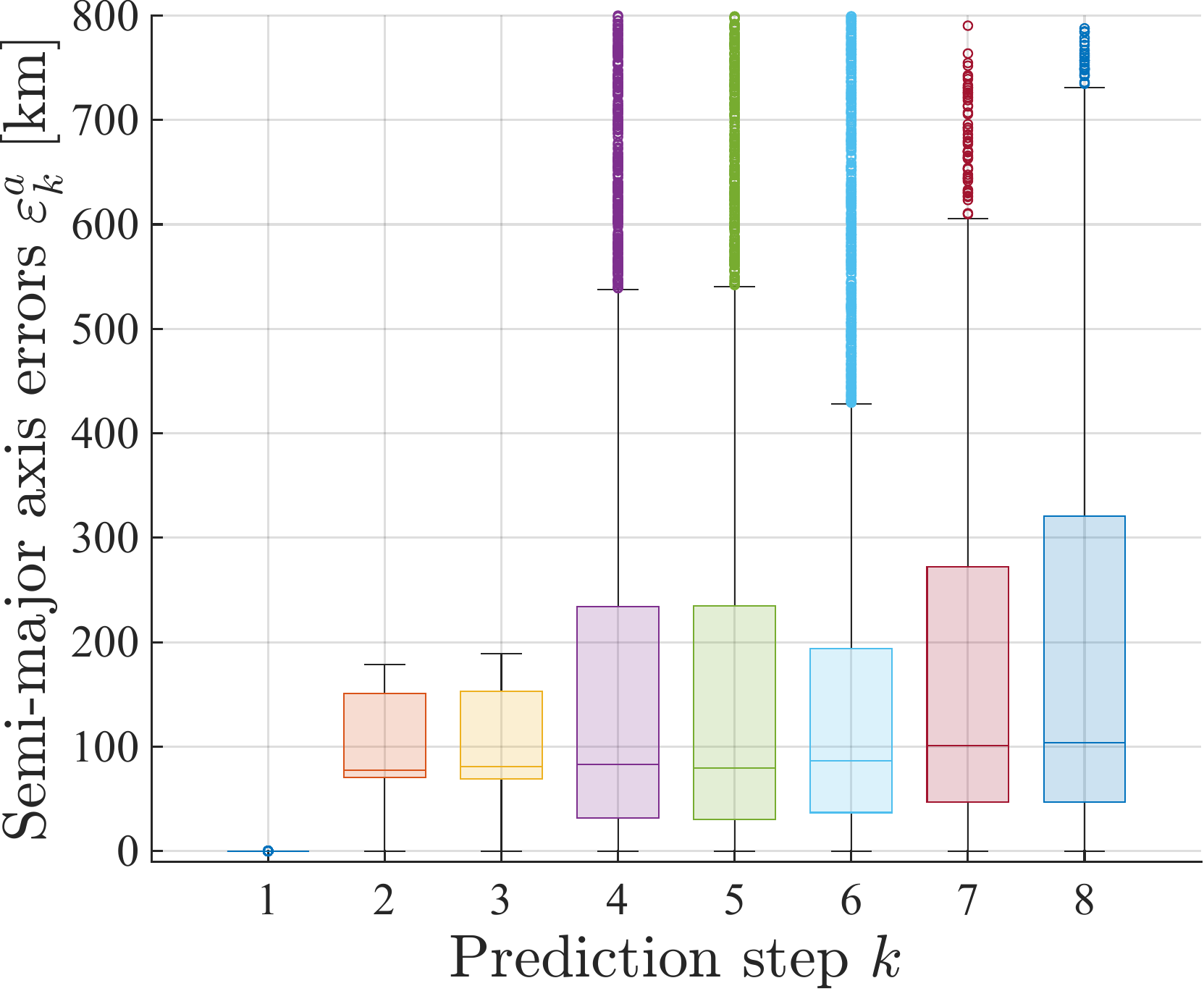}
    \subcaption{Error $\epsilon^{a}_{k=8}$ distribution.}
    \label{fig_ldmd_error_distribution_b}
    \end{minipage}
    \caption{Distribution of prediction errors for 5000 cases across each step.}
    \label{fig_ldmd_error_distribution}
\end{figure}

To assess the impact of this approximation, predictions for 5000 test cases are compared against numerical solutions under varying replacement distances \( d \). Figure~\ref{fig_ldmd_error_distribution} presents the distribution of prediction errors in semi-major axis and phase angle over steps $k =1,...,8$ for the 5000 cases. Errors generally increase with the prediction step, reflecting the accumulation of deviations over time. Phase angle errors (Fig.~\ref{fig_ldmd_error_distribution_a}) mostly remain below 8 degrees, with few outliers, whereas semi-major axis errors (Fig.~\ref{fig_ldmd_error_distribution_b}) typically remain under 800 km, with larger deviations occurring at higher steps. Overall, short-term predictions are more reliable, while error accumulation highlights the limitations in longer-term forecasts.

To compare the performance of LDMD and the state transition matrix (STM) approach, 100 initial-test point pairs were selected with a minimum phase-space distance of $2 \times 10^{-4}$. Each test point was numerically propagated to the 8th periapsis ($k=8$), and predictions were performed as follows:

\begin{itemize}
    \item LDMD prediction is performed as $\bm{x}_8^{\mathrm{LDMD}} = \bm{A}_{LDMD}^7 \bm{x}_1'$.
    \item STM prediction is computed using $\bm{x}_8^{\mathrm{STM}} = \bm{x}_1^{\mathrm{ref}} + \Phi(t_8)\delta \bm{x}$, where $\Phi(t_8)$ denotes the state transition matrix of the reference trajectory integrated up to the 8th periapsis.
\end{itemize}

\begin{figure}[h!]
  \begin{minipage}[b]{0.48\hsize}
    \centering
    \includegraphics[width=\textwidth]{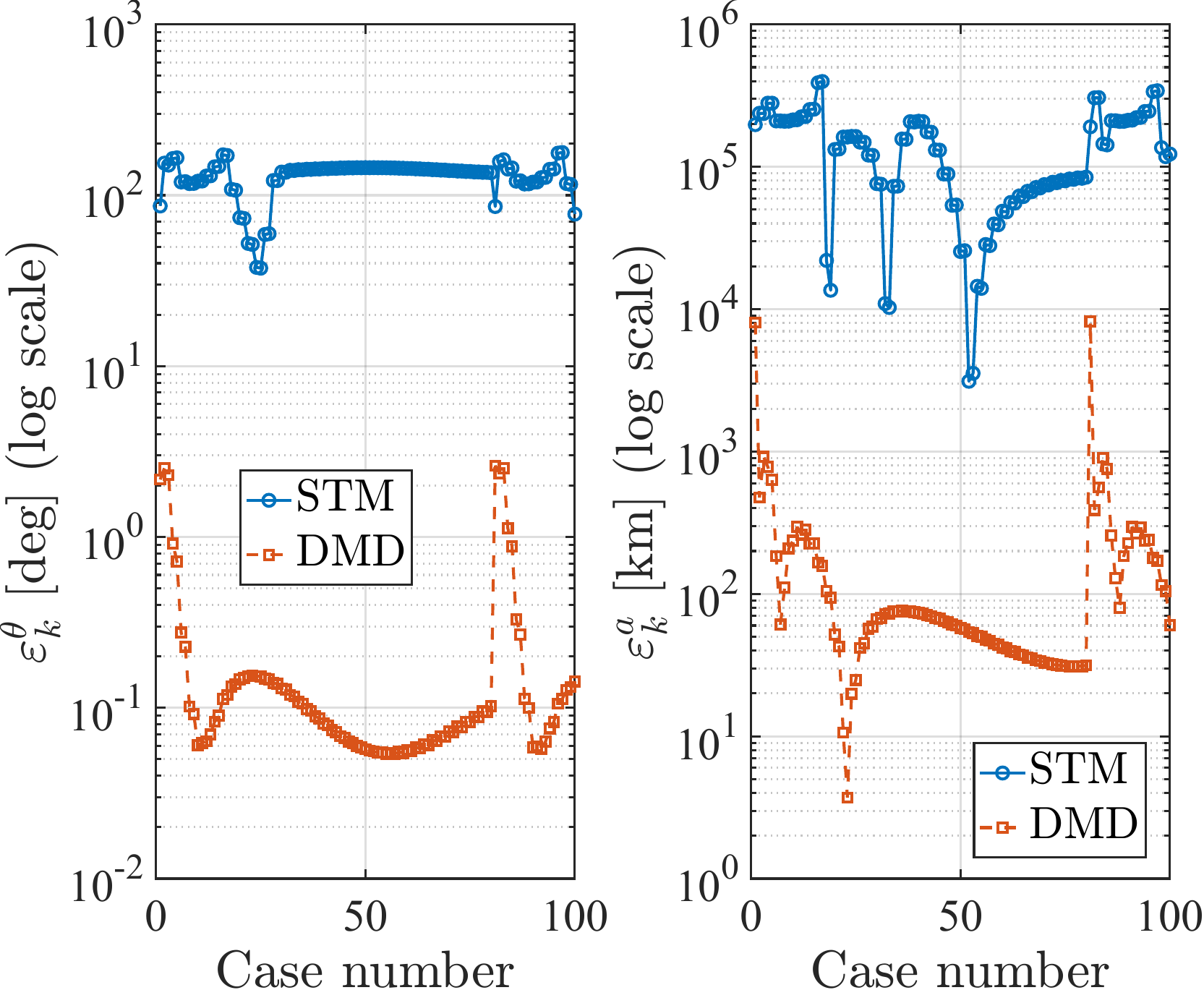}
    \subcaption{Approximation errors distribution.}
    \label{fig_ldmd_error_STM_DMD_compare_a}
    \end{minipage}
    \begin{minipage}[b]{0.48\hsize}
    \centering
    \includegraphics[width=\textwidth]{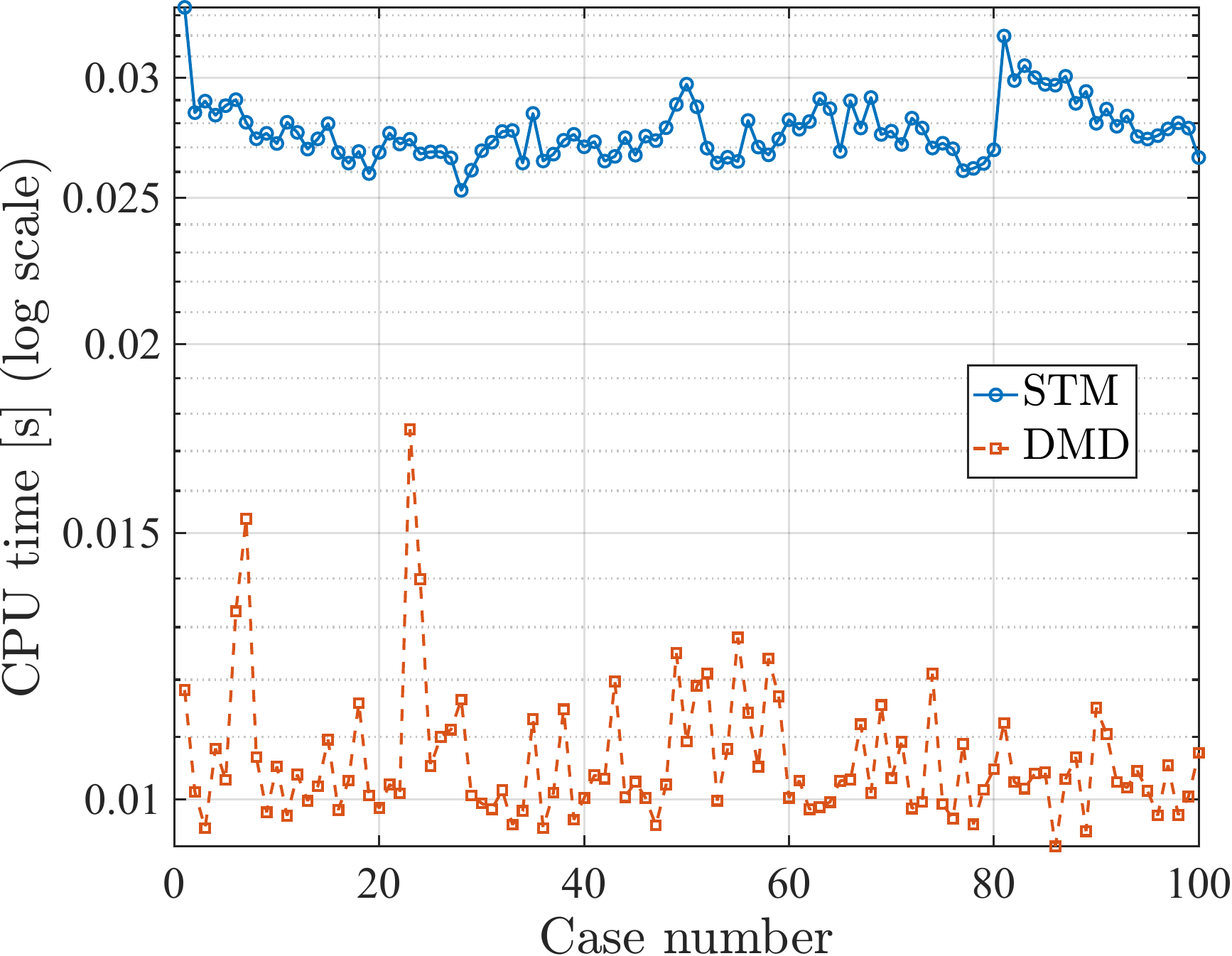}
    \subcaption{CPU times distribution.}
    \label{fig_ldmd_error_STM_DMD_compare_b}
    \end{minipage}
    \caption{\textcolor{black}{Comparison of STM- and LDMD-based trajectory predictions on the PPM. The case study considers initial and test points separated by $2 \times 10^{-4}$, using 100 data points.}}
    \label{fig_ldmd_error_STM_DMD_compare}
\end{figure}

Prediction errors in the $(\theta, a)$ space were evaluated against the numerically propagated results. As shown in Fig.~\ref{fig_ldmd_error_STM_DMD_compare_a}, the STM approach exhibits larger errors, primarily due to misalignment between the integration time and the perturbed trajectory’s periapsis. In contrast, Fig.~\ref{fig_ldmd_error_STM_DMD_compare_b} demonstrates that LDMD achieves lower computational cost by avoiding long-time propagation and integration of variational equations. Consequently, LDMD predictions require only matrix multiplications, providing an efficient framework for high-accuracy periapsis state estimation.

\subsubsection{Analysis of FTLE Field}
\label{sec_III_FTLE}

\begin{figure}[h!]
    \begin{minipage}[c]{0.48\hsize}
    \centering
    \includegraphics[width=\textwidth]{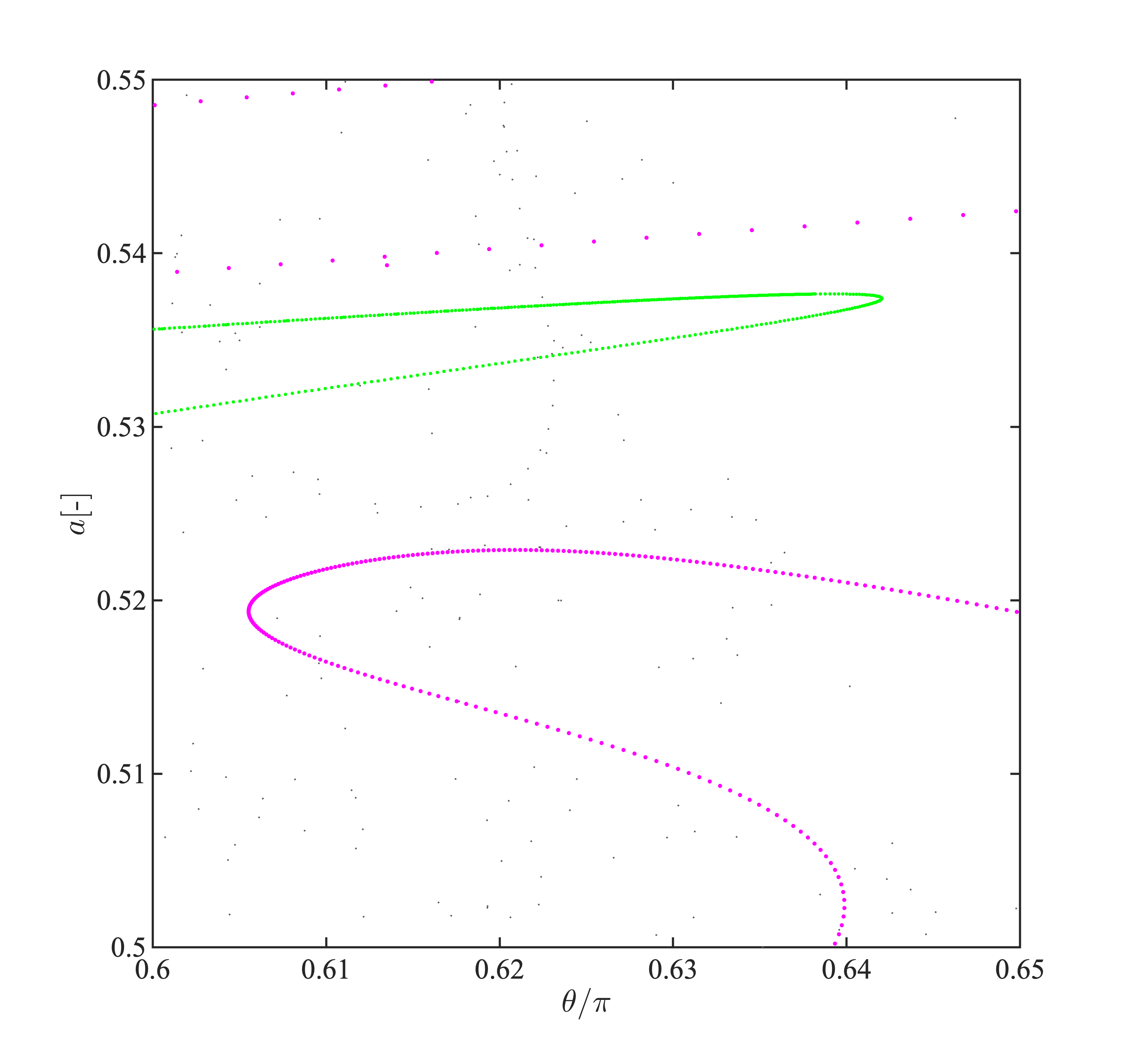}
    \subcaption{Region where manifold structures exist.}
    \label{fig_ftle_local_manifold_a}
    \end{minipage}
    \begin{minipage}[c]{0.48\hsize}
    \centering
    \includegraphics[width=\textwidth]{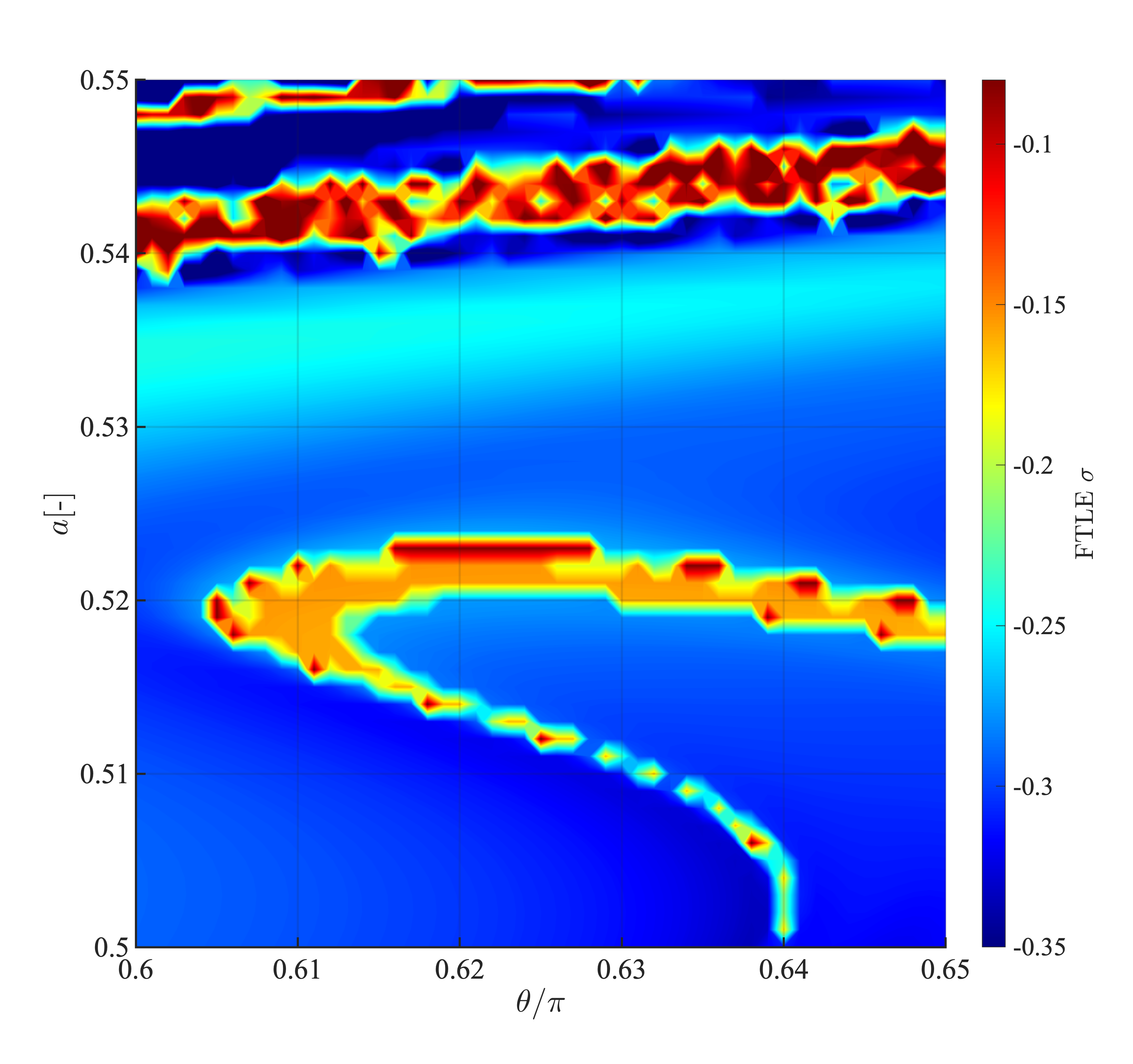}
    \subcaption{FTLE field in (a) region.}
    \label{fig_ftle_local_manifold_b}
    \end{minipage}
    \caption{\textcolor{black}{FTLE field in regions where the manifolds exist.}}\label{fig_ftle_local_manifold}
\end{figure}

The Finite-Time Lyapunov Exponent (FTLE) quantifies the exponential rate of separation between initially neighboring trajectories over a finite time interval~\cite{scheeres2016orbital}. As a scalar field defined over the phase space, it highlights regions of high local sensitivity to initial conditions, which are often associated with chaotic transport structures. It is widely used in dynamical systems to identify transport barriers and coherent structures. In the context of a discrete system governed by
\begin{align}
    \delta{x}(k+1) = A \delta{x}(k)
    \label{eq:risansystem}
\end{align}
The state transition matrix starting from step 1 is
\begin{align}
    \Phi(k, 1) = A^{k-1}
\end{align}
The finite-time evolution of an initial perturbation $\delta{x}_1$ is expressed as
\begin{align}
    \delta{x}_k = \Phi(k, 1) \delta{x}_1
\end{align}
The FTLE at point $x_1$ over $k$ steps is computed as
\begin{align}
    \sigma(x_1, k) = \frac{1}{k-1} \log \sqrt{ \lambda_{\max}(\Phi^T \Phi) }
\end{align}
where $\lambda_{\text{max}}$ denotes the largest eigenvalue of the matrix $\Phi^T\Phi$. Figure~\ref{fig_ftle_local_manifold} shows the FTLE field computed for the region surrounding the training data used in LDMD. The ridges in the FTLE field correspond to transport boundaries in phase space, indicating the presence of underlying manifold structures. Data-driven modeling offers an alternative pathway that does not require prior knowledge of periodic orbits—particularly beneficial for chaotic systems where identifying such orbits is challenging.

\subsection{Global Deformation Map}
\label{sec_IV_gdmd}

The Global Deformation Map-based DMD (GDMD) provides a global prediction framework that captures transitions across the entire PPM. In contrast to LDMD, which constructs discrete maps from locally clustered periapsis data, GDMD leverages data distributed throughout the phase space to model large-scale chaotic scattering dynamics. This broader coverage allows GDMD to represent global deformation patterns in the PPM more comprehensively.

\begin{figure}[h!]
  \centering
  \includegraphics[width=3.0in]{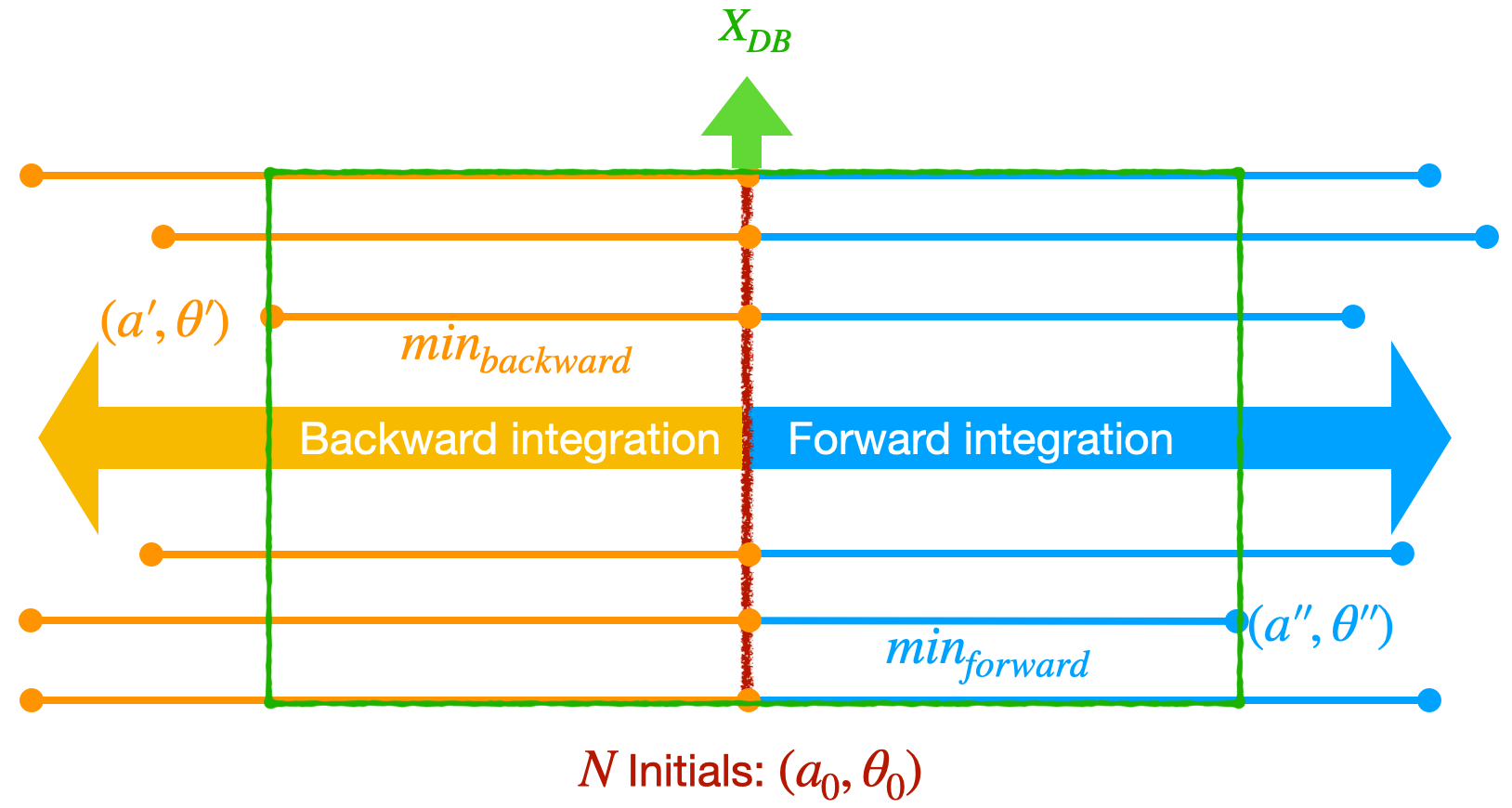}
  \caption{\textcolor{black}{Schematic diagram of GDMD database generation.}}
  \label{fig_GDMD_database_generation}
\end{figure}

\textcolor{black}{Figure~\ref{fig_GDMD_database_generation} illustrates the generation process of the GDMD dataset. The red line represents \( N \) initial conditions \((a_0, \theta_0)\), randomly sampled from the specified PPM region. Each initial condition is propagated both forward and backward in time. Integration terminates when the trajectory encounters either the Earth or the Moon, or exits the Earth-Moon region. The quantity $\texttt{min}_{\texttt{forward}}$ denotes the minimum number of periapses observed among all initial conditions during forward integration over a sufficiently long time interval. Similarly, $\texttt{min}_{\texttt{backward}}$ denotes the minimum number of periapses observed among all initial conditions during backward integration. To ensure sufficient data coverage, only trajectories satisfying the criterion:
\[
\texttt{min}_{\texttt{forward}} + \texttt{min}_{\texttt{backward}} + 1 \geq 500
\]
are retained, where the ‘+1’ accounts for all initial conditions. From the total of 533 sampled orbits, 500 suitable periapsis points are extracted to form the final dataset \( X_{DB} \). Table~\ref{table:compare} summarizes the key differences between LDMD and GDMD. LDMD uses \( m = 8 \) periapsis points per trajectory and samples \( n = 1681 \) orbits to ensure local resolution and prediction accuracy. In contrast, GDMD employs \( m = 500 \) periapsis points, relying on longer integration durations to cover broader regions of phase space.
While LDMD operates with fewer data points, it requires a higher-dimensional matrix \( A \), resulting in increased computational cost. Conversely, GDMD constructs a lower-dimensional matrix \( A \) from a larger dataset, thereby reducing computational complexity while capturing long-term chaotic behavior.}

\begin{figure}[htbp]
\centering
\includegraphics[width=\textwidth]{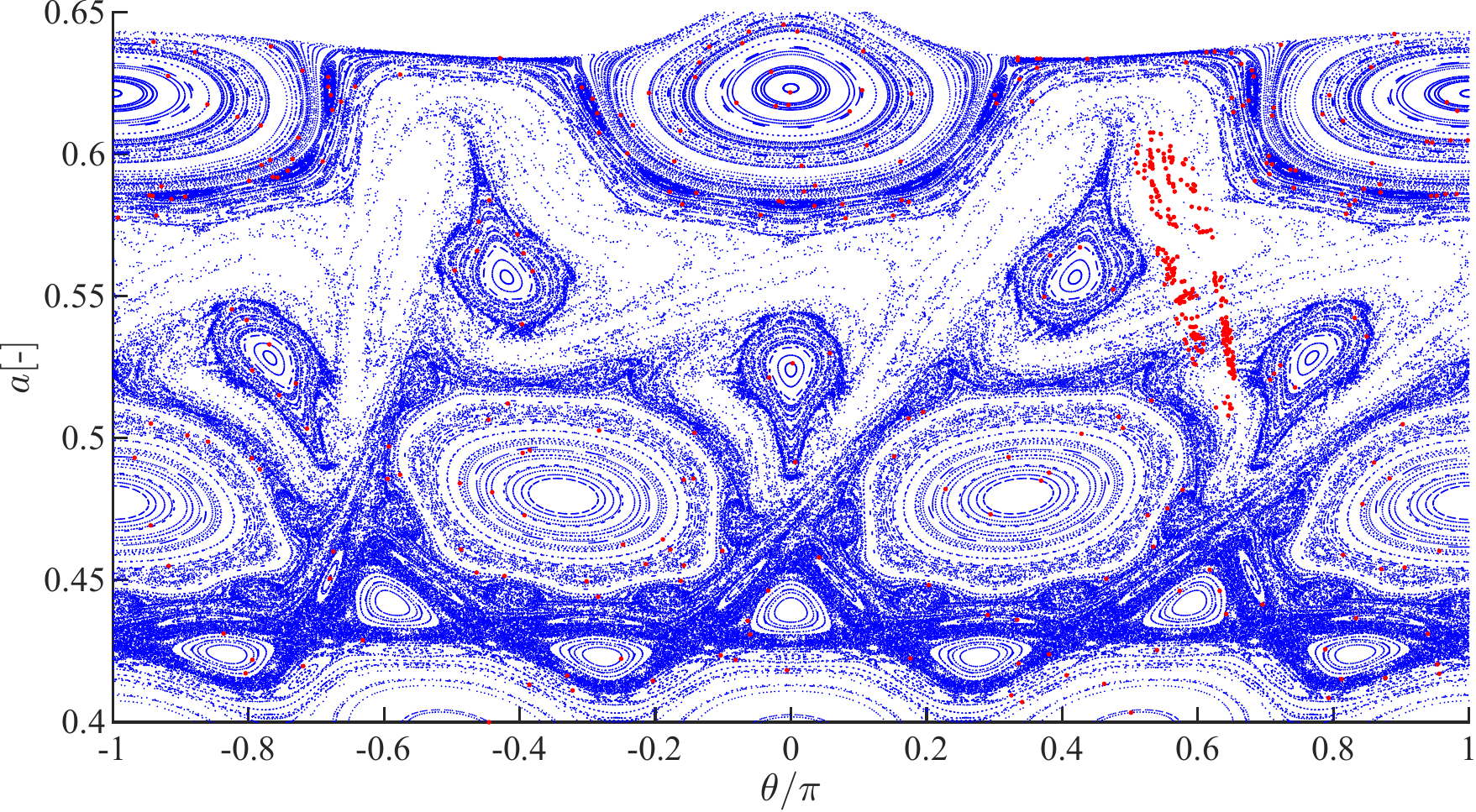}
\caption{Sampled periapsis data used in GDMD. Red dots indicate the initial states $\bm{x}_1$ used to construct the matrix $X_{DB}$, and blue dots represent all $(\theta, a)$ pairs in the database $X_{DB}$.}
\label{fig:global_database}
\end{figure}

Figure~\ref{fig:global_database} visualizes the sampled dataset. Red dots indicate the initial points \(\bm{x}_1\), and blue dots represent all \((\theta, a)\) pairs in \( X_{DB} \). 
From this database, the data matrices \( X \) and \( X' \) are constructed by selecting a consecutive index range from \( p \) to \( q \) (with \( p < q \)) for prediction. The matrices are given as:
\begin{align}
  X = \left[
  \begin{array}{cccc}
     \mid & \mid & \quad & \mid \\
    \bm{x}_p & \bm{x}_{p+1} & \ldots & \bm{x}_{q} \\
     \mid & \mid & \quad & \mid
  \end{array}
\right], \qquad 
  X' = \left[
  \begin{array}{cccc}
     \mid & \mid & \quad & \mid \\
    \bm{x}_{p+1} & \bm{x}_{p+2} & \ldots & \bm{x}_{q+1} \\
     \mid & \mid & \quad & \mid
  \end{array}
\right]
\end{align}

The discrete map \( A_{pq} \) is then computed via Eq.~\eqref{eq_dmd_compute_A}. By incorporating chaotic dynamics and long-term trajectory propagation, GDMD enables the reconstruction of global deformation structures in the PPM, revealing the influence of chaotic scattering dynamics across the phase space.


\begin{table}[htbp]
  \centering
  \caption{\textcolor{black}{Comparison of key data between LDMD and GDMD.}}
  \label{table:compare}
  \begin{tabular}{lll}
    \hline\noalign{\smallskip}
     &LDMD&GDMD\\
    \noalign{\smallskip}\hline\noalign{\smallskip}
    The number of periapsis $m$ & 8 & 500\\
    The number of orbits $n$ & 1681 & 533\\
    Total number of data $2mn$ & 26896 & 533000\\
    The dimension of $A$ $2n$ & 3362 & 1066\\ 
    \noalign{\smallskip}\hline
 \end{tabular}
\end{table}

\subsubsection{Data Recovery by GDMD}
\label{sec_III_gdmd_data_recovery}

\begin{figure}[htbp]
  \begin{minipage}[c]{0.48\hsize}
  \centering
  \includegraphics[width=1.0\linewidth,clip]{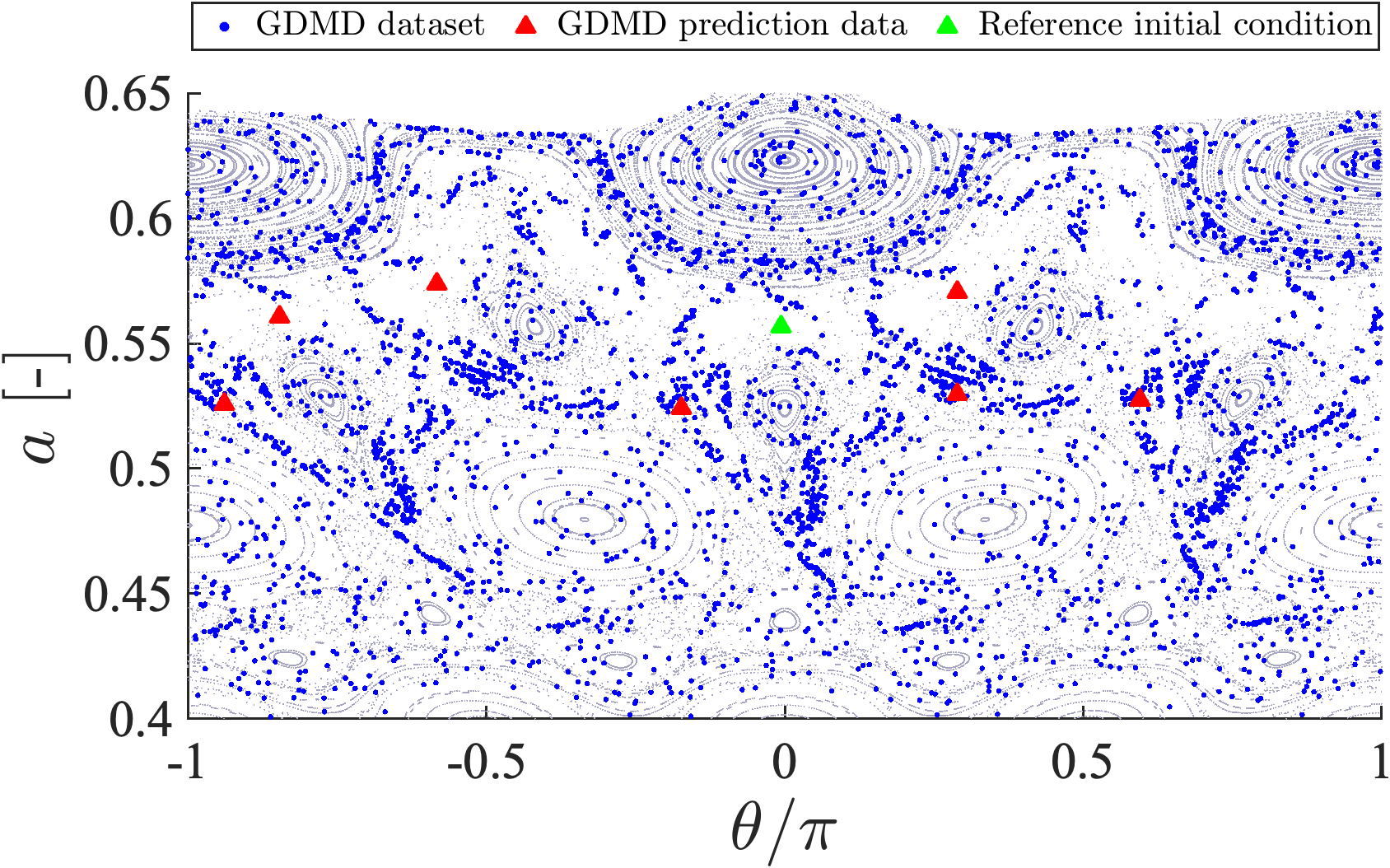}
  \subcaption{Data of periapsis extracted from $X_{DB}$.}
  \label{fig:global_data}
  \end{minipage}
  \begin{minipage}[c]{0.48\hsize}
  \centering
  \includegraphics[width=0.8\linewidth,clip]{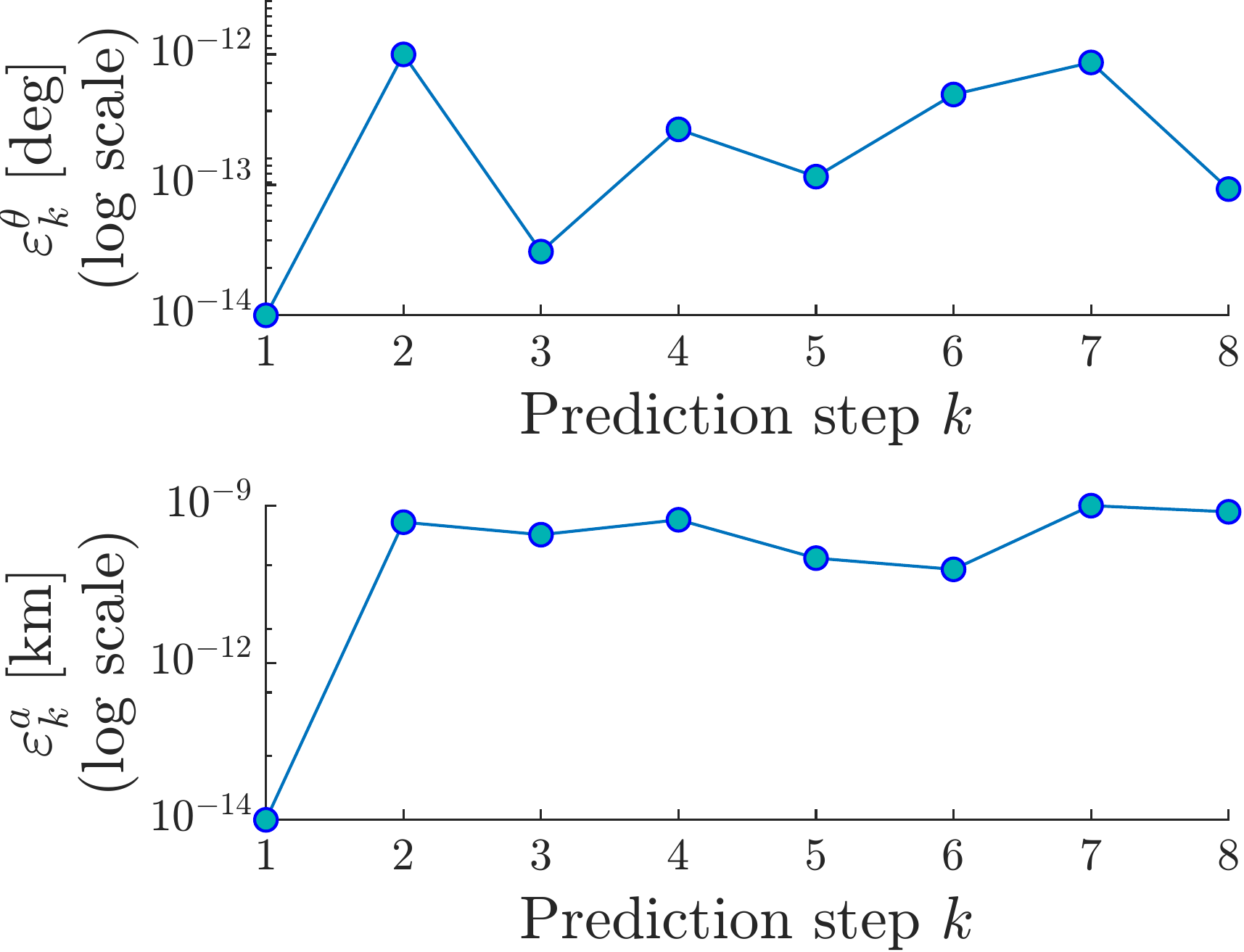}
  \subcaption{Approximation errors distribution in each step.}
  \label{fig:global_data_reconst}
  \end{minipage}
  \caption{Data recovery in GDMD.}
\end{figure}

Data recovery using GDMD is performed by constructing a discrete mapping $A_{pq}$, which leverages scattered data to reconstruct the temporal evolution of periapsis states. For $p=11$ and $q=18$, Figure~\ref{fig:global_data} shows the extracted periapsis data in blue, the initial data in green, and the GDMD-predicted data in red. Figure~\ref{fig:global_data_reconst} depicts the approximation error between GDMD predictions and numerical computations. These discrepancies are negligible, demonstrating that GDMD can accurately reconstruct orbital trajectories even from sparsely sampled data.

\begin{figure}[htbp]
    \begin{tabular}{cc}
      \begin{minipage}[t]{0.48\hsize}
        \centering
        \includegraphics[width=1.0\linewidth,clip]{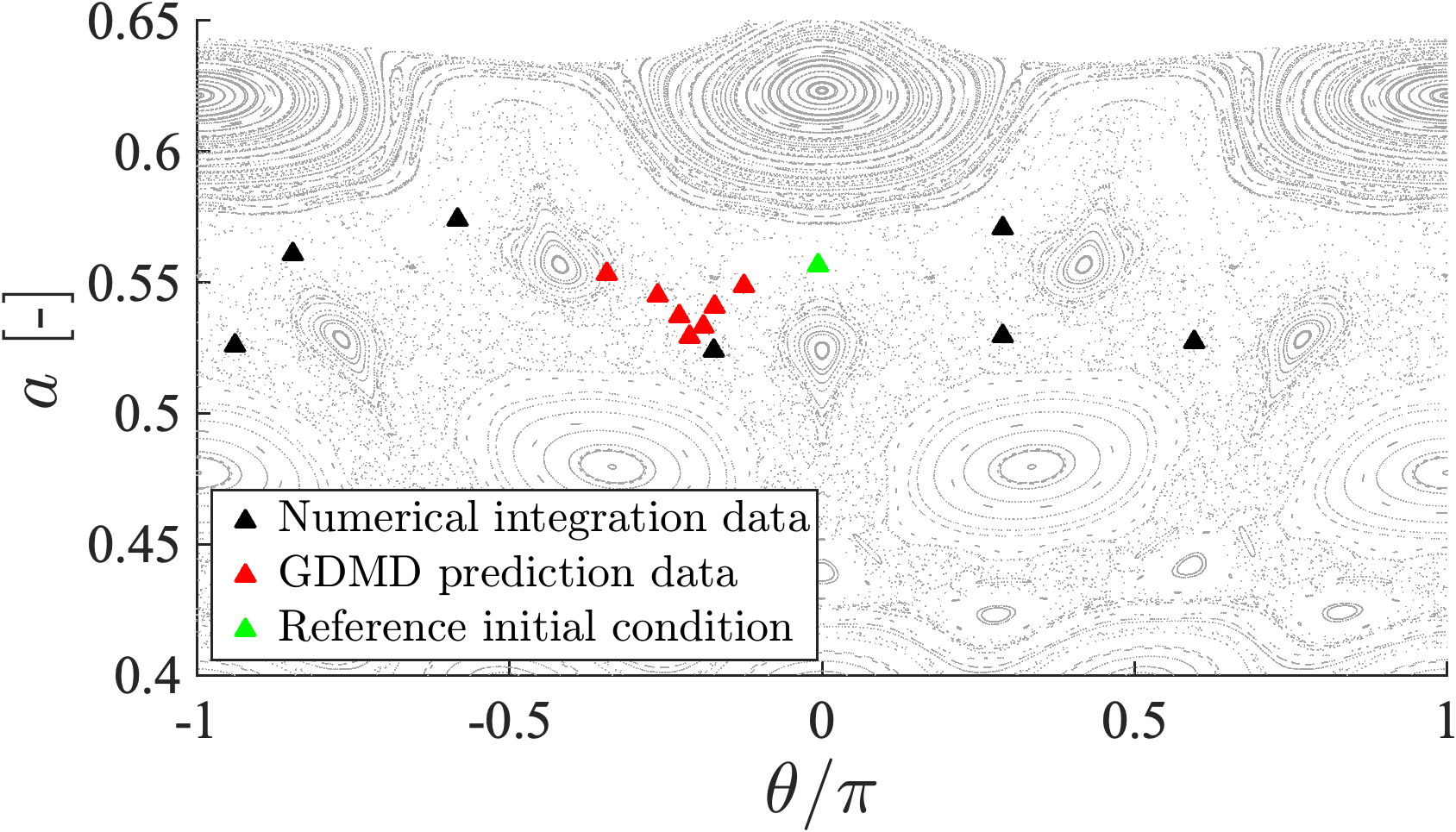}
        \subcaption{Periapsis evolution for $n=1$.} 
        \label{fig_data_recovery_ppm_n1}
      \end{minipage} &
      \begin{minipage}[t]{0.48\hsize}
        \centering
        \includegraphics[width=0.8\linewidth,clip]{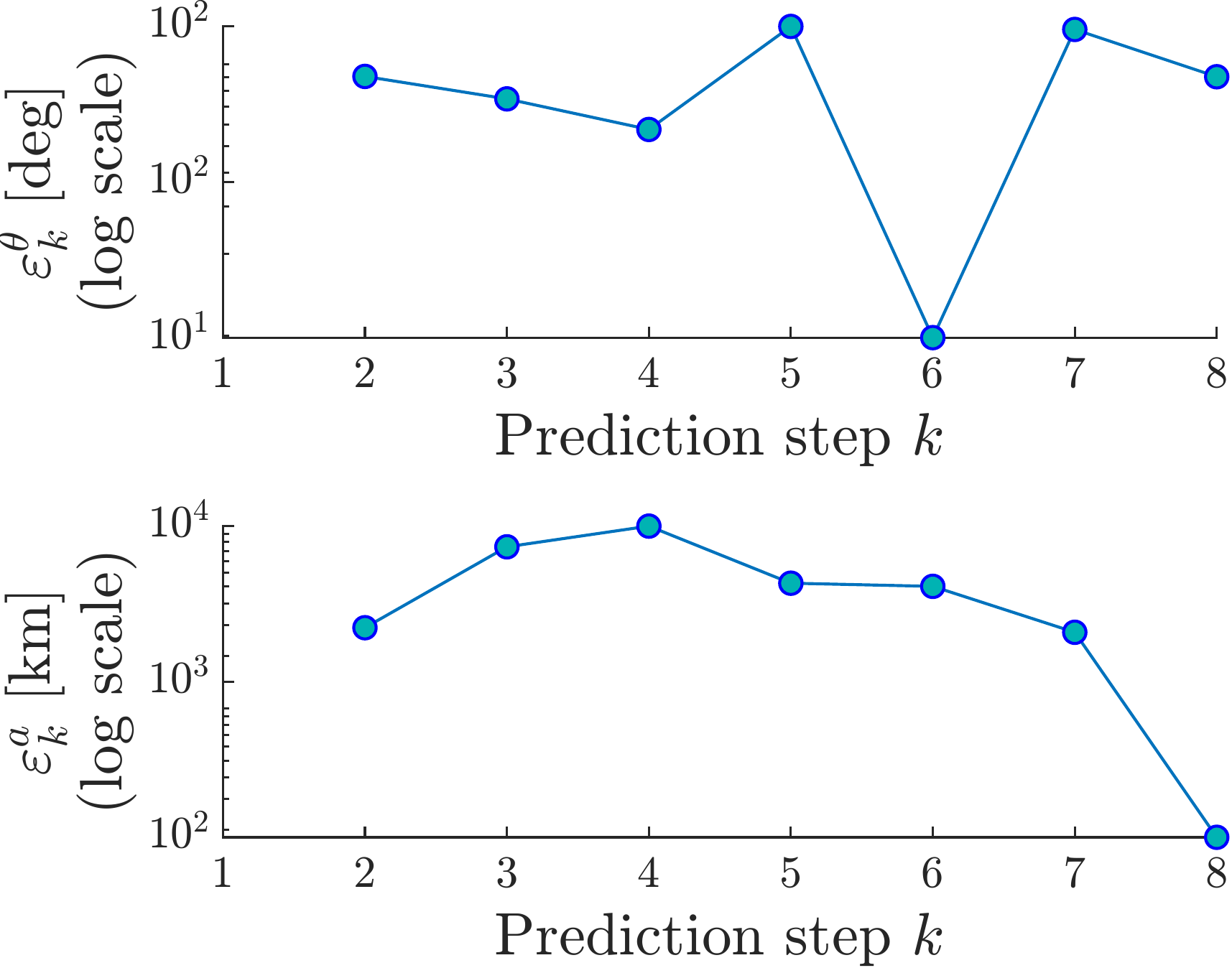}
        \subcaption{Data comparison for $n=1$.}
        \label{fig_data_recovery_atheta_n1}
      \end{minipage} \\
   
      \begin{minipage}[t]{0.48\hsize}
        \centering
        \includegraphics[width=1.0\linewidth,clip]{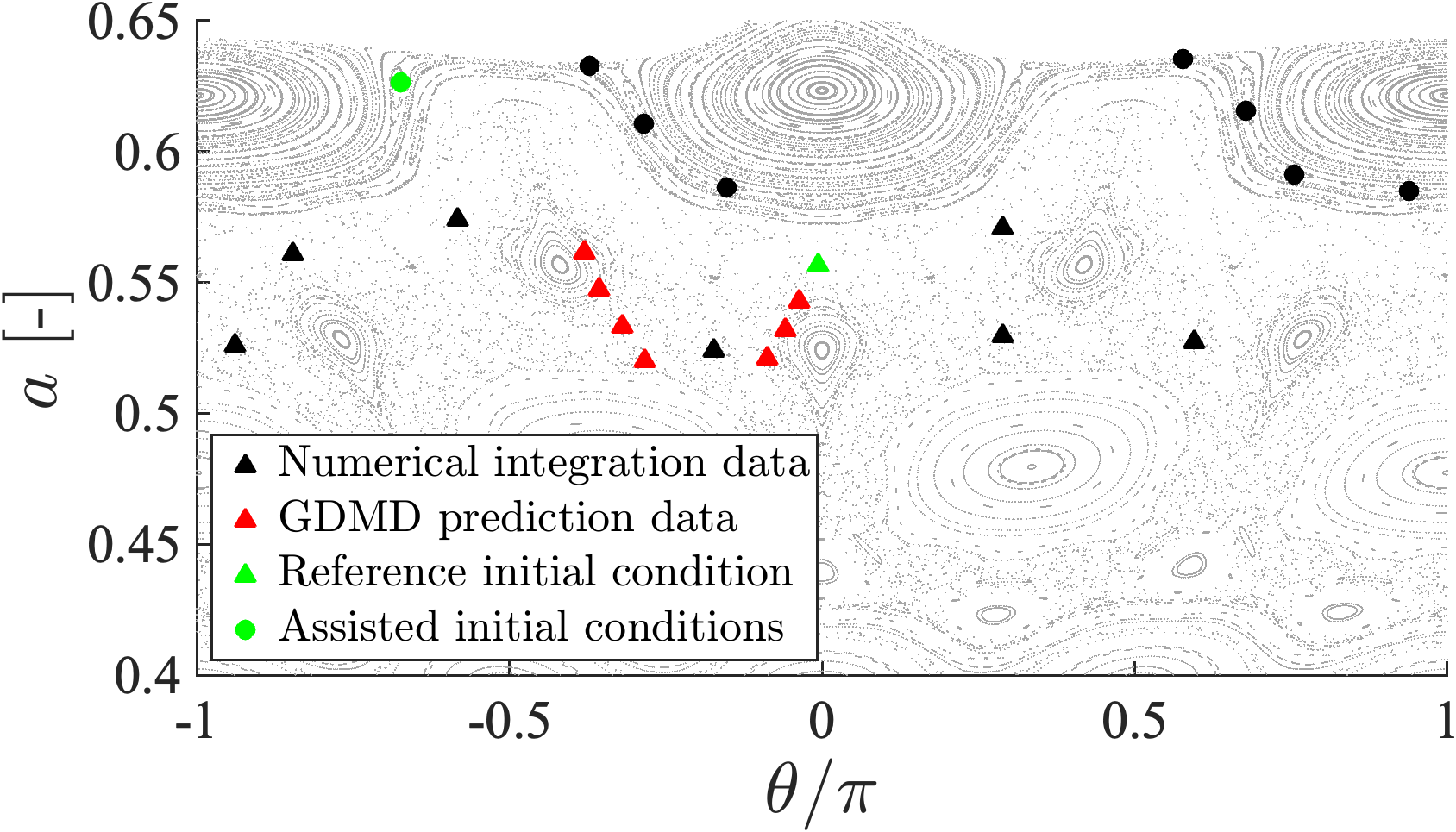}
        \subcaption{Periapsis evolution for $n=2$.}
        \label{fig_data_recovery_ppm_n2}
      \end{minipage} &
      \begin{minipage}[t]{0.48\hsize}
        \centering
        \includegraphics[width=0.8\linewidth,clip]{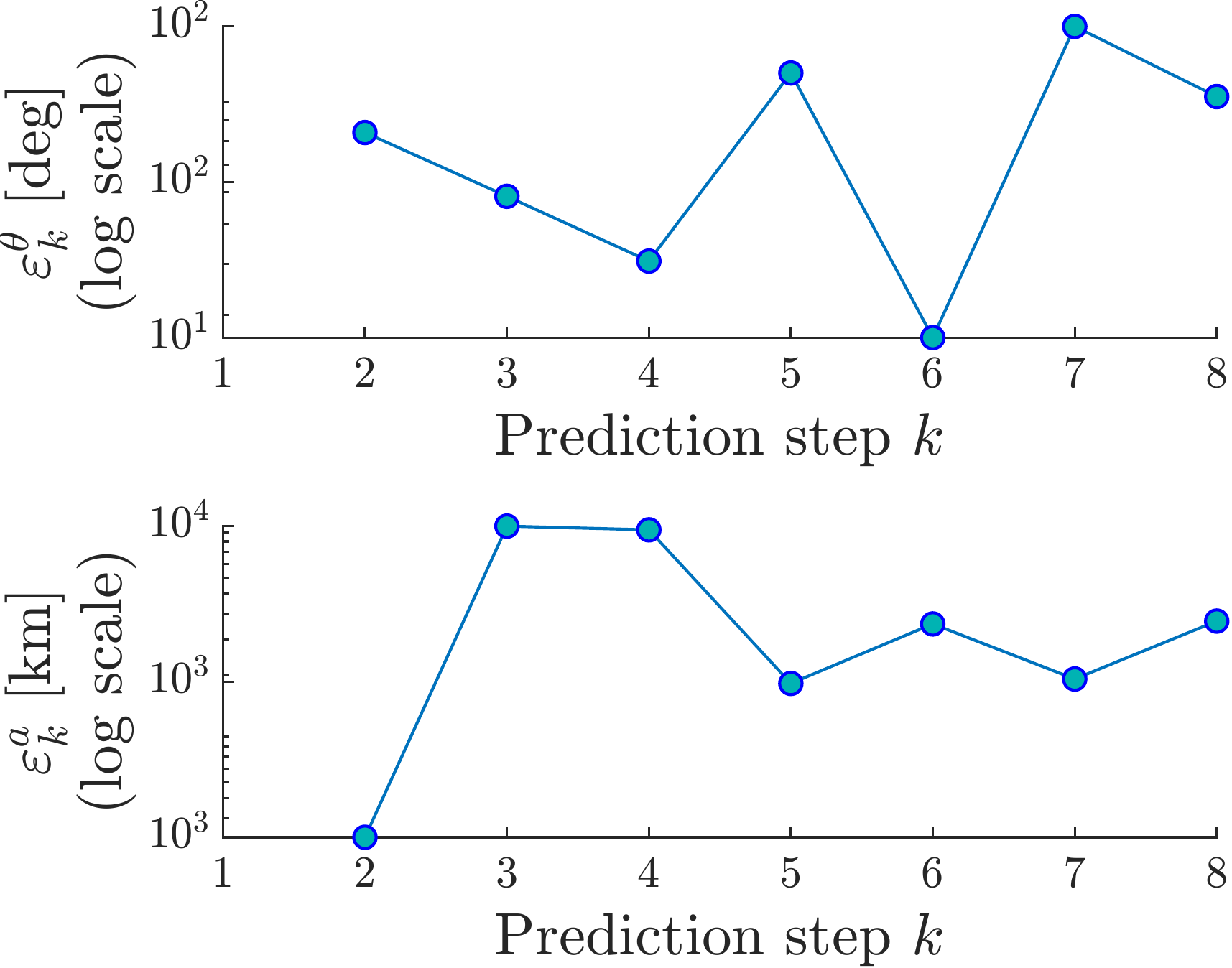}
        \subcaption{Data comparison for $n=2$.}
        \label{fig_data_recovery_atheta_n2}
      \end{minipage} \\

        \begin{minipage}[t]{0.48\hsize}
        \centering
        \includegraphics[width=1.0\linewidth,clip]{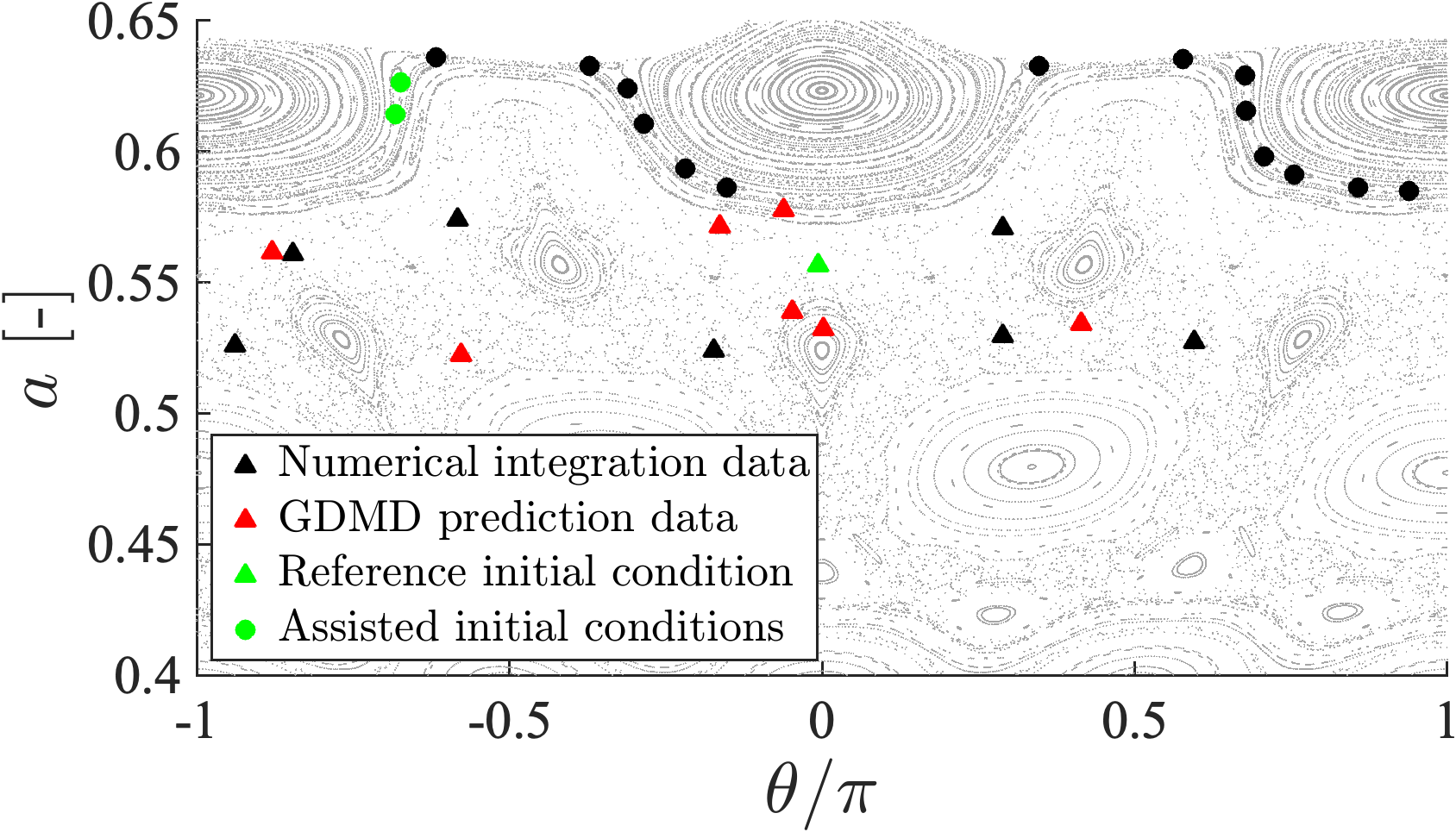}
        \subcaption{Periapsis evolution for $n=3$.}
        \label{fig_data_recovery_ppm_n3}
      \end{minipage} &
      \begin{minipage}[t]{0.48\hsize}
        \centering
        \includegraphics[width=0.8\linewidth,clip]{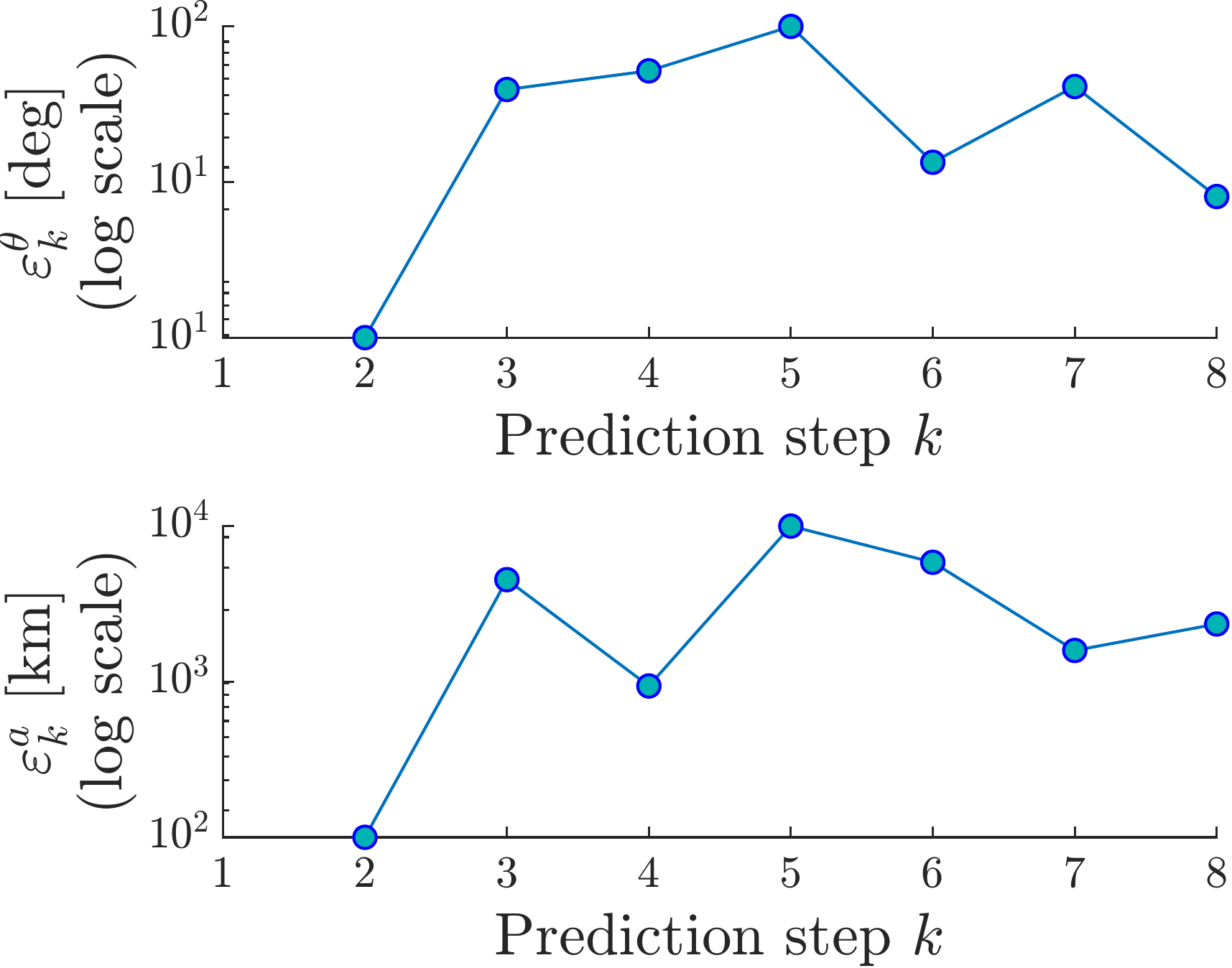}
        \subcaption{Data comparison for $n=3$.}
        \label{fig_data_recovery_atheta_n3}
      \end{minipage}\\
      
        \begin{minipage}[t]{0.48\hsize}
        \centering
        \includegraphics[width=1.0\linewidth,clip]{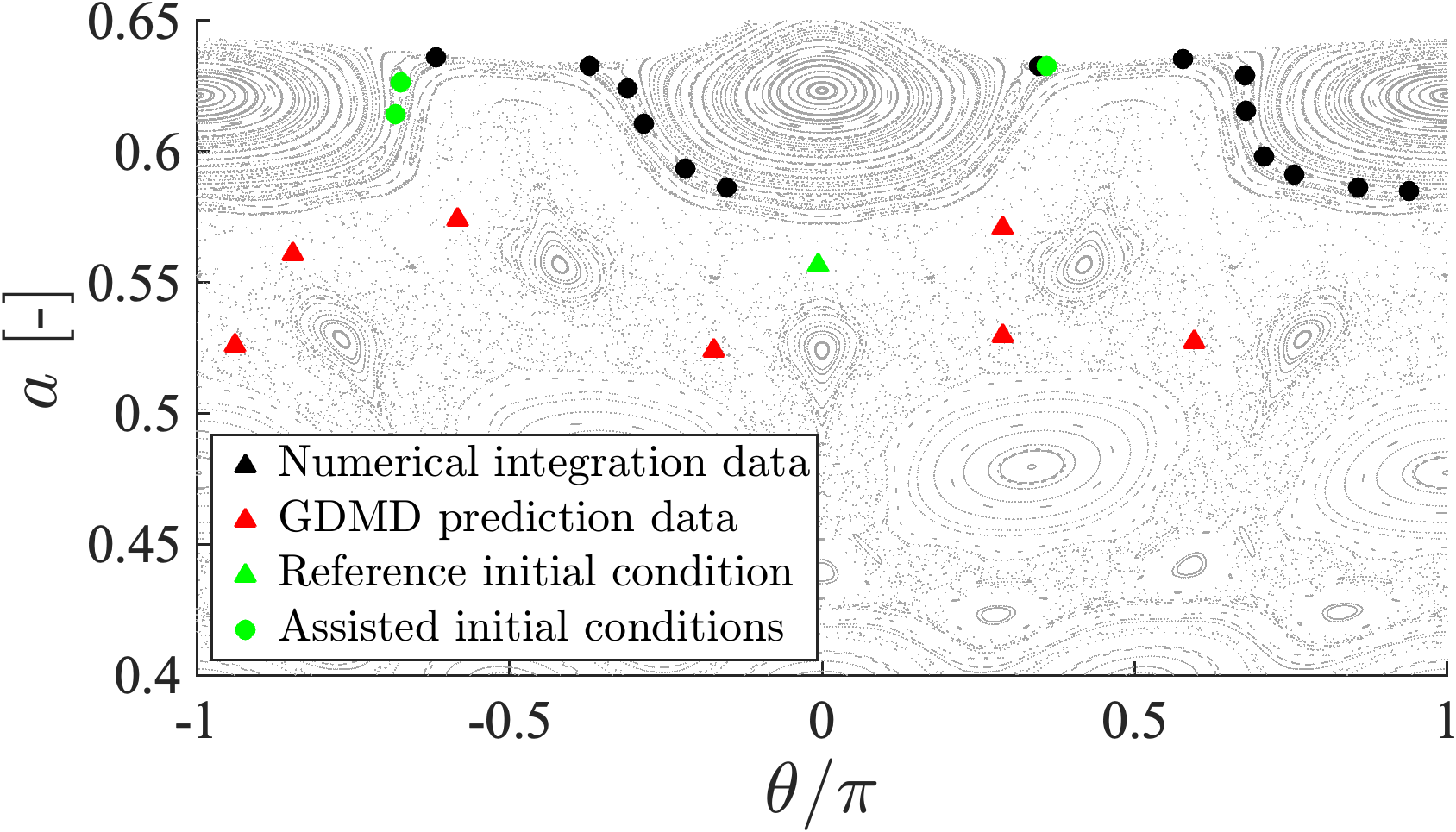}
        \subcaption{Periapsis evolution for $n=4$.}
        \label{fig_data_recovery_ppm_n4}
      \end{minipage} &
      \begin{minipage}[t]{0.48\hsize}
        \centering
        \includegraphics[width=0.8\linewidth,clip]{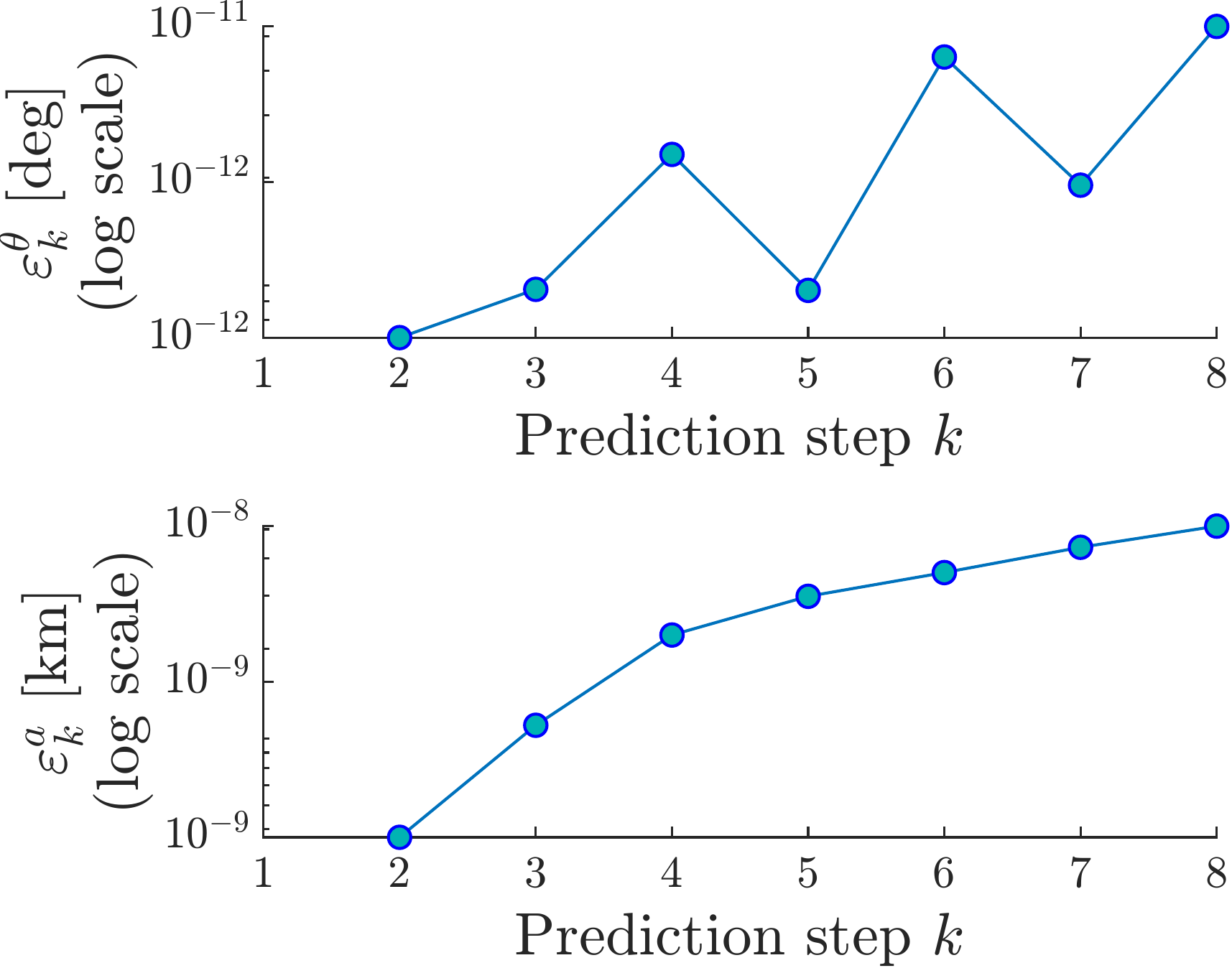}
        \subcaption{Data comparison for $n=4$.}
        \label{fig_data_recovery_atheta_n4}
      \end{minipage} 
    \end{tabular}
     \caption{\textcolor{black}{Data recovery results for $n=1, 2, 3, 4$.}}
  \end{figure}

To evaluate the required data volume, the test trajectory was initialized from random points within the chaotic region (green triangular marker in Fig.~\ref{fig_data_recovery_ppm_n1}). Approximation errors are presented in the second column; the first-step error is zero and therefore omitted. Using a single orbit ($n=1$) resulted in unsuccessful data recovery (Fig.~\ref{fig_data_recovery_atheta_n1}). Including a second orbit ($n=2$) near the 2:1 resonance yielded a slight improvement (Fig.~\ref{fig_data_recovery_atheta_n2}), but remained insufficient. With four orbits ($n=4$; Fig.~\ref{fig_data_recovery_ppm_n4} and~\ref{fig_data_recovery_atheta_n4}), accurate reconstruction is achieved. These results demonstrate that GDMD can construct reliable discrete maps using minimal initial points, without the need for carefully preselected orbits. Even with such limited and randomly chosen data, GDMD effectively captures chaotic transport structures in the proximity map, underscoring its robustness and practical utility as a dimensionality reduction tool for predicting chaotic trajectories.

\section{Discussion}
\label{sec_V_discussion}

The variation in the semimajor axis, $a$, within the PPM can be expressed as follows:
\begin{align}
     a_{k+1} = a_k + \delta a(a_k, \theta_k)    
\end{align}
where \(\delta a\) is referred to as the kick function~\cite{ross2007multiple}. 
The kick function captures the gravitational perturbation induced by the Moon near resonance, and plays a central role in explaining chaotic transport in the CRTBP.

\begin{figure}[tbph]
    \centering
    \includegraphics[width=0.8\linewidth,clip]{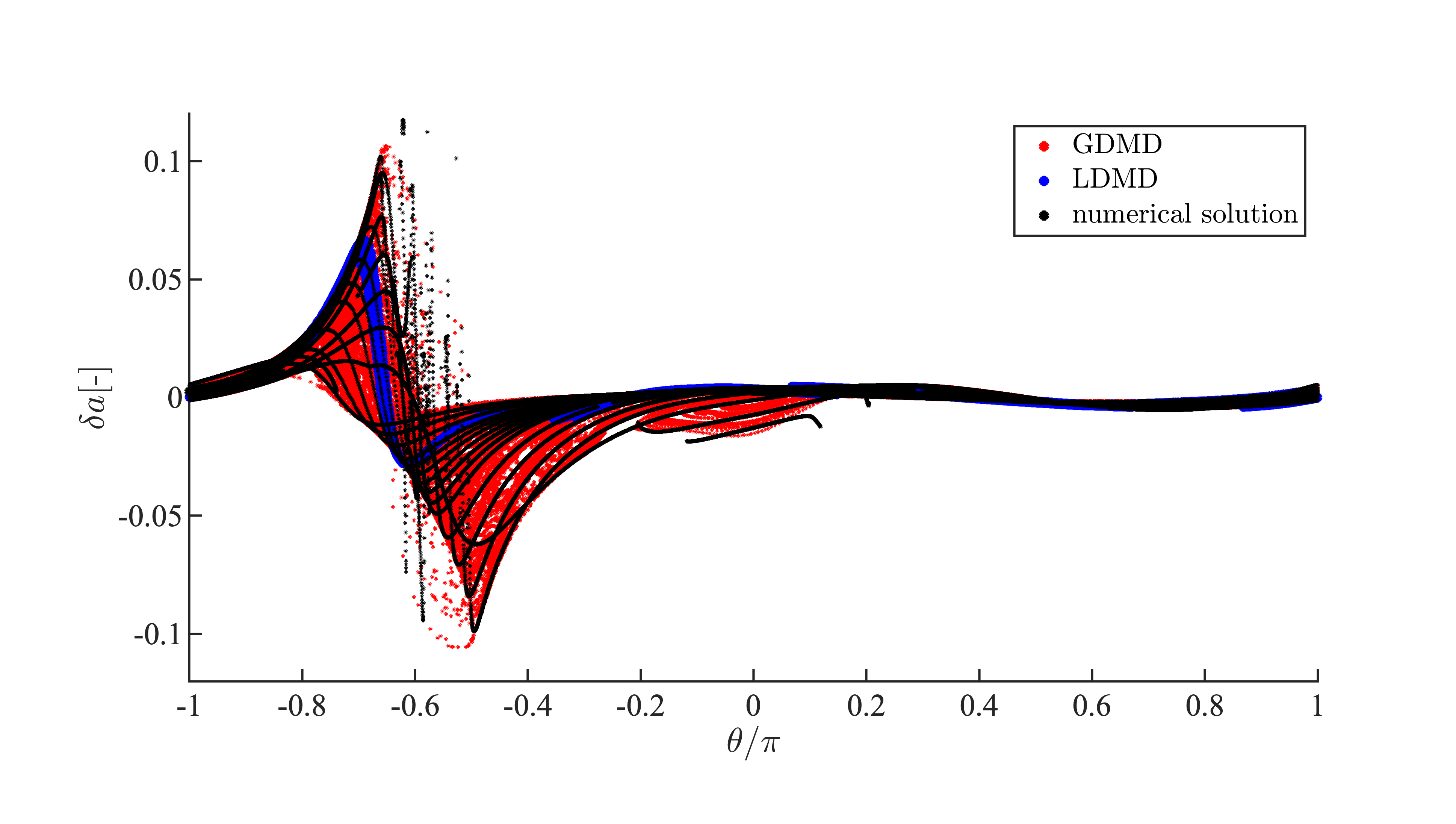}
    \caption{\textcolor{black}{Comparison of the kick function obtained from LDMD, GDMD, and the numerical solution.}}
    \label{fig:dynamics}
\end{figure}

Figure~\ref{fig:dynamics} presents the variation of \(\delta a\) as a function of the phase angle \(\theta\), where 17 curves obtained from numerical simulations are shown for different values of \(a \in [0.4, 0.65]\). 
Superimposed on these are the \(\delta a(\theta)\) functions reconstructed using LDMD and GDMD. 
The numerical results exhibit sharp peaks in the range \(-0.9\pi < \theta < -0.2\pi\), particularly around \(\theta = -0.7\pi\), which corresponds to close approaches to the Moon and is indicative of strong gravitational perturbations. 
The LDMD reconstruction captures this sharp variation around \(\theta = -0.7\pi\), reflecting the behavior of the kick function at a fixed value of \(a\). 
In contrast, the GDMD result reflects a superposition of multiple \(a\)-dependent kick functions, producing a broader structure that encompasses a wide range of dynamical behaviors.

These results demonstrate that both LDMD and GDMD capture the key transport mechanisms encoded in the kick function, and highlight the capability of data-driven models to reproduce resonance-driven chaotic dynamics in a reduced-order form.

\section{Application of discrete map to spacecraft trajectory design}
\label{sec_VI_application}

\begin{figure}[tbph]
  \centering
  \includegraphics[width=0.8\linewidth,clip]{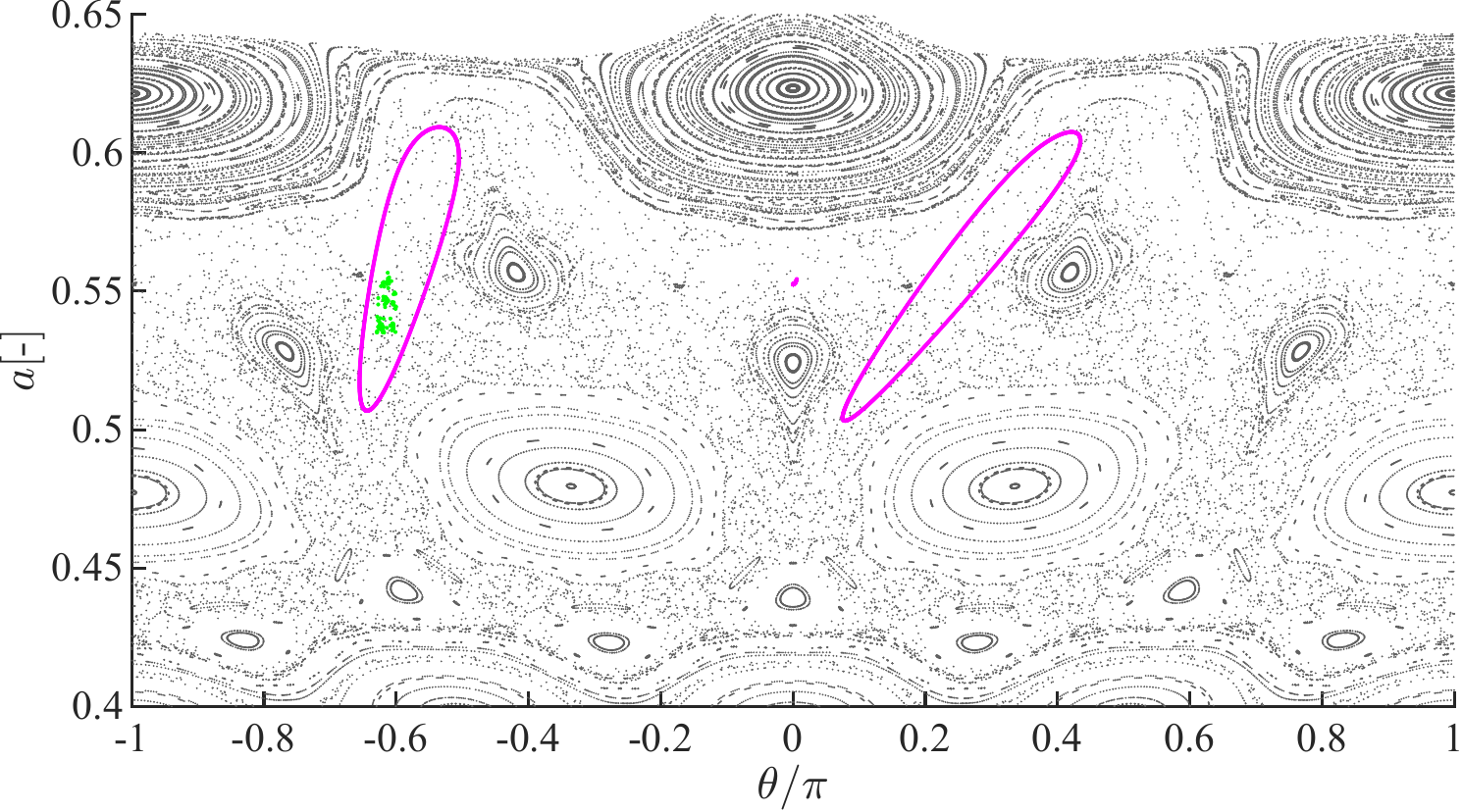}
  \caption{Target points in the PPM.}
  \label{fig_target_points_PPM}
\end{figure}

A method for targeting a desired periapsis using discrete mappings is presented. The targeting process involves determining an initial state that guides the trajectory to a specified location \cite{bollt1995targeting}. By selecting a target point within the manifold tube associated with the $L_1$ Lyapunov orbit, a transfer trajectory to the Moon can be designed. As an example, a tube structure formed by the backward propagation of the first periapses along the stable manifold of $L_1$ is considered, where multiple periapses exhibit similar dynamical characteristics. First, the periapses within the tube are sampled (see the green dots in Fig.~\ref{fig_target_points_PPM}). Then, periapses exhibiting similar dynamical behavior, i.e., those close to the target points in the PPM, are identified in the GDMD database ($X_{DB}$) and replaced accordingly. During target localization, the matrices $X$ and $X'$ are constructed from time-series data, where the time steps are reversed.
\begin{align}
    {X} = \left[
    \begin{array}{cccc}
       \mid & \mid & \quad & \mid \\
      \bm{x}_m & \bm{x}_{m-1} & \ldots & \bm{x}_{2} \\
       \mid & \mid & \quad & \mid
    \end{array}
  \right], \quad 
    X' = \left[
    \begin{array}{cccc}
      \mid & \mid & \quad & \mid \\
      \bm{x}_{m-1} & \bm{x}_{m-2} & \ldots & \bm{x}_1 \\
       \mid & \mid & \quad & \mid
    \end{array}
  \right]
\end{align}

\begin{figure}[tbph]
    \begin{minipage}[c]{1.0\columnwidth}
    \centering
    \includegraphics[width=0.65\columnwidth]{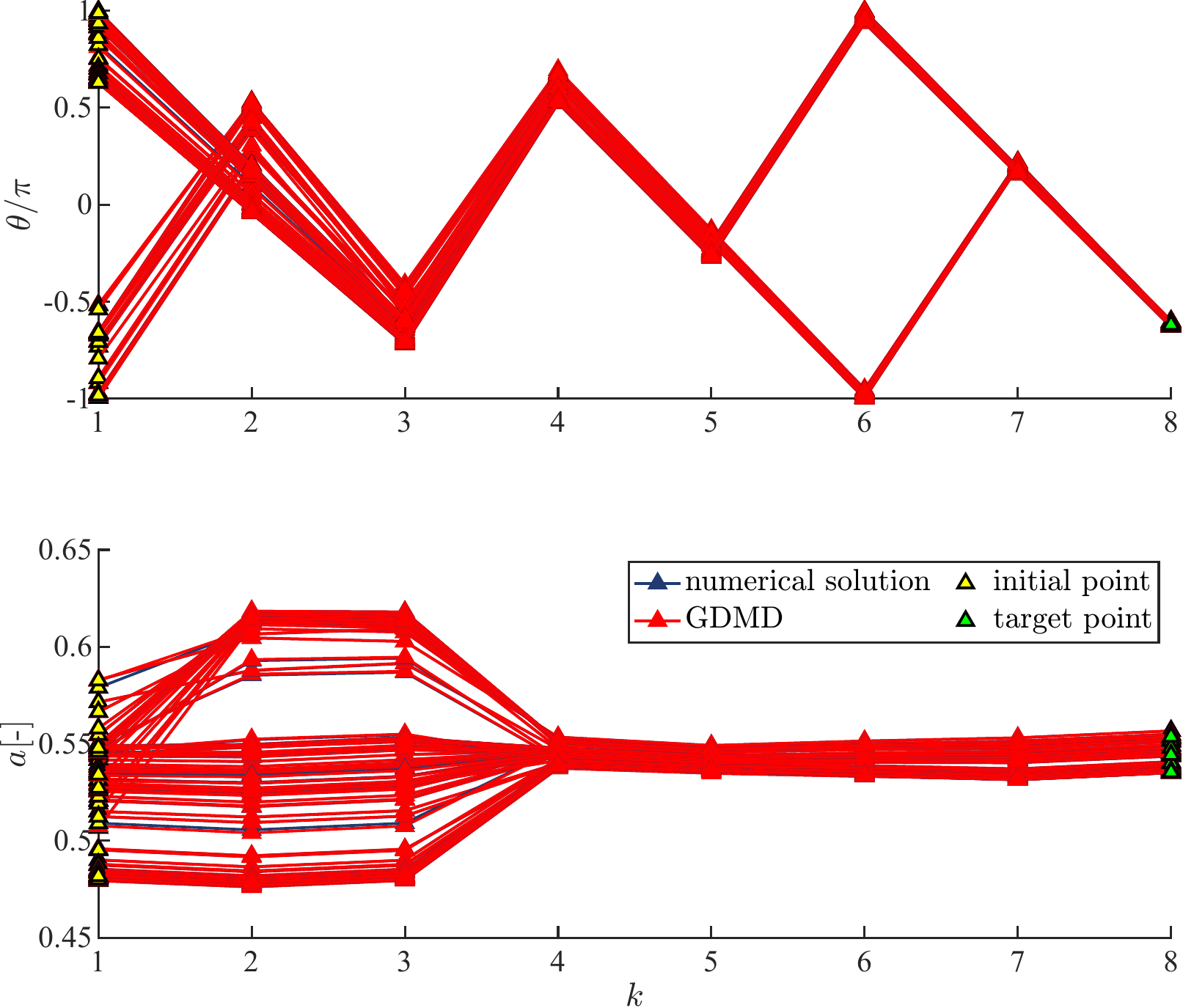}
    \subcaption{Targeting of chaotic trajectories by discrete mapping.}
    \label{fig_target_points_trajectory_a}
    \end{minipage}
    \begin{minipage}[c]{1.0\columnwidth}
    \centering
    \includegraphics[width=0.65\columnwidth]{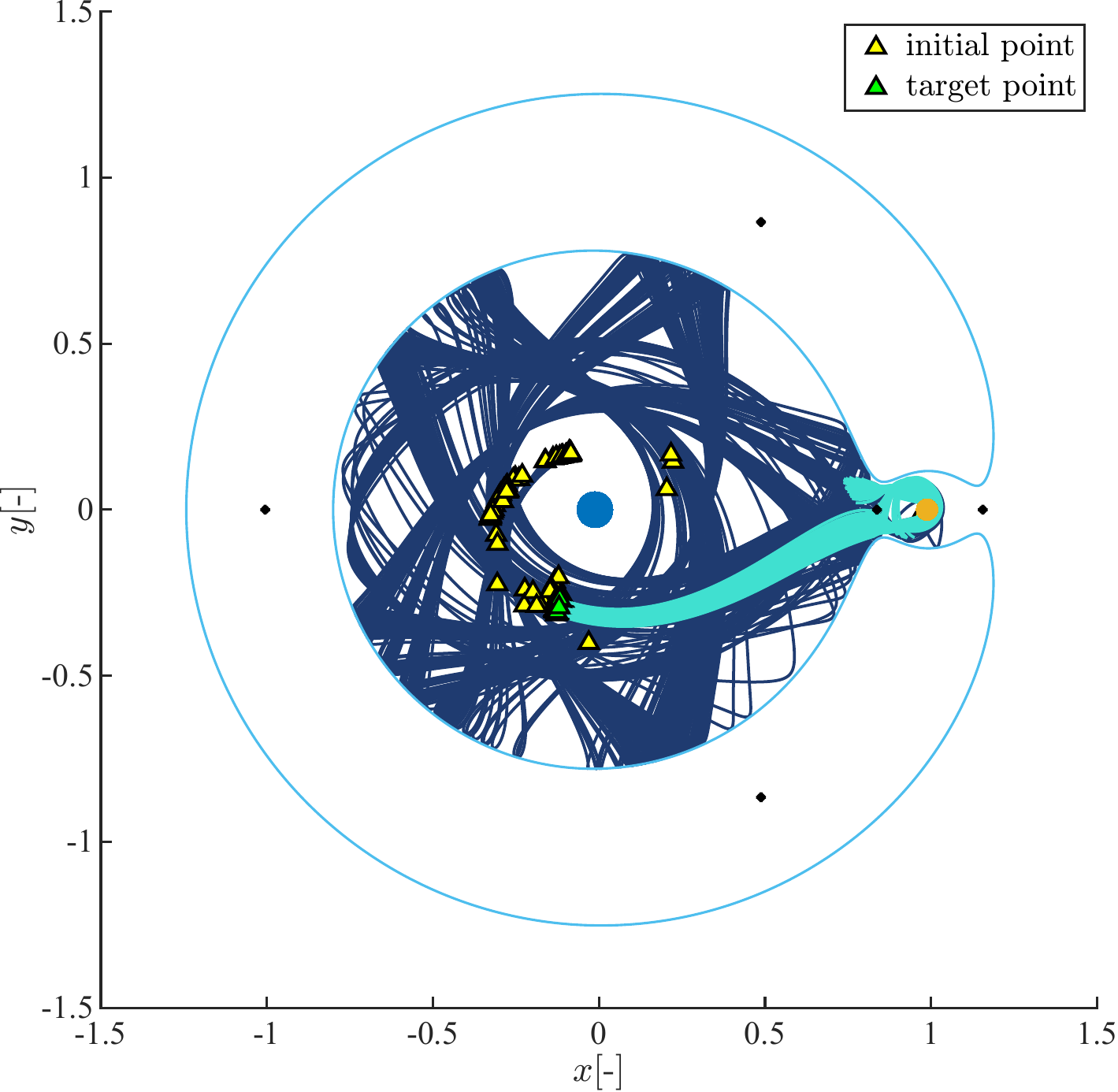}
    \subcaption{Trajectory in the rotating frame.}
    \label{fig_target_points_trajectory_b}
    \end{minipage}
    \caption{Orbits obtained from initial values by targeting.}
    \label{fig_target_points_trajectory}
\end{figure}

The discrete map, constructed from the back-propagated data, is then used to target the desired point $\bm{x}^*$. To compute this map, \(\bm{x}^*\) is substituted with its nearest neighbor \(\bm{x}_m\) in \(X_{DB}\), and the mapping is iterated backward in time. Figure~\ref{fig_target_points_trajectory_a} shows the results, showing the predicted values of $\theta$ and $a$ for the first seven steps of $\bm{x}^*$. The resulting initial state $(\theta, a)$ is then transformed into position and velocity and propagated forward to obtain the full trajectory using Eq.~\eqref{eq_crtbp_eoms}, as depicted in Fig.~\ref{fig_target_points_trajectory_b}. The results confirm that the trajectory successfully passes through the tube structure, successfully reaching the Moon, validating the effectiveness of the targeting method.

\section{Conclusion}
This study proposed two methodologies for predicting high-dimensional chaotic trajectories through low-dimensional discrete maps derived from data-driven modeling techniques. The LDMD method demonstrated the capability to predict periapsis transitions of chaotic trajectories within a localized region over a finite prediction horizon. However, relatively large prediction errors were observed at the boundaries of the manifold structure in the circular restricted three-body problem. This suggests that LDMD effectively captures the underlying manifold structure, as evidenced by its consistency with that derived solely from the discrete map. Consequently, the proposed method offers a potential tool for uncovering manifold structures from data. 
The GDMD approach utilizes time-series data distributed across the periapsis map, enabling the prediction of chaotic trajectories on a global scale. Notably, GDMD was shown to construct accurate discrete maps even with sparse data. To characterize the properties of these discrete maps, the variation of the semimajor axis, $\delta{a}$, as a function of the phase angle $\theta$ was analyzed and compared against results obtained from full trajectory propagation. The discrete maps generated via data-driven methods effectively approximate the underlying structure of chaotic behavior induced by the kick function, demonstrating their capability to predict periapsis transitions.
Finally, the proposed discrete mapping approach was applied to spacecraft trajectory design, demonstrating its practical utility in astrodynamics. By utilizing GDMD-based discrete maps, a targeting method was developed to identify initial conditions leading to a desired periapsis. This method successfully generated a trajectory that traverses the tube structure of the $L_1$ Lyapunov orbit and reaches the Moon. 

\textcolor{black}{Beyond predictive capability, the proposed framework offers clear advantages in computational efficiency, interpretability, and finite-time prediction stability. By replacing repeated numerical integration with algebraic map evaluations, chaotic transport can be analyzed and optimized at significantly reduced computational cost. Moreover, the learned linear operators admit a transparent spectral interpretation, providing direct geometric insight into phase-space deformation and transport pathways. By modeling transport through localized and global set deformations rather than single-trajectory linearization, the framework mitigates rapid error divergence in strongly chaotic regions and yields more stable finite-time predictions. Overall, these}
results highlight the potential of discrete maps not only for predicting chaotic transitions but also for optimizing trajectory design in multi-body dynamical environments.

\section*{Data availability}
The datasets supporting the findings of this study are available at Kyushu University Institutional Repository (QIR):  
\url{https://doi.org/10.48708/7348037}.

\section*{Acknowledgements}
The work of the first author was partially supported by JSPS KAKENHI Grant Numbers JP25K17776. 
The work of the third author was partially supported by Japan Science and Technology Agency, Fusion Oriented Research for Disruptive Science and Technology (JST FOREST Program), Grant Number JPMJFR206M and Grant-in-Aid for Scientific Research from the Japan Society for the Promotion of Science, Grant Number 22H03663.
The authors would like to thank Prof. Tomohiro Yanao for valuable discussions and insightful comments.

\section*{Author contributions statement}
T.U. and M.B. conceived the numerical method, and S.P. and M.B. conducted the dynamical system analysis. T.U. and S.P. performed the numerical analysis. Y.Y., H.C., and T.H. contributed to the discussion and interpretation of the results. All authors reviewed the manuscript.

\section*{Additional information}
\textbf{Competing interests}  
The authors declare no competing interests.

\bibliographystyle{spmpsci}
\bibliography{references}

@inbook{Prussing2013,
  title     = {Orbital Mechanics},
  edition   = {2nd},
  author    = {Prussing, John E. and Conway, Bruce A.},
  year      = {2013},
  publisher = {Oxford University Press},
  address   = {New York},
  chapter   = {4},
  pages     = {65--66}
}

@article{belbruno1993sun,
author = {Belbruno, Edward A. and Miller, James K.},
title = {Sun-perturbed Earth-to-moon transfers with ballistic capture},
journal = {Journal of Guidance, Control, and Dynamics},
volume = {16},
number = {4},
pages = {770-775},
year = {1993},
doi = {10.2514/3.21079}
}

@inproceedings{koon2000dynamical,
  author       = {Koon, Wang Sang and Lo, Martin W. and Marsden, Jerrold E. and Ross, Shane D.},
  title        = {Dynamical Systems, the Three-Body Problem and Space Mission Design},
  booktitle    = {International Conference on Differential Equations, Berlin, 1999},
  editor       = {Fiedler, B. and Gr{\"o}ger, K. and Sprekels, J.},
  series       = {Control and Dynamical Systems},
  publisher    = {World Scientific},
  year         = {2000},
  pages        = {1167--1181},
  note         = {Available at \url{https://www.gg.caltech.edu/~mwl/publications/papers/dynamicalThreeBody.pdf}}
}

@article{koon2001low,
author = {Koon, W. and Lo, Martin and Marsden, J. and Ross, Shane},
year = {2001},
month = {09},
pages = {},
title = {Low Energy Transfer to the Moon},
volume = {81},
isbn = {978-90-481-5865-2},
journal = {Celestial Mechanics and Dynamical Astronomy},
doi = {10.1023/A:1013359120468}
}

@article{Koon2000,
author = {Koon, Wang and Lo, Martin and Marsden, Jerrold and Ross, Shane},
year = {2000},
month = {07},
pages = {427-469},
title = {Heteroclinic Connections Between Periodic Orbits and Resonance Transitions in Celestial Mechanics},
volume = {10},
journal = {Chaos (Woodbury, N.Y.)},
doi = {10.1063/1.166509}
}

@inproceedings{Gomez2001,
  author       = {Gomez, G. and Koon, W. S. and Lo, M. W. and Marsden, J. E. and Masdemont, J. and Ross, S. D.},
  title        = {Invariant Manifolds, the Spatial Three-body Problem and Space Mission Design},
  booktitle    = {Proceedings of AIAA/AAS Astrodynamics Specialist Conference},
  series       = {AAS},
  number       = {01-301},
  address      = {Quebec City, Canada},
  month        = {August},
  year         = {2001}
}

@article{howell2001temporary,
  author    = {Howell, Kathleen C. and Marchand, Brian G. and Lo, Martin W.},
  title     = {Temporary Satellite Capture of Short-Period Jupiter Family Comets from the Perspective of Dynamical Systems},
  journal   = {Journal of Astronautical Sciences},
  volume    = {49},
  number    = {4},
  pages     = {539--557},
  year      = {2001},
  doi       = {10.1007/BF03546223},
  url       = {https://doi.org/10.1007/BF03546223}
}

@article{Villac2003,
author = {Villac, B. F. and Scheeres, D. J.},
title = {Escaping Trajectories in the Hill Three-Body Problem and Applications},
journal = {Journal of Guidance, Control, and Dynamics},
volume = {26},
number = {2},
pages = {224-232},
year = {2003},
doi = {10.2514/2.5062}
}

@article{Paskowitz2006A,
author = {Paskowitz, M. E. and Scheeres, D. J.},
title = {Robust Capture and Transfer Trajectories for Planetary Satellite Orbiters},
journal = {Journal of Guidance, Control, and Dynamics},
volume = {29},
number = {2},
pages = {342-353},
year = {2006},
doi = {10.2514/1.13761}}

@article{Paskowitz2006B,
author = {Paskowitz, Marci E. and Scheeres, Daniel J.},
title = {Design of Science Orbits About Planetary Satellites: Application to Europa},
journal = {Journal of Guidance, Control, and Dynamics},
volume = {29},
number = {5},
pages = {1147-1158},
year = {2006},
doi = {10.2514/1.19464}}

@article{ross2007multiple,
author = {Ross, Shane D. and Scheeres, Daniel J.},
title = {Multiple Gravity Assists, Capture, and Escape in the Restricted Three-Body Problem},
journal = {SIAM Journal on Applied Dynamical Systems},
volume = {6},
number = {3},
pages = {576-596},
year = {2007},
doi = {10.1137/060663374}
}

@article{hiraiwa2024designing,
  title = {Designing robust trajectories by lobe dynamics in low-dimensional Hamiltonian systems},
  author = {Hiraiwa, Naoki and Bando, Mai and Nisoli, Isaia and Sato, Yuzuru},
  journal = {Phys. Rev. Res.},
  volume = {6},
  issue = {2},
  pages = {L022046},
  numpages = {5},
  year = {2024},
  month = {May},
  publisher = {American Physical Society},
  doi = {10.1103/PhysRevResearch.6.L022046},
  url = {https://link.aps.org/doi/10.1103/PhysRevResearch.6.L022046}
}

@article{belbruno2008resonance,
title = {Resonance transitions associated to weak capture in the restricted three-body problem},
journal = {Advances in Space Research},
volume = {42},
number = {8},
pages = {1330-1351},
year = {2008},
issn = {0273-1177},
doi = {https://doi.org/10.1016/j.asr.2008.01.018},
url = {https://www.sciencedirect.com/science/article/pii/S027311770800080X},
author = {Edward Belbruno and Francesco Topputo and Marian Gidea}
}

@article{dellnitz2005transport,
author = {Dellnitz, Michael and JUNGE, OLIVER and KOON, WANG and LEKIEN, FRANCOIS and Lo, Martin and MARSDEN, JERROLD and Padberg-Gehle, Kathrin and PREIS, ROBERT and Ross, Shane and Thiere, Bianca},
year = {2005},
month = {03},
pages = {699-727},
title = {Transport in dynamical astronomy and multibody problems},
volume = {15},
journal = {International Journal of Bifurcation and Chaos},
doi = {10.1142/S0218127405012545}
}

@article{topputo2005earth,
  title={Earth-to-Moon Low Energy Transfers Targeting L1 Hyperbolic Transit Orbits},
  author={Topputo, Francesco and Vasile, Massimiliano and BERNELLI-ZAZZERA, FRANCO},
  journal={Annals of the New York Academy of Sciences},
  volume={1065},
  number={1},
  pages={55--76},
  year={2005},
  doi = {https://doi.org/10.1196/annals.1370.025}
}

@article{bonasera2021applying,
author = {Bonasera, Stefano and Bosanac, Natasha},
year = {2021},
month = {12},
pages = {},
title = {Applying data mining techniques to higher-dimensional Poincaré maps in the circular restricted three-body problem},
volume = {133},
journal = {Celestial Mechanics and Dynamical Astronomy},
doi = {10.1007/s10569-021-10047-3}
}

@inproceedings{urashi2021aas,
  title={Data-Driven Analysis of Chaotic Orbits in the Circular Restricted Three Body Problem},
  author={Urashi, Taiki and Bando, M and Hokamoto, S.},
  booktitle={AAS/AIAA Astrodynamics Specialist Conference},
  series={},
  volume={},
  year={2021},
  pages={AAS21--584},
  publisher={AAS},
  address={}
}

@article{liu1994chaotic,
  author    = {Liu, J. and Sun, Y. S.},
  title     = {Chaotic motion of comets in near-parabolic orbit: Mapping approaches},
  journal   = {Celestial Mechanics and Dynamical Astronomy},
  year      = {1994},
  volume    = {60},
  pages     = {3--28},
  doi       = {10.1007/BF00693090},
  url       = {https://doi.org/10.1007/BF00693090}
}

@article{malyshkin1999keplerian,
title = {The Keplerian Map for the Planar Restricted Three-Body Problem as a Model of Comet Evolution},
journal = {Icarus},
volume = {141},
number = {2},
pages = {341-353},
year = {1999},
issn = {0019-1035},
doi = {https://doi.org/10.1006/icar.1999.6174},
url = {https://www.sciencedirect.com/science/article/pii/S0019103599961742},
author = {Leonid Malyshkin and Scott Tremaine}
}

@article{chirikov1989chaotic,
  author  = {Chirikov, R. V. and Vecheslavov, V. V.},
  title   = {Chaotic dynamics of Comet Halley},
  journal = {Astronomy and Astrophysics},
  volume  = {221},
  pages   = {146--154},
  year    = {1989}
}

@article{boekholt2016origin,
   title={The origin of chaos in the orbit of comet 1P/Halley},
   volume={461},
   ISSN={1365-2966},
   url={http://dx.doi.org/10.1093/mnras/stw1504},
   DOI={10.1093/mnras/stw1504},
   number={4},
   journal={Monthly Notices of the Royal Astronomical Society},
   publisher={Oxford University Press (OUP)},
   author={Boekholt, T. C. N. and Pelupessy, F. I. and Heggie, D. C. and Portegies Zwart, S. F.},
   year={2016},
   month=jun, pages={3576–3584} }

@article{pathak2017using,
   title={Using machine learning to replicate chaotic attractors and calculate Lyapunov exponents from data},
   volume={27},
   ISSN={1089-7682},
   url={http://dx.doi.org/10.1063/1.5010300},
   DOI={10.1063/1.5010300},
   number={12},
   journal={Chaos: An Interdisciplinary Journal of Nonlinear Science},
   publisher={AIP Publishing},
   author={Pathak, Jaideep and Lu, Zhixin and Hunt, Brian R. and Girvan, Michelle and Ott, Edward},
   year={2017},
   month=dec }

@article{HOFMANN2025314,
title = {Analytical and data-driven Koopman operator for the perturbed Kepler and circular restricted three-body problems},
journal = {Acta Astronautica},
volume = {234},
pages = {314-328},
year = {2025},
issn = {0094-5765},
doi = {https://doi.org/10.1016/j.actaastro.2025.04.034},
url = {https://www.sciencedirect.com/science/article/pii/S0094576525002346},
author = {Christian Hofmann and Giovanni Lavezzi and Di Wu and Simone Servadio and Richard Linares}
}

@book{kutz2016dynamic,
  title={Dynamic mode decomposition: data-driven modeling of complex systems},
  author={Kutz, J Nathan and Brunton, Steven L and Brunton, Bingni W and Proctor, Joshua L},
  year={2016},
  publisher={SIAM}
}

@article{Wustner2025,
  author = {W{\"u}stner, Daniel and Gundestrup, Henrik Helge and Thaysen, Katja},
  title = {Dynamic mode decomposition for analysis and prediction of metabolic oscillations from time-lapse imaging of cellular autofluorescence},
  journal = {Scientific Reports},
  year = {2025},
  volume = {15},
  number = {1},
  article = {23489},
  doi = {10.1038/s41598-025-07255-4},
  url = {https://doi.org/10.1038/s41598-025-07255-4}
}

@misc{kutz2015,
      title={Multi-Resolution Dynamic Mode Decomposition}, 
      author={J. Nathan Kutz and Xing Fu and Steven L. Brunton},
      year={2015},
      eprint={1506.00564},
      archivePrefix={arXiv},
      primaryClass={math.DS},
      url={https://arxiv.org/abs/1506.00564}, 
}

@misc{grosek2014,
      title={Dynamic Mode Decomposition for Real-Time Background/Foreground Separation in Video}, 
      author={Jacob Grosek and J. Nathan Kutz},
      year={2014},
      eprint={1404.7592},
      archivePrefix={arXiv},
      primaryClass={cs.CV},
      url={https://arxiv.org/abs/1404.7592}, 
}

@article{Wustner2022,
  author       = {W{\"u}stner, Daniel},
  title        = {Dynamic Mode Decomposition of Fluorescence Loss in Photobleaching Microscopy Data for Model-Free Analysis of Protein Transport and Aggregation in Living Cells},
  journal      = {Sensors},
  year         = {2022},
  volume       = {22},
  number       = {13},
  pages        = {4731},
  doi          = {10.3390/s22134731}
}

@article{campagnola2012three,
author = {Campagnola, Stefano and Kawakatsu, Yasuhiro},
title = {Three-Dimensional Resonant Hopping Strategies and the Jupiter Magnetospheric Orbiter},
journal = {Journal of Guidance, Control, and Dynamics},
volume = {35},
number = {1},
pages = {340-344},
year = {2012},
doi = {10.2514/1.53334}
}

@article{Haapala2011,
author = {Haapala, Amanda and Howell, Kathleen},
year = {2011},
month = {01},
pages = {415-434},
title = {Trajectory design using periapse Poincaré maps and invariant manifolds},
volume = {140},
journal = {Advances in the Astronautical Sciences}
}

@article{Scott2010,
author = {Scott, Christopher J. and Spencer, David B.},
title = {Transfers to Sticky Distant Retrograde Orbits},
journal = {Journal of Guidance, Control, and Dynamics},
volume = {33},
number = {6},
pages = {1940-1946},
year = {2010},
doi = {10.2514/1.47792}
}

@ARTICLE{Oshima2015,
       author = {{Oshima}, Kenta and {Yanao}, Tomohiro},
        title = "{Jumping mechanisms of Trojan asteroids in the planar restricted three- and four-body problems}",
      journal = {Celestial Mechanics and Dynamical Astronomy},
         year = 2015,
        month = may,
       volume = {122},
       number = {1},
        pages = {53-74},
          doi = {10.1007/s10569-015-9609-4},
       adsurl = {https://ui.adsabs.harvard.edu/abs/2015CeMDA.122...53O}
}

@book{wiggins2013chaotic,
  title={Chaotic transport in dynamical systems},
  author={Wiggins, Stephen},
  volume={2},
  year={2013},
  publisher={Springer Science \& Business Media}
}

@book{scheeres2016orbital,
  title={Orbital motion in strongly perturbed environments: applications to asteroid, comet and planetary satellite orbiters},
  author={Scheeres, Daniel J},
  year={2016},
  publisher={Springer}
}

@article{bollt1995targeting,
  title={Targeting chaotic orbits to the Moon through recurrence},
  author={Bollt, Erik M and Meiss, James D},
  journal={Physics Letters A},
  volume={204},
  number={5-6},
  pages={373--378},
  year={1995},
  publisher={Elsevier}
}

@article{servadio2022dynamics,
  title={Dynamics near the three-body libration points via Koopman operator theory},
  author={Servadio, Simone and Arnas, David and Linares, Richard},
  journal={Journal of Guidance, Control, and Dynamics},
  volume={45},
  number={10},
  pages={1800--1814},
  year={2022},
  publisher={American Institute of Aeronautics and Astronautics}
}

@article{servadio2023koopman,
  title={Koopman-operator control optimization for relative motion in space},
  author={Servadio, Simone and Armellin, Roberto and Linares, Richard},
  journal={Journal of Guidance, Control, and Dynamics},
  volume={46},
  number={11},
  pages={2121--2132},
  year={2023},
  publisher={American Institute of Aeronautics and Astronautics}
}

\end{document}